\documentclass[11pt,a4paper]{article}

\usepackage[margin=1in]{geometry}
\usepackage[T1]{fontenc}
\usepackage[utf8]{inputenc}
\usepackage{lmodern}
\usepackage{amsmath,amssymb,amsthm}
\usepackage{booktabs}
\usepackage{longtable}
\usepackage{array}
\usepackage{xcolor}
\usepackage{enumitem}
\usepackage{listings}
\usepackage[obeyspaces,spaces]{xurl}
\usepackage[hidelinks]{hyperref}
\usepackage{microtype}
\usepackage{etoolbox}

\makeatletter
\@ifpackagelater{longtable}{2025/12/22}{}{%
  \patchcmd{\LT@output}{\vss}
    {\vskip\z@\@plus1fil\@minus\normalbaselineskip}{}%
    {\PackageError{stark-report}{Cannot patch longtable output}{Update longtable to v4.25 or later.}}%
  \patchcmd{\LT@output}{\vss}
    {\vskip\z@\@plus1fil\@minus\normalbaselineskip}{}%
    {\PackageError{stark-report}{Cannot patch longtable output}{Update longtable to v4.25 or later.}}%
}
\makeatother

\definecolor{linkblue}{HTML}{1F4E79}
\definecolor{lightgray}{HTML}{F5F5F5}

\hypersetup{
  pdftitle={Isabelle/STARK: A Formalization of a STARK-Style Protocol in Isabelle/HOL},
  pdfauthor={Diego Marmsoler},
  colorlinks=true,
  linkcolor=linkblue,
  citecolor=linkblue,
  urlcolor=linkblue
}

\lstdefinelanguage{Isabelle}{
  morekeywords={
    theory,imports,begin,end,definition,lemma,theorem,locale,assumes,shows,
    fixes,where,proof,qed,by,using,unfolding
  },
  sensitive=true,
  morecomment=[s]{(*}{*)}
}
\newcommand{\code}[1]{\begingroup\Urlmuskip=0mu plus 1mu\nolinkurl{#1}\endgroup}
\newcommand{\formalanchors}[1]{%
  \begingroup
  \scriptsize
  \begin{longtable}{
    >{\raggedright\arraybackslash}p{0.27\linewidth}
    >{\raggedright\arraybackslash}p{0.40\linewidth}
    >{\raggedright\arraybackslash}p{0.24\linewidth}}
  \toprule
  Concept & Isabelle theory & Item \\
  \midrule
  \endfirsthead
  \toprule
  Concept & Isabelle theory & Item \\
  \midrule
  \endhead
  #1
  \bottomrule
  \end{longtable}
  \endgroup
}
\newcommand{\theorysource}[1]{\code{#1}}
\newcommand{\anchorrow}[3]{#1 & \theorysource{#2} & \code{#3}\\}

\title{Isabelle/STARK: A Formalization of a STARK-Style Protocol in Isabelle/HOL}
\author{Diego Marmsoler}
\date{2026}

\begin{document}
\maketitle
\begin{abstract}
This report presents mechanized query-bounded soundness and honest completeness
for a STARK-style protocol in Isabelle/HOL. Native PMFs, SPMFs and CryptHOL
oracle computations support a thin probabilistic state interface. Independent
algebra/FRI and Merkle-collision libraries connect to a concrete Fiat--Shamir
compiler and STARK composition proof. An acceptance-preserving reduction carries
adaptive, privately randomized transcript producers to compiled staged
experiments; actual event coverage and component bounds then establish the
final acceptance bound. Exact modulo fibers and explicit query accounting
retain the specified sampler and count repeated calls. For a certified 192-bit
field, trace length 1024, scale 64 and 640 query repetitions, false square-workload
endpoints have acceptance probability at most $2^{-137}$ under
\code{fs_query_bound (2^20) P}. Separate cutoff proofs certify comparison against
every natural cutoff pair and the selected concrete pair $(10581,10581)$.
The actual absorbing-RO honest experiment accepts a correct endpoint with
probability one. The guarantees concern fixed statements in the classical,
typed, field-valued random-oracle model; they establish neither unrestricted
work-factor security, zero knowledge, knowledge extraction nor QROM security.
The current narrative and thirteen-statement manifest are separated from a
historical appendix. A deterministic executable companion has proved
fixed-hash correspondence, without a concrete-hash security claim.
\end{abstract}
\tableofcontents

\section{Introduction}
\label{sec:introduction}

Scalable transparent arguments, and STARKs in particular
\cite{ben-sasson2018stark}, reduce claims about program executions to claims
about low-degree polynomials over finite fields. A prover commits to trace,
composition and FRI evaluation tables; a verifier checks positions selected
by transcript-dependent challenges. Algebra, authentication, adaptive oracle
interaction and sampling therefore all contribute to the soundness argument.

This Isabelle/HOL development \cite{nipkow2002isabelle,isabelle2025}
formalizes a representative STARK-style protocol and proves query-bounded
soundness and honest completeness. The accompanying source snapshot is the
artifact described here; the repository reference
\cite{marmsoler2026isabellestark} identifies the project, without asserting a
particular hosting or release status. The report distinguishes current
theorems from earlier investigations retained in
Appendix~\ref{app:history}.

The principal concrete statement is
\code{square_192_fs_137_incorrect_endpoint}. For fixed endpoints $a,z$ in the
certified 192-bit field, its two premises are
$z\ne a^{2^{1023}}$ and \code{fs_query_bound (2^20) P}. It bounds the
producer's false-endpoint acceptance probability by $2^{-137}$. The concrete
interpretation discharges the generic field, domain and workload assumptions.
The profile has 1024 trace values, hence 1023 squarings, scale 64 and 640 query
repetitions; repetitions are not FRI folding depth.

The generic endpoint \code{soundness.fs_optimized_soundness} selects cutoffs
for the existing first-root error family. Separate proofs establish comparison
against every natural cutoff pair, including the parameter fallbacks, and
certify the lexicographically selected pair $(10581,10581)$ for the concrete
profile. These are optimality statements about a proved upper bound, not about
actual attack probability or optimal protocol parameters.

Honest correctness has two related interfaces. Generic
\code{verification.completeness} proves zero failure for the original
prover/verifier execution under honest trace validity. The separate
\code{square_ro_honest_experiment_acceptance_one} theorem proves acceptance one
for the actual absorbing-random-oracle honest experiment when
$z=a^{2^{1023}}$. The latter has no malicious query-cap premise or claim that
the honest callbacks satisfy a uniform all-state resource allowance.

The current development contributes:
\begin{itemize}[leftmargin=*]
  \item A precise prover/verifier and typed, domain-separated, field-valued
    random-oracle model, with cached answers and exact modulo query sampling.
  \item Independent algebra and FRI agreement/counting results, together with
    concrete proofs connecting accepted STARK executions to those interfaces.
  \item Generic collision extraction from inconsistent certified Merkle
    openings and its oracle-execution reduction, with explicit interaction costs.
  \item Native PMF/SPMF and CryptHOL-based execution, observation, replay and
    accounting contracts. A thin state interface replaces a parallel custom
    probability foundation.
  \item An acceptance-preserving reduction from adaptive, privately randomized
    FS producers to compiled staged experiments, followed by actual event-cover
    and component-bound proofs.
  \item A certified square application, automatic cutoff guarantees and a
    separate deterministic executable companion with proved native correspondence.
\end{itemize}

The four principal proof components are algebra/FRI, Merkle authentication,
Fiat--Shamir and composition. Native probability and deterministic evidence
libraries support these components. Isabelle and AFP libraries\cite{afp2026}
supply the underlying probability, oracle and algebraic infrastructure. Independence means that an interface does
not assume the concrete STARK locale; it does not mean that the concrete client
can omit authentication, chronology, domain or parameter obligations.

The guarantees are fixed-statement results in a classical random-oracle model.
Private samples in the restricted adversary interface have finite support,
and the syntactic cap bounds all continuations, counting repeated oracle calls.
The native probability libraries themselves also admit broader computations.
Neither zero knowledge, knowledge extraction, QROM security nor security of a
deployed bit-hash implementation is proved. The acceptance bound is not an
unrestricted 137-bit work-factor statement.

Sections~\ref{sec:architecture} and~\ref{sec:concrete} explain the reduction
and certified application. Section~\ref{sec:honest-ro} gives the actual honest
RO route. Appendix~\ref{app:statements} records the thirteen-statement manifest
with its exported premises. Source anchors identify the current theories and
declarations by name. Some qualified names reflect earlier theory ownership.
Appendix~\ref{app:history} describes earlier proof routes and diagnostics.

\formalanchors{
\anchorrow{Protocol definitions}{Core/Stark.thy}{stark; prover; verifier}
\anchorrow{Generic completeness}{Completeness/Completeness_Verifier.thy}{verification.completeness}
\anchorrow{Generic optimized soundness}{Soundness/FS_Optimized_Soundness.thy}{soundness.fs_optimized_soundness}
\anchorrow{Concrete 137-bit endpoint}{Soundness/FS_Square_192_137.thy}{square_192_fs_137_incorrect_endpoint}
\anchorrow{Publication statement checks}{Soundness/Publication_Manifest.thy}{thirteen exported statements}
}

\section{Mathematical and Cryptographic Background}

\subsection{The STARK reduction}

A STARK protocol starts with a computation trace.  In the algebraic view, this
trace is encoded as a polynomial over a finite field.  Correctness of the
computation is expressed by polynomial constraints: boundary constraints
describe initial or final values, and transition constraints describe how one
row of the trace relates to later rows.  Instead of checking the constraints at
all positions, the verifier samples positions and checks consistency there.

The protocol uses a composition polynomial to combine many constraints into one
object.  Constraint quotients divide constraint polynomials by vanishing
polynomials for their required roots.  If the trace is valid, the constraints
vanish on the declared roots, so the quotient is a polynomial.  Random linear
combination by Fiat-Shamir challenges then turns the family of constraints into
a single composition polynomial.  Soundness depends on the fact that an invalid
trace can satisfy these checks only with bounded probability.

Merkle trees are used to commit to evaluation tables.  The prover sends roots
of committed tables, and later reveals values and authentication paths at
sampled indices.  Binding is modeled through hash-map consistency and collision
events.  The verifier never receives full tables; it receives only sampled
openings and checks them against committed roots.

Low-degree testing is represented by a FRI-style folding protocol
\cite{ben-sasson2018fri,ben-sasson2018stark}.  The prover commits to successive
folded layers, the verifier samples positions, and each query checks a local
fold equation between adjacent layers.  At the final layer the verifier checks
that the committed value is constant.  The formal soundness theorem closes the
FRI contribution into adaptive challenge, remainder-aware residual-query, and
padding-consistency arithmetic terms; no semantic event probability remains in
the final bound.

The reduction can be viewed as a sequence of claims about increasingly compact
objects.  The original statement is a claim that there exists a valid trace.
Interpolation turns the trace into one or more low-degree polynomials.  The
transition and boundary constraints are then polynomial identities over a
finite domain.  After division by the appropriate vanishing polynomials, the
constraint checks become degree claims about quotient polynomials.  Finally,
random linear combination compresses these quotient polynomials into the
single composition polynomial used by the verifier.

The verifier cannot inspect this composition polynomial directly.  Instead, it
receives commitments to evaluation tables.  The soundness argument must then
connect three views of the same data: the algebraic view, in which a low-degree
polynomial has bounded agreement with any distinct low-degree polynomial; the
commitment view, in which Merkle roots bind the prover to table values except
with collision probability; and the transcript view, in which Fiat-Shamir
challenges are sampled after the data that they are supposed to test has been
committed.  The Isabelle proof follows this separation closely.  It first
extracts verifier-observed evidence from accepting executions, then classifies
acceptance of a false statement into bad events, and finally applies
probabilistic bounds to those events.

\subsection{Constraint quotients}

For a constraint polynomial \(C(X)\) that is required to vanish on a set
\(R\), the usual algebraic object is the quotient
\[
  Q(X) = C(X) / \prod_{r \in R} (X-r).
\]
This notation requires care in Isabelle/HOL. Polynomial division and field
division are both total operations; in particular, field division satisfies
\(a/0=0\). Nevertheless, evaluation of a polynomial quotient need not equal
pointwise division at a zero of the denominator. The formalization therefore
keeps the polynomial operation and the pointwise field operation separate.

If the vanishing condition holds, the denominator polynomial divides the
constraint polynomial, and the quotient is again a polynomial.  Once the
quotient polynomial has been constructed, it can be evaluated at any field
point.  What the verifier must avoid is evaluating the denominator expression
at a root while reconstructing quotient values from constraint values.  This
is one reason the protocol evaluates on a shifted coset disjoint from the
constraint roots.

\subsection{Random linear combination}

Let \(Q_1,\ldots,Q_m\) be quotient polynomials derived from the constraints.
The composition polynomial is a randomized linear combination such as
\[
  Q_1 + \alpha Q_2 + \alpha^2 Q_3 + \cdots
\]
or a closely related challenge-indexed combination.  A false trace may satisfy
some constraints at some points, but if the quotient family does not match a
valid low-degree object, then only a small set of challenge choices can hide
the disagreement.  The formal soundness proof expresses this by defining
bad-alpha sets and bounding the probability that the Fiat-Shamir output lands
in them.

In an interactive protocol, the verifier would sample alpha after receiving
the relevant commitments.  In the noninteractive Fiat-Shamir setting, alpha is
derived from the transcript by a random oracle \cite{fiat1986prove}.  The staged security
experiment preserves the essential ordering: the set of bad alpha values is
determined by a prefix of the transcript and oracle state before the alpha
output that is charged to the target-hit budget.

\subsection{Merkle binding}

Merkle commitments reduce table binding to collision resistance of the hash
function \cite{merkle1987signature}.  If a fixed root authenticates two
different values at the same table position under the same path shape, then
some internal hash node must collide.  The formalization represents this
property through hash-map output collisions rather than by assuming an
external cryptographic theorem.  This keeps the proof phrased in the same
probabilistic model as the Fiat-Shamir oracle.

The verifier sees only the opened positions.  Therefore Merkle soundness in
this development is not a theorem about reconstructing complete tables from
the proof.  It is a theorem about consistency of authenticated partial
openings.  This distinction is important throughout the soundness argument and
is revisited in Section~\ref{sec:proof-design}.

\subsection{FRI at a proof-interface level}

FRI is the low-degree test used by the protocol.  Each FRI round folds an
evaluation table into a smaller table using a challenge value.  A query checks
that two sibling values in one layer and one value in the next layer satisfy
the fold equation.  Repeating this over several rounds reduces a claim about a
large evaluation table to a final constant check.

The present formalization contains a detailed executable verifier for these
checks and closes the soundness endpoint using proved actual-query, target,
product, and relation-fiber bounds.  The report does not claim that the
constants are optimized.  The important point is that the final theorem has a
parameter-only right-hand side and the proof route uses verifier-observed
partial openings, challenge lists, query lists, and local fold checks rather
than an informal full-table reconstruction argument.

\subsection{Finite fields and Proth domains}

The protocol is formalized over fields satisfying a Proth-field class
constraint.  Proth fields are useful for STARKs because the multiplicative
group has a large power-of-two factor.  This allows the development to define
trace and evaluation domains as power-of-two multiplicative subgroups.  The
main primitive generator is denoted by \(\omega\).  From it, the formalization
derives a trace-domain generator \(g\) and an evaluation-domain generator
\(h\).

The trace domain is used for interpolation of the computation trace.  The
larger evaluation domain is used for low-degree testing and query checks.  The
evaluation points in the protocol are not just the subgroup generated by \(h\);
they are shifted by a nonzero field element.  Thus typical evaluation points
have the form
\[
  h^i \cdot \mathit{shift}.
\]
The shift is not cosmetic.  Constraint quotients have denominators whose roots
lie on the trace domain.  Evaluating a quotient expression directly on those
roots would expose division by zero at the level of field evaluation, even when
the polynomial quotient itself is well-defined.  Evaluating the same polynomial
on a disjoint shifted coset avoids these roots.  Since the low-degree property
is a property of the polynomial, not of a particular naming of the evaluation
domain, checking on a suitably large shifted domain is compatible with the
STARK reduction.

The important distinction is between polynomial division and pointwise field
division.  Once Isabelle has established that the denominator polynomial
divides the numerator polynomial, the quotient is a total polynomial object.
That polynomial may be evaluated at every field point.  By contrast, the
expression obtained by substituting a point into numerator and denominator and
then dividing in the field is also total, with value zero when the denominator
is zero. That convention does not justify recovering the evaluated polynomial
quotient there. Nonzero denominators on the shifted domain justify the identity
between committed quotient values and the verifier's pointwise divisions.

Power-of-two domain geometry also aligns with FRI.  Each FRI round halves the
logical domain, so the theory needs exact order, divisibility, and index-range
facts to justify every sibling lookup and next-layer lookup.  These facts are
mathematically routine but proof-relevant: Isabelle requires explicit evidence
that the queried list positions exist, that powers stay in the expected
subgroup, and that the fold equation refers to the same domain points as the
verifier's computation.

\formalanchors{
\anchorrow{Exact multiplicative order}{Core/Stark.thy}{exact_order}
\anchorrow{Trace-domain generator}{Core/Stark.thy}{g}
\anchorrow{Evaluation-domain generator}{Core/Stark.thy}{h}
\anchorrow{Shifted evaluation domain}{Core/Stark.thy}{eval_domain}
\anchorrow{Trace powers}{Core/Stark.thy}{trace_powers_of}
}

\section{Isabelle Formal Model}

\subsection{Sessions and theories}

The daily build selects twelve local Isabelle sessions, separating independent
support libraries from the concrete protocol, its proofs and the deterministic
executable companion. A thirteenth, optional session generates the native proof
document when explicitly requested.

\begin{center}
\scriptsize
\begin{tabular}{
  >{\raggedright\arraybackslash}p{0.43\linewidth}
  >{\raggedright\arraybackslash}p{0.49\linewidth}}
\toprule
Sessions & Responsibility \\
\midrule
\code{Polynomial_Products}, \code{Code_Proximity_Library},
\code{FRI_Mathematics} & Independent polynomial, domain, agreement, distance,
folding and counting mathematics.\\
\code{Deterministic_Evidence} & Finite-map, lookup-path and collision
certificates, declared in \code{Evidence/ROOT}.\\
\code{Evidence_Probability} & Native state interfaces, support and observation
transport, iteration, mixtures, replay and probability bounds.\\
\code{Oracle_Accounting} & CryptHOL oracle recording, replay, query accounting
and generic Merkle collision reduction.\\
\code{Stark_Core} & Typed hash/channel state, Merkle trees, the GF(5) field and
concrete prover/verifier definitions.\\
\code{Stark_Polynomial_Compatibility} & Bridge from the independent polynomial
library to the existing protocol vocabulary.\\
\code{Stark_Completeness} & Honest-valid-trace completeness for the core
prover/verifier execution.\\
\code{Stark_Examples} & Field certification, square workloads and the small-field
instance.\\
\code{Stark} & Concrete staged/FS soundness, cutoff optimization, the published
numerical profile, actual honest RO completeness and the statement manifest.\\
\code{Stark_Execution} & Separate pure deterministic interpreter, native
correspondence and generated SML for GF(5) and prime192.\\
\code{Stark_Publication} & Optional native proof document, declared separately
in \code{Publication/ROOT}.\\
\bottomrule
\end{tabular}
\end{center}

The protocol entry point is \theorysource{Core/Stark.thy}. It defines the main
locales and prover/verifier programs; completeness and soundness extend this
vocabulary with proof predicates. Historical qualified names sometimes remain
even when their physical owner is now an independent library.

The \code{Stark} session imports \code{Publication_Manifest}, which checks and
exports thirteen selected statements, including first-root and optimized
soundness, cutoff comparison and the concrete \(2^{-137}\) endpoints. The optional
\code{Stark_Publication} session generates the selected native proof document;
execution is a separate leaf session excluded from it. \code{ROOT} and
\code{Evidence/ROOT} determine daily build membership; \code{Publication/ROOT}
adds the explicitly requested document session.

\subsection{Native probability and state}

The probability foundation is Isabelle's native PMF/SPMF infrastructure.
PMFs describe total sampling; SPMFs describe computations that may fail.
Ordinary stateful computations use the thin interface in
\theorysource{Probability/Native_State_Interface.thy}:
\begin{lstlisting}
type_synonym ('a,'s) state_monad = "'s \<Rightarrow> ('a \<times> 's) spmf"
\end{lstlisting}
The familiar qualified combinator names are retained in the
\code{Prob_Monad} namespace. The theory
\theorysource{Core/Prob_Monad.thy} is a compatibility import, not a separate
distribution implementation. Return, sequencing, state reads and updates,
assertions, sampling, list traversal and iteration are implemented directly
with native PMF/SPMF operations. Oracle-interacting FS programs additionally
have a CryptHOL GPV interpretation; ordinary state programs need not be GPVs.

Native SPMFs are represented by PMFs over optional successful values. Thus
\code{m s} can also be observed as the full PMF over
\code{('a * 's) option}: \code{Some (x,t)} records a returned value and final
state, while \code{None} records failure. Its mass at \code{None} is the
missing mass of the successful SPMF. This is one failure layer, not a
successful computation that returns another optional result. This distinction
matters for statements about malformed transcripts, verifier rejection and
program equivalence that observes failure.

The weakest-precondition interface uses standard scalar types. For a
nonnegative post-expectation \(Q\),
\[
 \operatorname{wp}(m,Q,s)
   =\int^+ o.\,Q(o)\,d\operatorname{measure\_pmf}(m(s))
   \quad\text{has type }\mathit{ennreal}.
\]
For an event \(P\), \code{wp_event m P s} is its real-valued probability
under that same full option PMF. The real-valued
\code{failure_prob m s} is \code{pmf (m s) None}. In particular, acceptance
uses an event that explicitly excludes \code{None}; the integral of the
constant-one function over the full PMF is always one and does not measure
success. Error bounds and the final acceptance probabilities use \code{real},
without the earlier custom probability scalar or conversion layer.

The native types admit countably supported distributions. The original
finite-support restrictions therefore live at their intended interfaces,
rather than in a replacement distribution subtype. The state predicate
\code{finite_program m} requires finite successful support in every initial
state. The FS query restriction requires \code{finite (set_pmf d)} for each
private \code{FS_Sample d k}, and imposes the query restriction on every
continuation, including continuations outside the sample's support. Private
sampling itself is total because it uses a PMF; an adversary may nevertheless
fail explicitly. No blanket losslessness condition is imposed on attackers.

The retained WP rules mirror the program constructs. Bind rules expose an
intermediate value and state; assertion rules expose Boolean checks; and
specialized traversal and iteration rules carry the relevant support or
probability invariant through a list or loop. Completeness first excludes
\code{None} from the actual execution support and then obtains failure
probability zero. Soundness transports observed distributions and bounds
events in those executions. A successful-outcome inclusion alone is not
substituted for a distribution equality or a failure-sensitive observation.
This keeps the probabilistic interfaces separate from the subsequent algebra,
authentication and transcript-replay arguments.

\formalanchors{
\anchorrow{Native state and probability interfaces}{Probability/Native_State_Interface.thy}{state_monad; wp; wp_event; failure_prob; finite_program}
\anchorrow{Private sampling and uniform syntactic cap}{Soundness/FS_Program.thy}{fs_program; fs_query_bound}
\anchorrow{Native oracle interpretation}{Soundness/FS_Native_Execution.thy}{fs_gpv; fs_run; fs_gpv_interaction_bound}
}

\subsection{Channel, transcript, and random oracle}

The channel model records the transcript and the hash map used by Fiat-Shamir
queries.  The prover sends field elements to the transcript.  The verifier
reads the transcript in order.  Random field elements are obtained from
hash-map lookups or fresh hash extensions.  In the honest-completeness proof,
the central deterministic fact is transcript replay: when the verifier is run
on the transcript produced by the honest prover, each read obtains the value
previously sent by the prover.

In the current soundness experiment, recording a staged transcript message is
itself a random-oracle operation.  The function
\code{ro_record_staged_message} queries the oracle at
\code{TranscriptAbsorb (PState s) x}, updates the transcript state to the hash
output, and appends \code{x}.  The constructor \code{TranscriptAbsorb} provides
syntactic domain separation from the other oracle operations.  The verifier's
\code{ro_verify_monad} likewise uses the absorbing channel operations.  The
staged transcript budget counts these absorption queries, and the combined
verifier budget includes them in
\code{ro_absorb_checked_staged_security_hash_query_budget_for}.

This absorbing experiment must be distinguished from the older nonabsorbing
verifier route.  The latter remains useful as lower-level proof infrastructure,
but it is not the public security experiment documented here.  In the current
experiment an adversary constructs a transcript and extends the shared oracle;
the verifier then runs on that transcript with verifier-local counters reset.
The adversary controls the proof data, but the verifier's execution state is
not inherited wholesale from the adversary.

The channel state records more than a list of messages.  It also tracks the
current read position and the Fiat-Shamir bookkeeping needed to generate
challenge keys.  Resetting these verifier-local fields before verification is
therefore not just a cosmetic cleanup.  It expresses the security experiment:
the submitted transcript and random-oracle map are shared, while the verifier's
program counter, transcript cursor, and challenge counters start from their
canonical verifier values.

\subsection{Merkle commitments}

The Merkle-tree model provides creation and checking of authenticated paths.
The prover commits to trace, composition, and FRI evaluation tables by sending
Merkle roots.  The verifier later checks sampled openings against those roots.
The soundness proof works with authenticated partial openings, because those
are exactly what the verifier observes.

The Merkle interface is deterministic once the hash map is fixed.  Creation
extends the map with node hashes, and checking follows an authentication path
from a leaf to a claimed root.  Completeness uses the correctness direction:
paths produced from a created tree verify.  Soundness uses the binding
direction: incompatible authenticated openings under a fixed oracle map imply
a hash-output collision.  The latter is phrased as an event so that it can be
charged to the adversary's collision budget.

\subsection{Protocol locales}

The core \code{stark} locale fixes the domain, field, query, and specification
parameters.  The \code{prover} locale adds the concrete trace values and the
prover-side polynomial and commitment definitions.  The \code{verifier} locale
adds the verifier-side specification and defines the verifier monad.  The
completeness locale \code{verification} combines a prover and verifier, while
the soundness locale \code{soundness} is verifier-oriented and adds the
parameters needed for the staged security experiment.

The locale split is also a dependency-management device.  Domain facts that
both honest and adversarial executions need belong in \code{stark}.  Facts
about a particular honest trace belong in the completeness layer.  Facts about
adversarial budgets, target events, and bad-event partitions belong in the
soundness layer.  This keeps public theorem premises small: users of the final
soundness theorem do not need to supply intermediate event bounds that are
proved by the internal proof route.

The Core definitions fix the protocol vocabulary. The current FS and staged
experiments share \code{ro_verify_monad}; the generic core completeness
execution instead uses \code{verify_monad} with its pure-concat channel.
Actual-RO honest completeness separately connects the honest producer to
the shared absorbing verifier. The soundness layers describe evidence
extracted from that actual RO execution.  This is particularly visible in the soundness proof,
where accepted transcripts are related to partial Merkle openings, query
indices, FRI challenge lists, and final checks obtained from the actual
verifier program.

\formalanchors{
\anchorrow{State-monad assertion}{Probability/Native_State_Interface.thy}{assert}
\anchorrow{Event WP}{Probability/Native_State_Interface.thy}{wp_event}
\anchorrow{Failure probability}{Probability/Native_State_Interface.thy}{failure_prob}
\anchorrow{WP rules}{Probability/Native_State_Interface.thy}{wp_return; wp_bind}
\anchorrow{Transcript send}{Core/Channel_Core.thy}{send}
\anchorrow{Transcript read}{Core/Channel_Core.thy}{read}
\anchorrow{Domain-separated absorption}{Core/Channel_Core.thy}{TranscriptAbsorb}
\anchorrow{Staged message absorption}{Soundness/Staged_Security_Experiment_Core.thy}{ro_record_staged_message}
\anchorrow{Absorbing verifier}{Soundness/Staged_Security_Experiment_RO_Checked.thy}{ro_verify_monad}
\anchorrow{Transcript query budget}{Soundness/Staged_Security_Experiment_RO_Transcript_Budgets.thy}{ro_checked_staged_transcript_hash_query_budget_for}
\anchorrow{Merkle creation}{Core/Merkle_Tree.thy}{create}
\anchorrow{Protocol locale}{Core/Stark.thy}{stark}
\anchorrow{Prover locale}{Core/Stark.thy}{prover}
\anchorrow{Verifier locale}{Core/Stark.thy}{verifier}
\anchorrow{Prover program}{Core/Stark.thy}{prover_monad}
\anchorrow{Verifier program}{Core/Stark.thy}{verify_monad}
}

\section{Domain and Algebraic Well-Formedness}

The core locale contains the domain assumptions needed by both completeness and
soundness.  These assumptions specify that the primitive generator has the
right exact order, that the trace and evaluation lengths divide the field
group order, that the evaluation domain is nontrivial, and that the shift is
nonzero and separates the evaluation coset from the trace roots.

The current query sampler draws a field element, maps it through the
finite-field enumeration function \code{to_nat}, and reduces the resulting
natural number modulo the query-space size.  The locale assumption
\code{to_nat_range} identifies the raw encodings with \([0,N)\), where \(N\)
denotes \code{size}.  No divisibility assumption is imposed.  Instead, for
\(B\subseteq[0,q)\), where \(q\) denotes
\code{query_sample_space_size}, Isabelle proves the exact formula
\[
  |\{n<N\mid n\bmod q\in B\}|
  = \lfloor N/q\rfloor |B| + |B\cap[0,N\bmod q)|.
\]
The corresponding cardinality envelope for a set of size \(k\) is
\(\lfloor N/q\rfloor k+\min(k,N\bmod q)\).  Thus the formal protocol keeps the
concrete modulo sampler and accounts for its bias; exact uniformity is only a
diagnostic special case when \(q\) divides \(N\).

The algebraic specification is a list of constraints.  Each constraint is
paired with roots and a declared degree bound.  The core locale contains a
syntactic well-formedness condition ensuring that the degree of a constraint
applied to trace powers is bounded by the declared degree expression.  This
assumption is a specification-side condition: it says that the declared
constraint degrees are sound for all low-degree trace polynomials.

Several useful facts are derived from these assumptions.  The trace-domain and
evaluation-domain generators are aligned by construction.  The map from root
indices to field points is injective over the relevant range.  The shifted
evaluation domain is disjoint from the trace roots used in the constraint
quotients.  These facts are used repeatedly in the completeness and soundness
proofs.

The exact-order assumptions serve two roles.  Algebraically, they identify the
sets generated by powers of \(g\) and \(h\) with the intended finite domains.
Operationally, they justify finite-list encodings of those domains.  For
example, a table indexed by \([0..<n]\) is meaningful only if different
indices in that range correspond to different field points.  The derived
injectivity lemmas are therefore used both in polynomial arguments and in
list-index arguments.

The query-margin condition is a specification-sizing condition.  It states
that the query sample space is large enough relative to the degree bound after
clearing denominators.  Without such a margin, the usual argument that a
nonzero low-degree polynomial has only few roots would not give a nontrivial
query bound.  In the current model this condition lives in the soundness
locale layer because it is needed only for adversarial rejection probability,
not for executable honest runs.

The constraint-degree well-formedness condition belongs to the protocol
specification.  It says that if a trace polynomial has degree below the trace
length, then applying a declared constraint to the trace powers gives a
polynomial within the declared degree bound.  This is not a property of a
particular execution; it is a syntactic or semantic property of the constraint
language used in the specification.  The formalization currently states it as
a locale-level well-formedness condition and uses it to derive the degree
facts required by the composition and quotient arguments.

The composition polynomial is another key algebraic object.  It combines the
constraint quotients using verifier challenges.  The final soundness proof
uses both deterministic degree reasoning for the composition polynomial and
probabilistic reasoning for the random alpha challenges that define it.

The honest-completeness and soundness uses of the composition polynomial are
dual.  Completeness proves that if all constraints vanish on their declared
roots, then the prover's quotient construction agrees with the verifier's
evaluation checks and has degree at most \code{maxDegree}.  Soundness proves
that if no valid trace exists, then an accepting transcript must either
produce a low-degree object that would contradict falsity or fall into one of
the bounded bad events.

\formalanchors{
\anchorrow{Core protocol locale}{Core/Stark.thy}{stark}
\anchorrow{Maximum declared degree}{Core/Stark.thy}{maxDegree}
\anchorrow{Composition polynomial}{Core/Stark.thy}{cp}
\anchorrow{Query raw-preimage envelope}{Soundness/Soundness_Core_Base.thy}{query_raw_preimage_card_envelope}
\anchorrow{Exact modulo preimage arithmetic}{Mathematics/FRI/Modulo_Arithmetic.thy}{card_modulo_residue_preimage; card_modulo_set_preimage}
\anchorrow{Raw/natural sampler correspondence}{Soundness/Soundness_Core_Events.thy}{card_query_index_raw_preimage_eq_nat_preimage}
\anchorrow{Honest trace algebra}{Completeness/Completeness_Core.thy}{honest_trace_algebra}
\anchorrow{Honest trace validity}{Completeness/Completeness_Core.thy}{honest_trace_valid}
}

\section{Honest Completeness}

Generic completeness concerns the original pure-concat prover/verifier
execution. Acceptance in the absorbing-RO security experiment is established
separately in Section~\ref{sec:honest-ro}.  In this formalization it is phrased as a probabilistic failure
bound: if the honest trace satisfies the validity predicate, then the
probability that the verifier returns \code{None} is zero.  Since probabilities
are nonnegative, this gives the final inequality by simplification.

The validity predicate is deliberately narrow:
\code{honest_trace_valid} is definitionally equivalent to
\code{honest_trace_algebra}.  Its explicit content is that every declared
constraint vanishes at its declared roots.  Degree well-formedness, domain
alignment, root distinctness, and sampler facts are assumptions or consequences
of the enclosing locales; they are not bundled into the theorem premise.  The
premise is therefore about the concrete trace and specification, not about a
precomputed verifier transcript or a hidden no-failure obligation.

The proof is organized around a weakest-precondition reduction.  The first
step proves that, under \code{honest_trace_valid}, failure is not in the support
of the distribution produced by the combined prover/verifier execution.  This
deterministic support statement is then converted to a zero probability result
using the WP calculus.

The deterministic obligations follow the structure of the protocol:

\begin{itemize}[leftmargin=*]
  \item Transcript replay shows that verifier reads consume exactly the values
    sent by the honest prover.
  \item Merkle correctness shows that paths opened by the honest prover check
    against the roots produced during commitment.
  \item Query consistency shows that the verifier's queried values agree with
    polynomial evaluations of the honest trace powers.
  \item Composition-degree reasoning shows that the honest composition
    polynomial respects the verifier's degree bound.
  \item FRI consistency shows that each honest FRI layer satisfies the fold
    equation and that the final layer agrees with the committed constant.
\end{itemize}

The final theorem can be summarized as:

\begin{lstlisting}
lemma completeness:
  assumes "honest_trace_valid"
  shows "wp_event exec (\<lambda>a. Option.is_none a) init_state \<le> (0::real)"
\end{lstlisting}

The companion equality lemma \code{completeness_wp_reduction} states the exact
zero probability result before it is turned into the displayed inequality.

\subsection{Replay phase}

The replay phase proves that the prover and verifier traverse the same
transcript in compatible order.  This is not automatic in a state-monad model:
the prover program appends messages, while the verifier program later consumes
them through indexed reads.  The replay lemmas show that structured programs
with interleaved sends, reads, Fiat-Shamir calls, and hash-map extensions
preserve the expected correspondence.  After these lemmas are applied, many
verifier reads simplify to the concrete messages sent by the honest prover.

\subsection{Merkle and query phase}

For Merkle openings, the honest prover opens paths from trees it has just
created.  The proof threads the created-tree invariant through the query
rounds and FRI decommitments.  Each verifier check then reduces to the
deterministic correctness theorem for authenticated paths created from the
same tree.

For query values, the verifier asks for entries of the trace and composition
evaluation tables at sampled positions.  Honest correctness requires an index
calculation: a query index in the evaluation domain corresponds to several
trace-power evaluations at positions offset by the scaling factor.  The proof
uses domain-alignment and powers-scaled bounds to show that the list entries
opened by the prover are exactly the polynomial evaluations expected by the
verifier.

\subsection{FRI phase}

The honest FRI proof follows the executable FRI construction.  Each layer is
generated by the same folding operation that the verifier checks locally.  The
main deterministic fact is therefore that the sibling values opened from an
honestly generated layer and the next-layer value opened at the index reduced
modulo the next layer's length
satisfy the verifier's fold equation.  The final layer is constant, so the
final value check succeeds.  Combined with transcript replay and Merkle
correctness, this discharges the verifier's FRI assertions.

\formalanchors{
\anchorrow{Verification locale}{Completeness/Completeness_Core.thy}{verification}
\anchorrow{Honest trace validity}{Completeness/Completeness_Core.thy}{honest_trace_valid}
\anchorrow{WP completeness reduction}{Completeness/Completeness_Verifier.thy}{completeness_wp_reduction}
\anchorrow{Completeness theorem}{Completeness/Completeness_Verifier.thy}{completeness}
}

\section{Soundness Experiment and Adversary Interface}
\label{sec:experiment}

The original FS attack is \code{fs_security_experiment P}. It executes
\code{fs_run P}, transfers the returned transcript into the verifier state and
runs \code{ro_verify_monad}. Its acceptance event excludes failure, and its
probability at \code{adversary_initial_state} is
\code{fs_acceptance_probability P}. The shared oracle map and submitted
transcript survive the transfer; verifier-local cursors and challenge counters
start from their prescribed values. An attacker does not choose an arbitrary
hidden verifier state.

The producer syntax supports return, explicit failure, typed oracle queries
and private PMF samples. Continuations remain adaptive to the answers already
received. \code{fs_query_bound Q P} restricts every continuation and requires
finite support of private samples; cached queries count again. A private PMF
sample is lossless, but the entire producer need not be: \code{FS_Fail} remains
part of the interface. The proved translation to a CryptHOL GPV supplies the
needed interaction-bound implication, not an equivalence with every
interaction-bounded GPV.

The semantic false-statement premise is the negation of \code{exists_valid_trace}.
It says that no trace satisfies the declared constraints in the fixed public
instance. The final proof uses that premise after deriving low-degree and
constraint-consistent evidence from supported acceptance. It does not assume
that a malicious commitment has an honest complete-table representation.
The generic theorem also retains all assumptions of its enclosing
\code{soundness}, \code{verifier} and \code{stark} locales. The concrete square
interpretation proves these obligations rather than suppressing them.

The staged experiment is an internal proof interface. Its callbacks construct
a transcript against the shared oracle, and
\code{staged_adversary_controlled} records their semantic resource and
state-preservation obligations. Recorded messages are absorbed through
\code{TranscriptAbsorb}; the later verifier uses the same absorbing channel.
The FS compiler proves the required control contracts from the original
syntactic cap. Staged control is therefore not an extra public premise of the
concrete FS endpoint.

Parameter-dependent error definitions retain zero-round and out-of-regime
fallbacks. Stronger nonempty-round arguments apply only after their actual
domain, distance and schedule conditions have been established. The final
first-root and optimized theorems keep the same false-statement and query-cap
interface; an internal branch condition is not silently promoted to a new
public assumption.

\formalanchors{
\anchorrow{Producer syntax and restriction}{Soundness/FS_Program.thy}{fs_program; fs_query_bound}
\anchorrow{Native GPV execution and replay}{Soundness/FS_Native_Execution.thy}{fs_native_gpv_replay; fs_gpv_interaction_bound}
\anchorrow{Original attack and acceptance}{Soundness/FS_Adversary_Model.thy}{fs_security_experiment; fs_acceptance_probability}
\anchorrow{Staged transcript absorption}{Soundness/Staged_Security_Experiment_Core.thy}{ro_record_staged_message}
\anchorrow{Actual absorbing verifier}{Soundness/Staged_Security_Experiment_RO_Checked.thy}{ro_verify_monad}
\anchorrow{Uniform first-root endpoint}{Soundness/FS_First_Root_Soundness.thy}{fs_first_root_soundness}
}

\section{Architecture and the Current Proof Roadmap}
\label{sec:current-architecture}
\label{sec:architecture}

The current development separates four mathematical responsibilities, supported
by shared native probability and execution interfaces. Independence concerns
the premises and conclusions of an interface, not only its directory. The
general replay and accounting rules are independent libraries, while the
actual STARK event classification and its FS compiler remain concrete clients.

\begin{enumerate}[leftmargin=*]
  \item \textbf{Algebra and FRI.} Mathematical libraries prove polynomial
    evaluation, degree, agreement, distance, folding and counting bounds without
    the concrete hash model. They include affine reconstruction, paired and
    multiround folding, evaluation-code distance and exact modulo arithmetic.
    Concrete Soundness theories establish that tables, domains and indices
    obtained from accepted executions meet these contracts.
  \item \textbf{Merkle authentication.} Deterministic evidence libraries extract
    a collision from inconsistent certified paths. A generic CryptHOL reduction
    obtains certificates from oracle executions and accounts for its
    interactions. It retains the encoding, common-root, same-position and
    equal-path-length conditions; it does not reconstruct complete malicious
    tables from sampled openings.
  \item \textbf{Fiat--Shamir execution and accounting.} Native PMFs, SPMFs and
    CryptHOL GPVs support private randomness, failure and explicit oracle calls.
    Generic libraries transport projections and observations, combine private
    mixtures, replay cached queries and bound oracle events. The concrete
    compiler connects the restricted FS adversary to the staged STARK experiment
    and proves the query allowances needed by that connection.
  \item \textbf{Concrete composition.} The STARK layer proves that every
    supported accepting execution is covered by actual bad-oracle or
    mathematical events. It discharges the component bounds and combines them
    into first-root and optimized soundness, then instantiates the certified
    square workload. Event coverage and component estimates are obligations,
    not consequences of supplying a budget vector.
\end{enumerate}

The thin state interface has type
\(\mathtt{'s}\Rightarrow(\mathtt{'a}\times\mathtt{'s})\ \mathtt{spmf}\).
Its physical owner is \theorysource{Probability/Native_State_Interface.thy};
the historical \code{Prob_Monad} namespace remains for compatibility.
\code{Core/Prob_Monad.thy} is a compatibility import, not a second probability
implementation. Shared execution, relation, potential, replay and accounting
contracts live in Probability, Evidence and CryptHOL. Their clients retain
concrete authentication, chronology, domain and algebraic premises.

\subsection{From the original attack to a compiled experiment}

The source-level route is:
\[
\begin{aligned}
  &\text{original FS attack}
    \longrightarrow \text{private mixture of fixed strategies}\\
  &\longrightarrow \text{compiled staged experiment}
    \longrightarrow \text{covered bad-oracle and mathematical events}\\
  &\longrightarrow \text{proved component ledger}
    \longrightarrow \text{first-root and optimized soundness}.
\end{aligned}
\]

The original experiment \code{fs_security_experiment P} runs the producer,
transfers its returned transcript to the verifier and runs
\code{ro_verify_monad}; its acceptance probability is
\code{fs_acceptance_probability P}. The adversary language has return,
failure, oracle-query and private-sampling constructors. Native execution
\code{fs_run} uses a CryptHOL GPV and the typed cached oracle. The restriction
\code{fs_query_bound Q P} requires finite support for private samplers and
bounds every continuation, including continuations outside the support of a
particular execution. Cached queries still count as calls.

Private-randomness normalization is an exact PMF mixture over strategies from
\code{fs_normalize}. Every supported strategy is fixed with respect to private
sampling and inherits the query restriction. It can still adapt its next
oracle query to previous answers. Fixing a strategy therefore does not identify
the original randomized adversary with one deterministic attack.

The transformation \code{fs_compile_replay A T} replays a fixed producer at
the staged callbacks and serves its returned transcript through the existing
sliced adversary. \code{fs_compiled_mixture A P} mixes these experiments over
the normalization PMF. The observation correspondence is equality of projected
whole-outcome distributions: it retains failure, returned data and observed
final state. Its observation must ignore the final residual
\code{PTranscript}. Acceptance meets this condition, giving
\code{fs_compiled_mixture_acceptance}. This does not assert equality of the
unprojected final states.

Witnessed experiments additionally expose first-root and query evidence for
authentication and chronology proofs. Canonical projections erase this evidence
while preserving the plain experiment and final state. Events inspecting
returned data use the projected predicate. An erased witness is recovered only
with the corresponding supported preimage evidence, not from projection
equality alone.

\subsection{Actual events and final composition}

In \theorysource{Soundness/Stark_Soundness_Composition.thy},
\code{stark_acceptance_components} derives a ten-event cover from a supported
accepting execution, using actual execution extension, prefix preservation and
the MCA acceptance classification. The cover includes the final collision,
trace first-root target, composition prefix target, clean builder target,
trace and composition bad challenges, query-phase builder target, decoded
all-query consistency, initial-state target and clean mathematical residual.
These events need not be disjoint or independent.

The theorem \code{stark_component_bound} applies a generic ledger rule only
after this cover and the individual probability estimates are available.
In \theorysource{Soundness/FS_Soundness_Assembly.thy},
\code{fs_composed_security_bound} proves the common builder, initial-target
and residual estimates and combines them with seven supplied estimates.
Its callers discharge those estimates through the actual event-accounting
theorems. \code{fs_compiled_witnessed_bound_suffices} then uses witness erasure
and the private-strategy correspondence to return the bound to the original
FS adversary.

The first-root route retains the parameter guards and fallback to prefix/coarse
accounting, including the separate zero-round treatment.
\code{fs_first_root_soundness} is the resulting bound for the original producer.
The optimized endpoint instantiates it at \code{fs_optimal_cutoffs}; separate
theorems prove comparison with every natural cutoff pair and the concrete
selected pair. Optimization is over the existing error-bound family. It proves
neither a matching attack nor optimality among all protocols or proof techniques.

\subsection{Query costs describe different executions}

Each replay callback receives the original syntactic allowance. The proved
total staged attacker allowance is
\[
 \bigl(\lceil\log_2\mathtt{clength}\rceil
       +\lceil\log_2(\mathtt{maxDegree}+1)\rceil
       +\mathtt{rounds}+4\bigr)Q,
\]
which is \(664Q\) for the published square profile. This is a sufficient
allocation, not an exact call count or lower bound. Builder and verifier hashing
add their own overheads.

The direct experiment executes one original producer and one verifier, with
\code{fs_direct_hash_budget Q = Q + ro_verifier_hash_query_budget}.
Proved correspondences transport this cost to the particular compiled events
for which it is valid. Prefix, first-root, builder and query-phase estimates
retain their own guards, cache bounds and chronology conditions. The proof
does not replace every replay cost by \(Q\), count only fresh calls, or equate
the syntactic restriction with an unrestricted GPV interaction bound. The
required implication to the native bound is proved for the translated adversary.

\formalanchors{
\anchorrow{Independent folding and agreement}{Mathematics/FRI/Multiround_Fold_Soundness.thy}{multiround folding contracts}
\anchorrow{Certified-path collision extraction}{Evidence/Lookup_Path_Collision.thy}{lookup_path_collision_complete}
\anchorrow{Oracle-execution Merkle reduction}{CryptHOL/Merkle_Collision_Reduction.thy}{merkle_attack_collision_reduction; merkle_reduction_queries}
\anchorrow{Native adversary execution}{Soundness/FS_Native_Execution.thy}{fs_run; fs_gpv_interaction_bound}
\anchorrow{Private strategy normalization}{Soundness/FS_Private_Randomness_Normalization.thy}{fs_normalize_security; fs_normalize_fixed; fs_normalize_bound}
\anchorrow{Compiled experiment and correspondence}{Soundness/FS_Adaptive_Staged_Embedding.thy}{fs_compile_replay; fs_compiled_mixture_acceptance}
\anchorrow{Replay allowances}{Soundness/FS_Replay_Allowance.thy}{fs_replay_budgets; fs_compile_replay_controlled}
\anchorrow{Direct experiment cost}{Soundness/FS_Probability_Accounting.thy}{fs_direct_hash_budget}
\anchorrow{Actual event coverage and ledger}{Soundness/Stark_Soundness_Composition.thy}{stark_acceptance_components; stark_component_bound}
\anchorrow{Concrete bound assembly}{Soundness/FS_Soundness_Assembly.thy}{fs_composed_security_bound; fs_compiled_witnessed_bound_suffices}
\anchorrow{Optimized final bound}{Soundness/FS_Optimized_Soundness.thy}{fs_optimized_soundness}
}

\section{Agreement, Sampling and Weighted Amplification}

The separate binary correlated-agreement bridge changes the internal event
decomposition, not the verifier. On $n$ distinct successor points, for degree
bound $k$ and missing-set allowance $t$ with $3t+k<n$, it proves that at most
$\max(1,2t)$ challenges permit low-degree agreement on some set missing at
most $t$ points without low-degree component agreement on that same set.
Every challenge, including zero, is covered. This is a derived binary result,
not the full all-arity correlated-agreement hypothesis or its full radius range.

The actual-fold bridge reconstructs parent agreement on both siblings. The
query-support induction uses the existing authenticated global residual and
the actual constant terminal table, not an assumption of honest full-domain
folding. With $M$ query indices, define the existing modulo envelope
$E(M,n,k)=\lfloor M/n\rfloor k+\min(k,M\bmod n)$.
For initial distance greater than $r$, the resulting conditional bound is
\[
 B=\max\left(|D_0|-r-1,\quad
       \max_{i<m}E(M,|D_{i+1}|,|D_{i+1}|-t_i-1)\right),
\]
where the subtractions are natural-number subtraction and the maximum includes
zero. It requires exclusion of the new same-subset bad event at every layer.
The raw sampler preimage is bounded by the unchanged field-to-index modulo
envelope, not by silently replacing the sampler with uniform indices.

At $|D_0|=65536$, $M=65472$, $d=1023$, $r=13544$, and
$t_i=\lfloor |D_{i+1}|/4\rfloor$, kernel-checked arithmetic gives support
threshold $49150$ and $B=51991$. The sum of the ten per-layer bad-challenge
cardinality envelopes is $32736$; it is not an adaptive probability bound.
Each per-prefix bad set is strictly smaller than the field under the stated
existing rate and round-fit conditions. Prefix stability uses the existing
no-new-Merkle-path-target condition and is not asserted unconditionally.
The theorem \code{soundness.fri_mca_scale64_authenticated_ceiling} applies
to the existing authenticated-chain evidence with explicit initial distance
and the new event complement. It is neither a bound on the old balanced-clean
branch nor a new public security profile. Old-clean near-64k attainability
remains open; public-classification transport is supplied by the
subsequent integration below. This intermediate argument alone is not a numerical security certificate.

The subsequent bounded adaptive-oracle audit proves separate trace and
composition joint-event bounds
\[
 \Pr[\text{clean builder output and new MCA chain event}]
 \ \leq\ \frac{q(C+7q+2)}{|\mathbb F|},\qquad
 C=\max(1,2t_0),
\]
where $q$ is the unchanged full staged transcript hash-query budget, including
opening-stage attacker queries and transcript work. The composition event
also includes the exact decoded-degree bound $d\leq\texttt{maxDegree}$.
These are not probabilities conditional on the guards. The checked builder
itself tests only the logarithmic round-count bound; it is not silently
strengthened to the exact degree guard.

The auxiliary lemmas retain the power-of-two domain, quarter-rate and
one-extra-level fit conditions. At $|D_0|=65536$, kernel arithmetic gives
$C=16384$. A domain-separated challenge insert leaves the conceptual table
unchanged. Absorption-prefix uniqueness on the collision-free/no-initial-output
branch determines one root at a trace key and both degree and root at a
composition key. Later activation
of an old queried output is covered by the existing drift targets, with
cardinality at most $7|\operatorname{dom} M|+2$.
The uniform local cap, drift union bounds and replacement of every map size
by $q$ are conservative proof steps; the quadratic charge is not an inherent
adversarial lower bound. A union bound could combine the two separately proved
events, but this auxiliary charging layer alone provides no public acceptance
theorem. The subsequent integration below supplies the builder/query/replay
transport and trace/composition distance split, including the near-code case.

\paragraph{End-to-end correlated-agreement integration.}
The theorem \code{soundness.ro_absorb_stark_soundness_mca} now bounds the
existing adversary acceptance probability with precisely the existing three
premises: false statement, well-formed staged budgets, and a controlled staged
adversary. Its parameter-only error tests nonempty trace folding, a power-of-two
domain, the existing one-extra-level fit conditions, and the two quarter-rate
conditions. In that regime it takes the minimum of the new complete expression
and the existing conditioned explicit all-round bound. Outside the regime it
uses that existing bound, including its separate zero-round case. Previously
proved stronger conditional endpoints also remain available.

The proof extracts both recorded chains and their actual terminal values,
preserving the exact decoded composition degree. It transports tables across
the prefix, query-start, builder and verifier states only under the existing
charged target complements. The residual query sets include active-chain and
distance guards; in particular, an empty global intersection is never treated
as a finite sampling space. A zero-fold composition candidate is constant and
has zero code distance. The near/near case uses arbitrary distance cutoffs and
the existing two-decoder semantic trichotomy, retaining its decoded-alpha event.
No far-from-code premise is added to public soundness.

\paragraph{Prefix-local target budgets.}
The actual saved query-head program has hash-range budget $H$, defined in
\theorysource{Soundness/Soundness_FRI_Query_Head_Range.thy} by the pre-query attacker
budgets plus the unchanged commitment/challenge header work. Write $T$ for the
query/opening tail budget, $V$ for the verifier budget, $q=H+T$ for the full
builder budget and $n=q+V$ for the security budget. On the existing charged
clean branch, the saved query-head map has at most $H$ entries. This is proved
from supported execution outcomes and hash extension, not introduced as an
assumption on the adversary.

The composition-prefix and query-phase target charges therefore use
$(T+V)(1+2H)/|\mathbb F|$ and
$T(m_T+m_C+2H)/|\mathbb F|$, respectively, where $m_T$ is the trace-fold count
and $m_C$ the maximum composition-fold count. Their actual events and tail
programs are unchanged. The exact saving relative to their earlier charges is
\[
 \Delta_{\rm target}=\frac{2(T+V)^2+2T^2}{|\mathbb F|}.
\]
The local event lemmas require nonempty trace folding, already checked inside
the MCA regime; the public zero-round fallback is retained unchanged. A
zero-fold composition still has the existing empty prefix target. The later
builder-to-verifier target, both collision charges, and the adaptive MCA and
decoded-alpha transition charges retain their former budgets. No improvement
in the fresh-sample exponent or asymptotic scaling is asserted by this
target-only refinement, and these
upper-bound savings do not settle near-$64000$ execution realizability.

Let $p(B)$ denote the existing raw modulo-preimage envelope divided by field
size, and let $k=\texttt{rounds}-\texttt{staged\_attacker\_query\_budget}$
use natural-number subtraction. In the historical reference geometry
($\texttt{clength}=1024$, scale $64$, degree bound $1023$, powers $2$ and
1024 constraint roots), kernel-checked expression evaluation gives
\[
 S_{\rm old}=2p(64320)^k+p(60667)^k,\qquad
 S_{\rm MCA}=2p(54954)^k+p(54953)^k
\]
at the two cutoffs $10581$. The earlier fixed-table ceiling $64263$ is not
the integrated old public residual envelope. The old balanced schedule at
$C=54$ has radius $13438$ and direct challenge cap $13441$, whereas the new
direct cap is $16384$. Consequently the two new adaptive charges add
$5886q/|\mathbb F|$ to the reference expression. The quadratic drift costs
and all collision, target and decoded-alpha charges are retained.

The exact comparison theorem
\code{soundness.ro_mca_complete_strict_comparison_iff} cancels only common charges:
the new expression is smaller exactly when the sampled savings together with
the prefix-local target credit exceed the
increased challenge charges. There is no unconditional strict improvement or
new concrete security profile. This expression calculation neither constructs
a field/specification instance nor bounds the old balanced-clean survivor
maximum by $54954$. Near-64k attainability and trusted reconstruction of the
large U7/U6 numerical certificates remain separate, deferred questions.

\paragraph{Weighted all-repetition amplification.}
The additional theorem
\code{soundness.ro_absorb_stark_soundness_mca_weighted} in
\theorysource{Soundness/Soundness_FRI_Weighted_Soundness.thy} proves a complete
alternative for the same acceptance experiment and exactly the same three
public premises. Its error takes the minimum with the previous public MCA
error inside the existing MCA regime with positive query repetitions, and
returns the previous error otherwise. Separate identities cover zero query
repetitions and zero trace-fold depth. No positive composition-fold premise
is introduced; the existing conditioned fallback is retained as well.

To distinguish attacker complexity from total experiment work, write $Q$ for
the attacker-only oracle-query budget, $q_B$ for the full builder hash budget,
$q_V$ for the verifier hash budget, $q=q_B+q_V$, $F=|\mathbb F|$, and $R$ for
query repetitions. Let $p_S,p_T,p_C$ be the existing exact modulo-preimage
envelopes divided by $F$, for the decoded-semantic, trace-FRI and
composition-FRI residual branches, respectively. Put
\[
 P(n)=p_S^n+p_T^n+p_C^n,\qquad
 \Gamma=\frac{5q(q-1)}{2F}.
\]
With $C$ denoting the unchanged sum of all other collision, target,
MCA-challenge and decoded-alpha charges, the complete nonempty expressions
are
\[
 E_{\rm weighted}=C+\frac{q}{F}+qP(R)+\Gamma,\qquad
 E_{\rm MCA}=C+\frac{q_B}{F}+P(\max(R-Q,0)).
\]
The connection charge $\Gamma$ is shared once across the semantic and both
FRI branches. The final-map initial-value charge $q/F$ replaces the old
builder-only charge; it is not silently omitted or charged twice. The theorem
\code{soundness.ro_mca_weighted_increment_comparison} proves exactly
\[
 E_{\rm weighted}\leq E_{\rm MCA}
 \quad\Longleftrightarrow\quad
 qP(R)+\Gamma+\frac{q_V}{F}\leq P(\max(R-Q,0)).
\]
The public minimum is therefore no worse than the previous public bound,
but this does not assert an unconditional strict improvement.

The proof uses result-preserving oracle logging and weighted query paths,
accounts for adaptive candidate births, and averages only unrevealed FRI
challenge answers while respecting cached entries. Execution replay connects
these auxiliary scores to the original guarded residual events; no
challenge-before-query restriction or adversary weakening is assumed.
Consequently, subtracting the entire $Q$ from the sampling exponent is not
necessary in this alternative argument. Budget dependence remains in $q$
and the exception charges, whose optimality is not established. The residual
envelopes themselves are unchanged. This result neither selects a concrete
cryptographic profile nor resolves decoded-semantic tightness or near-$64000$
execution realizability. Its current theorem and definitions provide the source anchors for this comparison.

\section{Main Soundness Reductions}

\subsection{Bad-event decomposition}

The soundness proof proceeds by covering accepting executions into bad
events.  If a false statement is accepted, then some proof obligation must have
failed: a hash collision occurred, a random alpha challenge hit a bad set, a
sampled query missed the disagreement, Merkle openings were inconsistent, or a
FRI low-degree check accepted a bad candidate.  The final theorem combines the
probability bounds for these events by repeated union bounds.

The deterministic part of this decomposition lives close to the verifier's
operational semantics.  It extracts from accepting verifier runs the transcript
fields, Merkle roots, query indices, authenticated openings, and FRI opening
transcripts that later proof layers reason about.  The probabilistic part then
charges the extracted bad events to hash-collision budgets, target-hit budgets,
query-list bounds, and FRI accounting terms.

\subsection{Composition randomization}

Composition soundness depends on the alpha challenges used to combine
constraints.  For a false trace, the set of bad alpha assignments is small
unless side events such as hash collisions or Merkle inconsistencies occur.
The formal proof uses staged target-hit accounting: the target set may depend
on the transcript prefix and prefix oracle state, but it is fixed before the
corresponding challenge is sampled.  This ordering is essential for a valid
probabilistic bound.

\subsection{Query soundness}

The query argument checks whether sampled positions detect disagreement between
the committed tables and any low-degree candidate consistent with the evidence.
The important formal distinction is that the verifier receives partial
openings, not full tables.  Therefore the proof works with partial table
candidates and query-index events tied to the transcript and verifier branch.
This avoids treating unopened table entries as if they had been authenticated.

\subsection{Merkle consistency}

Merkle consistency is represented through authenticated openings.  A partial
opening records a root, length, index, value, and path.  If two authenticated
openings force inconsistent values under the same commitment structure, the
proof reduces that inconsistency to a hash-collision event.  The soundness
bound then charges the collision probability to the existing hash budget.

\subsection{FRI reductions}

The FRI portion separates trace and composition low-degree testing.  Both sides
use sampled query indices, Fiat-Shamir challenges, committed layer roots, local
fold equations, and final-value checks.  The formalization contains several
layers that refine broad FRI events into verifier-tied and query-aware events.
The current endpoint closes those events with actual-query products,
raw-preimage cardinalities, staged prequery terms, and relation-fiber hash
bounds for both trace and composition.  No unbounded FRI event probability is
exported by the public result.

\subsection{Event layers}

The proof uses several layers of events because different stages require
different information.  Execution events mention concrete verifier outcomes.
Extraction events mention data read from accepting runs.  Partial-opening
events mention authenticated Merkle evidence.  Algebraic bad events mention
candidate polynomials, disagreement sets, degree predicates, and challenge
sets.  Finally, budget events mention hash collisions or random-oracle outputs
falling into selected target sets.

Moving between these layers is the core of the soundness proof.  A deterministic
lemma usually has the form: every accepting execution of a false statement is
contained in a union of more structured bad events.  A probabilistic lemma then
bounds one of those structured events.  The final theorem is obtained by
composing these inclusions and applying union bounds in the WP calculus.

\subsection{Composition path}

The composition path begins with the event that the randomized composition
does not faithfully expose constraint failure.  The proof defines bad sets of
alpha challenges for fixed transcript prefixes.  These sets are small by
polynomial-degree reasoning.  Because the target set is prefix-dependent, the
proof uses dynamic target-budget lemmas: the adversary may influence the prefix,
but once the prefix is fixed, the next random-oracle output must avoid a small
set except with the target-hit probability plus controlled side events.

The deterministic degree component is discharged by specification
well-formedness.  The probabilistic component is charged to the alpha
target-hit bound and to hash-collision or Merkle-side events that explain
unclean prefixes.  This avoids adding a public theorem assumption for
composition soundness.

\subsection{Query path}

The query path is subtle because the verifier checks only sampled positions.
The proof therefore avoids a global claim that sampled openings identify a
unique complete table.  Instead, it defines prefix-fixed and transcript-indexed
query targets.  These targets are fixed before the query challenge is sampled
and are therefore compatible with the random-oracle budget lemmas.

The size of a query target is controlled by low-degree disagreement: two
distinct low-degree candidates can agree on only a bounded number of domain
points.  The sampled query detects disagreement except when the query lands in
this bounded set.  The formal proof connects the sampler to the finite query
space through the exact quotient/remainder count of its raw modulo preimages,
derived from \code{to_nat_range}; it does not assume that the query-space size
divides the field-encoding size.

\subsection{FRI path}

The FRI path is divided into trace and composition sides.  Both sides use the
same conceptual structure, but their low-degree predicates and degree bounds
differ.  The trace side checks whether a trace table is compatible with a
low-degree trace polynomial.  The composition side checks whether a composition
table has degree at most \code{maxDegree}.  In both cases, verifier-observed
evidence consists of sampled openings, layer roots, challenge lists, local fold
checks, and a final value check.

The endpoint theorem uses proved actual-query, target, product, and fiber bounds
rather than symbolic FRI assumptions.  For query indices, repeated sampling is
counted over the raw field encodings: exact raw-list preimage cardinalities are
divided by the field-encoding size to the number of rounds, and closed bounds
use the maximum singleton-fiber envelope.  Other field challenges retain their
usual inverse-field-size factors.  Staged prequery terms account for attacker
queries made before verifier sampling, and relation-fiber bounds turn oracle
target relations into explicit hash charges.  The public theorem therefore
asks the user for neither a trace nor composition FRI reduction as an external
premise.

\formalanchors{
\anchorrow{Trace low-degree predicate}{Soundness/Soundness_Core_Base.thy}{trace_table_low_degree}
\anchorrow{Composition low-degree predicate}{Soundness/Soundness_Core_Base.thy}{composition_table_low_degree}
\anchorrow{Decoded all-query agreement event}{Soundness/Soundness_FRI_Robust_Security_Classification.thy}{ro_absorb_checked_staged_security_robust_decoded_all_queries_consistent}
\anchorrow{Authenticated partial-table evidence}{Soundness/Soundness_Partial_Merkle.thy}{authenticated_opening_in; partial_authenticated_table}
\anchorrow{Compatibility with observed values}{Soundness/Soundness_Partial_Merkle.thy}{table_agrees_with_authenticated_openings}
\anchorrow{Inconsistent openings imply collision}{Soundness/Soundness_Partial_Merkle.thy}{inconsistent_authenticated_openings_imp_hash_collision}
\anchorrow{Authenticated FRI query evidence}{Soundness/Soundness_FRI_RO_Actual_Query_Accepted_Evidence.thy}{ro_recorded_query_fri_accepted_evidence}
\anchorrow{Exact modulo query fibers}{Mathematics/FRI/Modulo_Arithmetic.thy}{card_modulo_residue_preimage; card_modulo_set_preimage}
\anchorrow{Concrete raw-preimage envelope}{Soundness/Soundness_Query_Index_Modulo_Bounds.thy}{card_query_index_raw_preimage_le_query_envelope}
\anchorrow{Query-list probability envelope}{Soundness/Soundness_FRI_Query_List_Exact_Product.thy}{query_index_raw_list_preimage_probability_bound}
\anchorrow{Query relation fibers}{Soundness/Soundness_FRI_Query_List_Prequery.thy}{query_index_raw_list_relation_fiber_bound}
\anchorrow{Hash collision budget}{Soundness/Soundness_Oracle_Budgets.thy}{hash_collision_budget_value}
\anchorrow{Hash target budget}{Soundness/Soundness_Oracle_Target.thy}{hash_target_budget_value}
}

\section{Proof-Design Constraints and Rejected Direct Routes}
\label{sec:proof-design}

This section records proof-design constraints that explain the final proof
architecture.  It is not a chronology of development attempts.  Each point is
a formal reason why a simpler-looking direct proof route would be too strong.

\subsection{Complete-table reconstruction}

A tempting direct route is to reconstruct complete committed tables from
sampled Merkle openings.  This is too strong.  Sampled openings constrain only
the queried positions. Many mathematical completions of the observed values
can agree on all opened positions and differ elsewhere. This does not assert
that two fully authenticated trees can have the same root without a collision:
a completion of the sampled values is not itself a fully authenticated tree.
The formalization therefore reasons about authenticated partial openings and
compatible conceptual tables.

\subsection{Singleton full-table candidates}

Another tempting route is to prove that sampled openings determine a unique
full-table candidate.  This is not derivable from the verifier evidence alone.
Uniqueness can hold for a fully opened table, but the verifier samples only a
small number of positions.  The final proof therefore uses prefix-fixed or
transcript-indexed query targets rather than global singleton full-table
covers.

\subsection{Broad existential events}

Some broad query and Merkle events existentially quantify over verifier
continuations or witnesses.  Such events are useful for deterministic
classification, but they are too coarse for direct probabilistic accounting:
the event may depend on choices that were not fixed before the corresponding
challenge was sampled.  The proof therefore introduces branch-tied and
verifier-local events.

\subsection{FRI witnesses}

For FRI, the verifier observes committed roots, sampled openings, local fold
checks, and final values.  It does not observe complete FRI layer tables.
Therefore the final route phrases FRI reasoning over verifier-observed
partial-opening evidence and refined layer-chain targets, rather than assuming
complete-table witnesses can be reconstructed from sampled data. Conceptual
completions used as proof objects do not authenticate the unopened entries.

\formalanchors{
\anchorrow{Authenticated partial-table evidence}{Soundness/Soundness_Partial_Merkle.thy}{partial_authenticated_table; table_agrees_with_authenticated_openings}
\anchorrow{Authenticated accepted evidence}{Soundness/Soundness_FRI_RO_Actual_Query_Accepted_Evidence.thy}{ro_recorded_query_fri_accepted_evidence}
\anchorrow{Recorded fold-chain evidence}{Soundness/Soundness_FRI_RO_Actual_Query_Fold_Evidence.thy}{ro_fri_fold_recorded_chain_evidence}
\anchorrow{Supported authenticated query evidence}{Soundness/Soundness_FRI_RO_Actual_Query_Accepted_Evidence.thy}{ro_verifier_query_round_program_recorded_fri_accepted_evidence}
\anchorrow{Explicit security classifier}{Soundness/Soundness_FRI_Conditioned_Explicit_Security_Classification.thy}{ro_absorb_checked_staged_security_clean_conditioned_explicit_classification}
}

\section{Certified Square Application and Optimized Soundness}
\label{sec:concrete}
\subsection{Scalable square-sequence workload bridge}

The theory \theorysource{Examples/Square_Sequence_Workload.thy} generalizes the small
example to a trace length $L$ and arbitrary field endpoints $a,z$. Its
polynomial constraints are $q_0-a$, $q_0-z$, and $q_1-q_0^2$, with root-index
lists $[0]$, $[L-1]$, and $[0,\ldots,L-2]$ and degree annotations $1,1,2$.
The scalar specification is proved to agree with polynomial evaluation for
every list of length at least two and every evaluation point, not just for
one selected trace. Degree bounds and root well-formedness are also proved.

The companion theory \theorysource{Soundness/Soundness_Square_Sequence_Workload.thy}
discharges the existing soundness locale over prime residue fields. For
power-of-two $L,S$, $L\geq2$, $LS\mid F-1$, $LS<F-1$, and positive repetitions,
its remaining specification margin is exactly $2L<(L-1)S$. No new locale or
public soundness assumption is introduced. One generator and shifted domain
works for every pair of endpoints. The compatibility operation and inherited
error parameters remain arbitrary; no adversarial budget is restricted.

The actual \code{exists_valid_trace} predicate implies
$z=a^{2^{L-1}}$: evaluating the shifted trace polynomial at a trace point
gives the next trace value, and the constraints imply the squaring recurrence.
The same domain supports genuine true and false cases: $(a,z)=(1,1)$ has the
constant-one trace polynomial, whereas $(1,0)$ has no valid trace. This generic
layer alone is not interpolation completeness or an execution construction.
The concrete core-completeness layer below supplies the converse on the
certified field, without a machine-code correctness claim.

At $L=1024,S=64$, the certified metadata are maximum degree $1023$, root-list
length sum $1025$, deduplicated root count $1024$, and query-space size $65472$.
The deduplicated count uses injectivity of the actual trace-domain powers;
it is not substituted for the larger sum in the locale margin. The associated
\code{query_agreement_bound} is $2047$, not a complete FRI or public security
bound. These facts connect a real computation family to the parameters used by
the accounting proof. The following layers discharge a concrete prime-field
certificate, connect the first-root ledger and certify the numerical target.

\formalanchors{
\anchorrow{Verifier specification obligations}{Examples/Square_Sequence_Workload.thy}{square_workload_mod_ring_verifier}
\anchorrow{Actual valid-trace implication}{Soundness/Soundness_Square_Sequence_Workload.thy}{square_workload_valid_trace_endpoint}
\anchorrow{Reference metadata}{Soundness/Soundness_Square_Sequence_Workload.thy}{square_workload_reference_metadata}
\anchorrow{Shared-domain true and false cases}{Soundness/Soundness_Square_Sequence_Workload.thy}{square_workload_prime_field_nonvacuity}
}

\subsection{Certified 192-bit prime field and workload domain}

The concrete cardinality is
\[
 \begin{aligned}
 F&=3531942672373065915328229302366199208109272268385770536961\\
  &=10379446647006781035\cdot 2^{128}+1.
 \end{aligned}
\]
\code{Prime_192_Certificate} proves $\operatorname{prime}(F)$ using
Pocklington with factored part $2^{128}$, whose square exceeds $F$.
Its only prime divisor is two; the odd cofactor need not be factored.
The certificate reconstructs
\[
 \operatorname{mod\_exp}(13,(F-1)/2,F)=F-1.
\]
The exponent uses exact natural division. A separate congruence argument
handles natural subtraction in the predecessor-coprimality obligation,
including positivity of the reduced half-power. Doubling the exponent reuses
the half-power equality for the full-power condition.

The supporting \code{Prime_192_Replay} theory uses the library's original
\code{mod_exp_aux} recurrence. ML computes candidate products and
quotient/remainder decompositions, but HOL proofs check every product,
decomposition and strict remainder bound. Equality composition checks the
initial expression, every recurrence step, terminal case and final modulo;
proof-dependency checks reject oracle-backed or assumption-bearing results.
Numerical audit files are not proof inputs. This is a fixed certificate
reconstruction, not a general primality algorithm or an executable STARK.

\code{Prime_Field_192} constructs a finite index type of cardinality $F$,
proves its \code{prime_card} instance and obtains the existing residue-field
and encoding infrastructure. It proves $2^{191}\leq F<2^{192}$,
$65536\mid F-1$ and $65536<F-1$. A full generator chosen by Hilbert choice
from proved existence supplies both the evaluation root and a nonzero shift
outside its subgroup. The primality witness 13 is not certified as a full
generator, and the chosen generator is not supplied as an executable numeral.

\code{square_192_soundness} discharges the existing soundness-locale predicate
for $L=1024$, $S=64$, arbitrary endpoints and every positive repetition count
on this one fixed domain. \code{square_192_semantics} proves the actual
valid-trace implication $z=a^{2^{1023}}$ and same-domain true $(1,1)$ and false
$(1,0)$ instances. The proof uses a positive repetition count only to interpret
the existing locale internally; its valid-trace conclusions do not constrain
any adversarial budget or select a security profile.

This is a concrete field/domain and workload interpretation, rather than
a probability estimate by itself. It discharges the field and specification
conditions needed by the numerical soundness result. The separate
\code{square_192_valid_trace_iff} theorem supplies the converse computation
relation by interpolation: validity is exactly $z=a^{2^{1023}}$, including
zero input. Generic pure-concat completeness and honest acceptance in the
absorbing-RO experiment remain distinct execution results; the latter is
proved for the actual final verifier used by the FS experiment.

\formalanchors{
\anchorrow{Exact primality certificate}{Examples/Prime_192_Certificate.thy}{certificate_prime_192_is_prime}
\anchorrow{Concrete field cardinality}{Examples/Prime_Field_192.thy}{CARD_field_192}
\anchorrow{Fixed proper evaluation domain}{Examples/Prime_Field_192.thy}{field_192_domain}
\anchorrow{Concrete soundness-locale obligations}{Soundness/Soundness_Square_Sequence_192.thy}{square_192_soundness}
\anchorrow{Actual valid-trace and nonvacuity results}{Soundness/Soundness_Square_Sequence_192.thy}{square_192_semantics}
}

\subsection{Prefix witnesses within the current accounting proof}

The current first-root theorem retains the useful prefix decomposition of
the compiled experiment. The witnesses contain saved states, raw query
samples and transcript chunks from their actual supported executions.
Conditioning on a supported head retains the producer source as ghost data;
the query-start target is fixed there, rather than before the oracle-dependent
producer runs. The projection contracts retain failure and the returned
witnesses required by the subsequent event proof.

Put $h=Q+h_0$, where $h_0$ is the checked head budget with zero fixed-selector
allowances, and put $t_0=R(1+C)$, where $C$ is the unchanged per-round transcript
envelope. Let $v$ denote the verifier hash budget, and write
\[
 D=\mathit{ceil\_log}(\mathit{clength})+
   \mathit{ceil\_log}(\mathit{maxDegree}+1).
\]
The retained composition-prefix and query-phase charges are
\[
 T_{\mathrm{composition}}=\frac{(t_0+v)(1+2h)}{F},
 \qquad
 T_{\mathrm{query}}=\frac{t_0(D+2h)}{F}.
\]
The fixed-source selectors require no further adversarial oracle calls, but
internal challenges, transcript absorptions and verifier hashing remain
charged. This is neither a restriction that the original adversary spends
zero queries on openings nor a claim that cached calls are free.

The sufficient operational replay allocation is still
\[
 (\mathit{ceil\_log}(\mathit{clength})+
   \mathit{ceil\_log}(\mathit{maxDegree}+1)+R+4)Q,
\]
which is $664Q$ at the square profile. It describes the replayed staged
execution. The direct original-producer-plus-verifier budget
$Q+\mathit{ro\_verifier\_hash\_query\_budget}$ describes a different
execution; event correspondences justify the particular accounting
transports that use it. Neither quantity is an exact call count or a
distinct-key count. The original query cap remains \code{fs_query_bound Q P}.

\formalanchors{
\anchorrow{Witness-preserving replay}{Soundness/FS_Prefix_Witness_Replay.thy}{fs_prefix_security_witness_elimination; fs_prefix_head_budget; fs_prefix_tail_budget}
\anchorrow{Composition-prefix and query-phase charges}{Soundness/FS_Prefix_Target_Accounting.thy}{fs_compiled_composition_prefix_bound; fs_compiled_query_phase_bound}
\anchorrow{Direct producer/verifier allowance}{Soundness/FS_Probability_Accounting.thy}{fs_direct_hash_budget; fs_direct_range_target_collision_budget}
\anchorrow{Total staged replay allowance}{Soundness/Publication_Reduction_Facts.thy}{fs_replay_total_allowance}
}

\subsection{First-root soundness and the concrete ledger}

The current complete symbolic theorem is \code{fs_first_root_soundness}.
For a false statement and an original query-bounded FS producer, it bounds
acceptance by \code{fs_first_root_parameter_error N rT rC Q}. Adaptive
oracle queries and private randomness remain part of that producer model;
the final theorem does not ask the caller to supply a staged adversary or
separate component-event inequalities.

The first-root accounting uses the actual producer output and saved
first-root state for the outer and nested-alpha target events. Targets are
not assumed fixed before the adaptive producer runs. All ten classification
events remain covered and charged by their corresponding proofs. Their union
bound requires neither disjointness nor independence, and no event-overlap
subtraction is performed across a possibly failing verifier.

The improved nonempty-round sum is used inside the existing parameter-regime
gate. The complete parameter error keeps the preceding prefix error as a
minimum there and as the outside-regime fallback. Consequently, zero-round
and other fallback cases are preserved: the unrestricted parameter theorem
does not acquire the internal fit, rate or nonempty-round premises used by
its stronger branch. Separate comparison lemmas establish that the eligible
sum is no worse than the preceding one.

For the square profile with $R=640$ and cutoffs $r_T=r_C=10581$, the
source-connected sufficient ledger is
\[
\begin{aligned}
 E_{\mathrm{first}}(Q)
  &=\frac{21Q^2+2464458513Q+8773058386762}{2F}\\
  &\quad +(Q+914610)
       \bigl(p(54953)^{640}+2p(54954)^{640}\bigr),
\end{aligned}
\]
where the exact modulo envelope is
\[
 p(B)=\frac{\lfloor F/65472\rfloor B+
                  \min(B,F\bmod65472)}{F}.
\]
The query sampler is still field sampling followed by natural decoding and
modulo reduction. The quotient and remainder terms retain its bias;
$p(B)$ is not replaced by the idealized ratio $B/65472$.

The retained comparison ledger is
\[
\begin{aligned}
 E_{\mathrm{prefix}}(Q)
  &=\frac{10613Q^2+2470872425Q+8773058386762}{2F}\\
  &\quad +(Q+914610)
       \bigl(p(54953)^{640}+2p(54954)^{640}\bigr).
\end{aligned}
\]
The native algebraic comparison is stated additively:
\[
 E_{\mathrm{first}}(Q)+\frac{5296Q^2+3206956Q}{F}
     =E_{\mathrm{prefix}}(Q).
\]
It proves the first-root ledger no greater than the prefix ledger. This is an
identity between sufficient arithmetic bounds, not an equality for actual
acceptance probabilities or an unqualified identity for every branch of the
public minimum. Quadratic dependence remains in other charges, without a
claim that those charges are necessary.

\formalanchors{
\anchorrow{Actual first-root target bounds}{Soundness/FS_First_Root_Accounting.thy}{fs_first_root_builder_bound; fs_first_root_security_bound}
\anchorrow{Nested-alpha target bound}{Soundness/FS_First_Root_Alpha.thy}{fs_compiled_first_root_alpha_security}
\anchorrow{Complete native FS theorem}{Soundness/FS_First_Root_Soundness.thy}{fs_compiled_first_root_nonempty_bound; fs_first_root_soundness}
\anchorrow{Eligible full-sum comparison}{Soundness/FS_Cutoff_Optimality.thy}{fs_first_root_nonempty_error_le_prefix}
\anchorrow{Exact source-connected ledger and acceptance bound}{Soundness/FS_Square_First_Root_Ledger.thy}{square_fs_first_root_error_ledger; square_fs_first_root_soundness}
\anchorrow{Native ledger comparison}{Soundness/FS_Square_First_Root_Comparison.thy}{square_fs_first_root_ledger_balance; square_fs_first_root_ledger_le_prefix}
}

\subsection{The certified 137-bit endpoint}

At $Q=2^{20}$, the prefix ledger is exactly
\[
 \frac{14268787487890250}{2F}
 +1963186\bigl(p(54953)^{640}+2p(54954)^{640}\bigr).
\]
The closed power certificates bound both modulo bases by $68/81$, and the
retained block-power estimate bounds each resulting $640$th power by
$2/(5\cdot2^{160})$. The final rational certificate establishes
\[
 \frac{14268787487890250}{2F}
 +1963186\cdot3\cdot\frac{2}{5\cdot2^{160}}\le2^{-137}.
\]
The factor three retains the semantic branch and the two FRI residual
branches. The source equality, power estimate and final rational inequality
are HOL proofs, not external floating-point computations.

The current acceptance proof follows the first-root route, then applies the
proved first-root/prefix comparison and this closed prefix inequality.
For fixed endpoints $a,z$ in the certified field, the resulting corollary
\code{square_192_fs_137_incorrect_endpoint} has exactly the endpoint condition
$z\ne a^{2^{1023}}$ and \code{fs_query_bound (2^20) P} as its two public
premises. Concrete field, domain, degree, specification and positive-round
obligations are already discharged. The existing first-root error itself can
be smaller than the prefix certificate used to prove this integer-bit target;
no separately certified $138$-bit endpoint follows merely from that fact.

The supported result remains a false-acceptance bound in the fixed-statement,
classical, typed field-valued random-oracle model. It is not an unrestricted
work-factor claim, QROM theorem, adaptively chosen-statement guarantee,
zero-knowledge or knowledge-extraction result, or a concrete-hash security
theorem. Shared \code{FS_Square_135_Arithmetic} facts remain prerequisites of
the numerical argument; the earlier separate $128$- and $135$-bit security
endpoints belong to the historical development, not the current supported
publication profile. Historical decimal estimates and adjacent-query
thresholds are likewise distinguished from this certified target.

\formalanchors{
\anchorrow{Retained small power certificates}{Soundness/FS_Square_135_Arithmetic.thy}{fs_135_block; fs_135_blocks; fs_135_power}
\anchorrow{Closed rational certificate}{Soundness/FS_Square_137_Arithmetic.thy}{fs_137_total}
\anchorrow{Exact specialization and numerical target}{Soundness/FS_Square_192_137.thy}{fs_137_ledger; square_fs_137_prefix_target}
\anchorrow{Concrete false-endpoint theorem}{Soundness/FS_Square_192_137.thy}{square_192_fs_137_incorrect_endpoint}
}

\subsection{Automatic cutoffs and their separate optimality proof}

The canonical optimized interface packages selection of $r_T,r_C$ without
changing the verifier or adversary. For fixed analysis exponent $N$ and cap
$Q$, \code{fs_optimal_cutoffs N Q} chooses a pair minimizing the complete
\code{fs_first_root_parameter_error} on
\[
  \{0,\ldots,\mathit{clength}\cdot\mathit{scale}\}^{2}.
\]
\code{fs_optimized_parameter_error N Q} is the value of the existing error
at that chosen pair. The theorem \code{fs_optimized_soundness} instantiates
the uniform first-root soundness theorem at those cutoffs. Its theorem-local
premises are still just \code{fs_query_bound Q P} and a false statement;
the enclosing locale assumptions remain part of the exported statement.

The separate optimality development justifies using the finite grid.
Clipping each arbitrary natural cutoff at
$\mathit{clength}\cdot\mathit{scale}$ cannot increase the complete error,
including its nested minima and fallback branches. The finite argmin and
clipping lemmas therefore give the global comparison
\[
 \begin{aligned}
 &\operatorname{fs\_optimized\_parameter\_error}(N,Q)\\
 &\qquad\le
   \operatorname{fs\_first\_root\_parameter\_error}(N,r_T,r_C,Q)
   \qquad(r_T,r_C\in\mathbb N).
 \end{aligned}
\]
The tie-breaking rule makes the selected pair lexicographically least among
all natural pairs attaining that minimum. No additional mathematical premise
is added to the soundness theorem by this optimization.

For the certified square profile, $N=16$ and $Q=2^{20}$, the complete
comparison and tie proofs establish
\[
 \begin{aligned}
 \operatorname{fs\_optimal\_cutoffs}(16,2^{20})&=(10581,10581),\\
 \operatorname{fs\_optimized\_parameter\_error}(16,2^{20})
      &=E_{\mathrm{first}}(2^{20}).
 \end{aligned}
\]
These facts concern the complete error, rather than only its sampling-power
subexpression. The concrete optimized endpoint
\code{square_192_fs_optimized_incorrect_endpoint} consequently certifies the
same $2^{-137}$ bound under the same two public premises.

The specialized \code{square_192_optimized_cutoffs_code} and
\code{square_192_optimized_error_code} equations expose the proved pair and
ledger directly to code generation. They avoid evaluating the generic
$(65536+1)^2$-entry argmin grid for this instance; they are consequences of
the optimality proof, not an external search certificate. This is optimality
within the existing upper-bound family at fixed protocol parameters and query
cap. It is not a tightness theorem for attacks, an optimization over all
protocol parameters, or a proof that a smaller acceptance probability is
impossible.

\formalanchors{
\anchorrow{Optimized definitions and canonical soundness theorem}{Soundness/FS_Optimized_Soundness.thy}{fs_optimal_cutoffs; fs_optimized_parameter_error; fs_optimized_soundness}
\anchorrow{Clipping the complete error}{Soundness/FS_Cutoff_Optimization.thy}{fs_first_root_parameter_error_clip; fs_first_root_parameter_error_fallback}
\anchorrow{Global comparison and lexicographic rule}{Soundness/FS_Cutoff_Optimality.thy}{fs_optimized_parameter_error_le; fs_optimal_cutoffs_lex_least}
\anchorrow{Concrete complete-error optimum}{Soundness/FS_Square_Optimized.thy}{fs_square_optimal_cutoffs; fs_square_optimized_ledger}
\anchorrow{Certified code shortcuts and endpoint}{Soundness/FS_Square_Optimized.thy}{square_192_optimized_cutoffs_code; square_192_optimized_error_code; square_192_fs_optimized_incorrect_endpoint}
}

\section{Honest Acceptance in the Absorbing Random-Oracle Experiment}
\label{sec:honest-ro}

Generic core completeness and absorbing-RO acceptance concern different
executions. The former uses the original pure concatenation/send--read channel
within \code{exec}. The latter uses the typed random oracle to absorb each
message and derive the challenges used by the final security verifier.
The actual square-workload theorem therefore proves the whole absorbing
experiment directly rather than substituting the core theorem for it.

For fixed $a,z$ in the certified field, let $A_{\rm honest}$ be
\code{square_ro_honest_callbacks}, let $s_0$ be
\code{adversary_initial_state}, and let $\mathit E$ be
\code{ro_absorb_checked_staged_security_experiment}. The result is
\[
 z=a^{2^{1023}}\quad\Longrightarrow\quad
 \operatorname{wp\_event}(\mathit E(A_{\rm honest}),
       \lambda o.\ \neg\operatorname{is\_none}(o),s_0)=1.
\]
The endpoint equation is the only exported premise. Field/domain/workload
obligations are discharged in the concrete interpretation. The final verifier
is shared with the FS experiment, as established by the verifier-factorization
contract; the honest callbacks and malicious producer syntax are distinct
interfaces.

\subsection{Hash-only reconstruction and recorded commitments}

The callbacks rebuild the honest commitment data using typed hash queries,
without inspecting the protocol-local verifier cursor or counters. A recorded
lookup chain identifies each absorption key, input message and resulting
cursor. In an extending oracle map, querying a recorded key returns its stored
answer. Shadow replay therefore reconstructs the same trace commitments,
alpha challenges, composition commitments and terminal scalars.

The replay contracts are exact native weakest-precondition equalities, including
failure-sensitive postconditions. They are not implications conditioned on
eventual verifier success. The supporting hash-only programs preserve the
protocol-local fields; replay reuses the recorded map and the actual keys.
Repeated locations and already-cached answers are retained rather than assumed
fresh. None of these replay arguments needs collision freedom.

The trace has ten folding rounds in this profile. Composition depth is derived
from the actual decoded degree and can be zero; it is not replaced with a
fixed ten-round composition chain. The commitment proof constructs the
corresponding Merkle trees and their terminal constant words, establishing
their lengths and authentication witnesses from the actual callbacks.

\subsection{Openings, query values and FRI chains}

Honest opening callbacks reconstruct the committed words and serialize the
same leaf/path data at the actual sampled positions. The deterministic Merkle
lemma states that a path from a created power-of-two tree authenticates its
root for an in-range index. The concrete proof discharges those size and
index conditions for both siblings of every FRI opening.

The opening format and successful algebraic checks are proved separately.
Serialization alone would not show acceptance. The query-value lemmas
identify the trace openings and randomized composition evaluation. Polynomial
fold identities then identify each opened successor with the verifier's fold
expression. The original natural-number modulo operations are retained: each
successor index is the preceding index modulo the smaller layer length.
The terminal trace and composition checks use the proved constant terminal
polynomials, including composition depth zero.

The actual query-round contract combines the counted query challenge, the two
trace-power decommitments, all FRI opening steps and terminal assertions. Its local
premises describe a supported producer prefix, recorded query/opening outcomes,
oracle extension and an aligned replay state. Those conditions are derived
along the whole execution; they are not added to the public endpoint theorem.

\subsection{Chronological verifier replay}

Header replay starts \code{ro_verify_monad} from the prescribed
\code{verifier_state_from_adversary}. It consumes the serialized header,
recovers the actual roots, challenges, alphas and terminal values, and leaves
the original query continuation. The degree check uses the honest degree
theorem rather than substituting the builder's weaker round-count test.

The all-query induction decomposes a supported checked-query sequence into
its challenge, opening, message recording and tail. Map extension and exact
cursor/counter preservation establish the next replay state. It composes all
640 repetitions, consuming their transcript while retaining the actual oracle
map and the matching final cursor/counters. No challenge value is excluded,
and the proof does not infer missing evidence from an erased execution.

For an actual supported complete builder outcome, the resulting verifier WP
equals the postcondition of the successful unit-result list and its precise
final replay state. Combining that equality with the independently proved
honest builder no-failure result proves no failure of the whole experiment.
The local verifier replay contract observes full optional outcomes, not merely
inclusion of successful supports. The whole-experiment conclusion is no failure
and acceptance one; the random-oracle experiment is not a single point mass.

\subsection{Failure and resource scope}

Acceptance explicitly excludes \code{None}. Native SPMF missing mass is that
failure outcome in the underlying option PMF; no second failure layer is
introduced. Integrating the constant one over the full option PMF would count
failure too and would not express successful acceptance.

The current publication manifest checks this acceptance-one theorem with its
one endpoint premise, no hidden hypotheses or residual constraints, and no
proof oracles or skipped proofs. The generic replay lemmas retain their own
explicit local conditions; these are supplied by the concrete execution proof.

This correctness result supplies neither a malicious FS query-cap proof for
the honest callbacks nor a uniform all-state callback budget. Historical
resource upper bounds and the separate allowance obstruction are recorded in
Appendix~\ref{app:history}, with their archived scope. They are not additional
current premises of honest acceptance or the 137-bit false-endpoint theorem.

\formalanchors{
\anchorrow{Recorded hash-only replay}{Soundness/RO_Honest_Shadow_Replay.thy}{shadow_absorb_known_wp; replayable_ro_bind}
\anchorrow{Commitment replay}{Soundness/RO_Honest_Query_Reconstruction.thy}{shadow_commitment_prefix_replay}
\anchorrow{Actual honest composition}{Soundness/Square_Sequence_192_RO_Composition.thy}{square_ro_composition_chain_authenticated}
\anchorrow{Actual opening witnesses}{Soundness/Square_Sequence_192_RO_Openings.thy}{square_ro_query_opening_authenticated_trees}
\anchorrow{Created-path authentication}{Soundness/RO_Honest_Verifier_Openings.thy}{protocol_created_path_exact_wp}
\anchorrow{Indexed FRI replay}{Soundness/RO_Honest_Verifier_FRI_Chains.thy}{mfold_indexed_exact_wp}
\anchorrow{Concrete polynomial fold}{Soundness/Square_Sequence_192_RO_FRI_Algebra.thy}{square_ro_fold_expression_next_value}
\anchorrow{Whole query replay}{Soundness/Square_Sequence_192_RO_Query_Verifier.thy}{square_ro_actual_query_round_wp}
\anchorrow{Header replay}{Soundness/Square_Sequence_192_RO_Verifier_Header.thy}{square_ro_actual_header_wp}
\anchorrow{All queries and actual acceptance}{Soundness/Square_Sequence_192_RO_Completeness.thy}{square_ro_all_queries_replay_wp; square_ro_honest_experiment_acceptance_one}
}

\section{The Separate Deterministic Executable Companion}
\label{sec:execution}

Native PMF/SPMF programs specify probabilistic behavior; the executable companion
does not enumerate those distributions. The separate \code{Stark_Execution}
session defines a pure instruction interpreter with return, failure, oracle
query, state read and state write operations. Its structural bind propagates
failure and sequences the same transcript accumulator, transcript and four
counters. A probabilistic interpretation connects these instructions to the
existing native state interface.

\subsection{Whole-outcome and cache correspondence}

The channel, Merkle, prover and verifier layers prove exact interpretation
equalities, including returned data, failure and final state. In the relevant
locales, \code{run_ro prover_program = prover_monad} and
\code{run_ro verifier_program = verify_monad}. The interpreter reuses the same
polynomial, domain, index, encoding and authentication-path functions.

For a fixed pure hash function $H$ and program $p$, the fixed-oracle native
execution is a point mass at the deterministic cached-reference outcome:
\[
 (\operatorname{run\_fixed}\ H\ p)\ s
   =\operatorname{return\_pmf}(\operatorname{run\_cached}\ H\ p\ s).
\]
Here \code{return_pmf None} is failure. It is not
\code{return_spmf None}, which would introduce an additional successful
optional value. The equality covers the complete optional result/final-state
outcome and supports observations that inspect failure.

If \code{cache_consistent H s}, removing the reference interpreter's cache
gives the direct deterministic execution on the correspondingly erased state.
Consistency means that each stored answer equals $H$ at its key; it holds for
the empty cache and is preserved. The premise matters: the proved inconsistent
cache counterexample distinguishes replay of a stored answer from recomputing
$H$. This erasure theorem does not equate a fixed hash with the random-oracle
distribution used in soundness.

\subsection{Examples and assurance boundary}

Ordinary SML code generation supplies GF(5) and prime192 wrappers, using AFP's
SHA-256 implementation. The small example serves the same pedagogical role as
compact educational STARK presentations such as STARK101~\cite{starkware101};
the mechanized model makes its assumptions and proof obligations explicit. Their parameters are:
\begin{center}
\begin{tabular}{lrr}
\toprule
Parameter & GF(5) & Prime192\\
\midrule
Trace length &2&32\\
Scale &1&4\\
Query repetitions &1&8\\
Generator and shift &2&23\\
Reference fixture elements &21&1105\\
\bottomrule
\end{tabular}
\end{center}
The prime192 modulus is the certified field used by the security application,
but this workload is not the 1024/64/640 security profile. The companion proves
the order of its explicit generator 23; the security application retains its
choice-defined generator. Encoding, constructor separation, counters and the
modulo query behavior remain part of the executable correspondence.

The public generated interfaces are \code{prove_square},
\code{verify_square}, \code{prove_square_192} and \code{verify_square_192}.
The driver, fixture and Python comparison sources are retained in the execution
directory. From the repository root,
\code{tools/run-stark-execution.sh} runs both generated examples, their complete
standalone runtime suites and Python fixture comparisons. It uses native SML
exports and Isabelle's bundled Poly/ML, and works after a cache-hit build without
loose generated code. Each run writes its code, transcripts and logs to a fresh
ignored directory under \code{Execution/build/}. The normal build checks the
retained execution proofs, code generation and the included GF(5) in-theory
checks; the complete standalone prime192 runtime suite requires the explicit
runner. An \code{external_file} declaration alone does not execute that suite.

The generic/native and prime192 correspondence proofs have no proof-oracle
dependencies. The GF(5) wrappers retain the existing evaluation-oracle
dependency of their example-locale proofs, separately from the thirteen
oracle-free publication statements. Runtime checks do not prove HOL theorems,
compiler correctness or concrete SHA-256 security.

\formalanchors{
\anchorrow{Pure instruction interpreter}{Execution/Stark_Execution_Program.thy}{execution_program; run_det}
\anchorrow{Native probabilistic interpretation}{Execution/Stark_Execution_Semantics.thy}{run_prob; run_prob_bind}
\anchorrow{Cache and fixed-hash correspondence}{Execution/Stark_Execution_Cache.thy}{run_fixed_pointmass; fixed_hash_erases; inconsistent_cache_counterexample}
\anchorrow{Prover interpretation}{Execution/Stark_Execution_Prover.thy}{run_ro_prover_program}
\anchorrow{Verifier interpretation}{Execution/Stark_Execution_Verifier.thy}{run_ro_verifier_program}
\anchorrow{Explicit large-field generator}{Execution/Stark_Execution_Generator_192.thy}{execution_generator_192_order}
\anchorrow{Large-field executable instance}{Execution/Stark_Execution_Square_192.thy}{prove_square_192; verify_square_192}
}

\section{Related Work}
\label{sec:related-work}

The following works provide the main mathematical and formal-verification context.
\begin{itemize}[leftmargin=*,nosep]
  \item \textbf{IOPs, FRI and STARKs.}
    Ben-Sasson, Chiesa and Spooner introduce interactive oracle proofs and
    their compilation into noninteractive proofs~\cite{ben-sasson2016iop}.
    The original FRI paper develops Reed--Solomon proximity testing
    \cite{ben-sasson2018fri}, while the STARK construction combines algebraic
    encodings and cryptographic techniques for transparent verifiable
    computation~\cite{ben-sasson2018stark}.
  \item \textbf{Simplified FRI soundness and its mechanization.}
    Garreta, Mohnblatt and Wagner give an MCA-based round-by-round soundness
    analysis~\cite{garreta2025simplifiedfri}. The \code{simple-rbr-fri}
    project reports Lean~4 proofs of FRI soundness and perfect completeness;
    its cited snapshot leaves a supporting instantiation lemma unformalized
    \cite{simple_rbr_fri_code}.
  \item \textbf{WHIR.}
    Arnon et al. develop proximity testing for constrained Reed--Solomon codes
    with fast verification~\cite{arnon2024whir}. The work also supplies the
    instantiation lemma referenced by \code{simple-rbr-fri}.
  \item \textbf{Fiat--Shamir security of FRI.}
    Block et al. analyze FRI, batched FRI and related SNARKs through
    round-by-round soundness and knowledge soundness~\cite{block2023fsfri}.
    Block and Tiwari examine concrete noninteractive FRI parameters and the gap
    between provable and conjectured security~\cite{block2024concretefri}.
  \item \textbf{ArkLib.}
    This Lean project develops modular infrastructure for interactive oracle
    reductions, composition and lifting, with FRI, Merkle commitments and
    Fiat--Shamir among its development targets~\cite{arklib_code}.
  \item \textbf{Mechanized SNARK soundness.}
    Bailey and Miller formalize soundness proofs for six pairing-based SNARK
    constructions, including Groth16, in Lean~3~\cite{bailey2023snarks}.
  \item \textbf{Cairo and S-two AIR verification.}
    Avigad et al. verify the algebraic encoding of Cairo execution in Lean~3
    \cite{avigad2021cairo} and the implication from S-two AIR satisfiability to
    the computational claim in Lean~4~\cite{avigad2026stwo}.
  \item \textbf{VCVio.}
    Tuma and Hopper formalize oracle computations, a forking lemma and the
    Fiat--Shamir transform for sigma protocols in Lean~\cite{tuma2024vcvio}.
  \item \textbf{CryptHOL.}
    Basin, Lochbihler and Sefidgar provide an Isabelle/HOL framework for
    probabilistic games, oracle access and mechanized cryptographic security
    arguments~\cite{basin2020crypthol}. This development uses its native
    SPMF/GPV and oracle infrastructure directly; the concrete STARK compiler
    and event-cover proof remain protocol-specific.
\end{itemize}

This work contributes an integrated Isabelle/HOL development connecting
workload semantics, honest completeness and cryptographic soundness to the same
executable probabilistic verifier. It complements AIR-correctness results with
acceptance bounds and carries FRI reasoning through authenticated openings,
transcript absorption and adaptive Fiat--Shamir execution. Its technical
connections include acceptance-preserving replay for oracle-adaptive producers
with private randomness, explicit replay allowances, exact modulo-sampler
analysis, and weighted-path and event-sensitive probability accounting.

Alongside these reusable proofs, the development supplies a discharged concrete
instance: correct square-sequence endpoints have honest acceptance probability
one, while incorrect endpoints have acceptance probability at most $2^{-137}$
for modeled producers making at most $2^{20}$ oracle calls, using the certified
192-bit field and 640 query repetitions. The contribution is scoped to a fixed
statement in a classical field-valued random-oracle model. Comparisons with
general frameworks and other protocols concern the verified layers and their
connections; they do not establish priority or superiority of security parameters.

\section{Current Limitations and Future Work}

The main guarantees concern a fixed statement in a classical, typed,
domain-separated, field-valued random-oracle model. Adversaries may make adaptive
oracle queries, use private randomness and fail, but the public query restriction
retains finite private support and bounds every syntactic continuation. Native
PMF/SPMF and GPV types admit more general behavior; using them does not enlarge
the adversary class covered by the STARK theorem. Extending that class, bounding
only supported executions or admitting quantum queries requires further proofs.

The concrete \(2^{-137}\) result is an acceptance-probability bound under the
stated \(2^{20}\) call restriction, not an unrestricted 137-bit work factor.
The cutoff selector minimizes the existing complete upper-bound family,
including its fallbacks. Global cutoff comparison and the certified
\((10581,10581)\) pair establish neither a matching malicious execution nor
optimal protocol parameters. The historical residual/tightness investigations
in the appendix are separate results and open questions, not unproved premises
of the current endpoint.

The foundation uses Isabelle's native infrastructure, but the whole proof is
not protocol-independent. Concrete domain/table bridges, authenticated opening
extraction, transcript chronology, event coverage and numerical accounting
remain STARK clients. The FS compiler proves a transformation for this verifier
and attacker interface, not a universal Fiat--Shamir theorem for arbitrary
interactive proofs. Further extraction requires an explicit interface and
instantiation proof rather than a directory move or presentation wrapper.

Generic Merkle collision extraction and an oracle-execution reduction already
exist. They retain encoding, path, cached-oracle and interaction premises; they
do not identify unopened entries of a malicious commitment. A collision-resistant
concrete hash instantiation requires a separate cryptographic model and
reduction. Neither the random oracle nor the SHA-256 runtime supplies it.

The sampler retains field-to-natural projection followed by modulo reduction.
Exact quotient/remainder fibers and their envelope account for bias without
assuming divisibility. A uniform or rejection sampler might simplify terms or
improve constants, but changes the protocol experiment and needs a new
preservation and soundness argument. Likewise, replacing the proved union of
trace/composition events by an intersection or maximum requires a genuine
execution-level relation, not a numerical rearrangement.

The replay allocation and direct producer-plus-verifier budget describe
different computations. Costs are transported only through the established
correspondences. Cached calls are not free, and the sufficient \(664Q\) staged
allocation is not proved minimal. Improvements require exact accounting for
the transformed programs and the particular events being charged.

The executable companion proves full-outcome native correspondence and
consistent-cache erasure for fixed hashing. Its prime192 fixture has 32 trace
entries, scale 4 and 8 query repetitions, with a separately certified explicit
generator. It is not an execution of the 1024/64/640 security profile, whose
generator remains choice-defined. Runtime assertions and fixture comparisons
are tests, not HOL proofs; SML compiler correctness and deployment behavior are
outside the artifact. The GF(5) correspondence retains its documented ordinary
Isabelle evaluation-oracle dependency, distinct from the oracle-free selected
publication statements.

No zero-knowledge simulator or knowledge extractor is formalized. Neither
property follows from honest completeness and false-statement soundness.
Concrete-hash hardness, QROM security, adaptive-statement or multi-instance
security, independent clean-machine reproduction and a verified compiler remain
separate goals, each requiring its own model and validation.

\section{Checking the Current Artifact}
\label{sec:current-validation}

The formal artifact uses Isabelle2025-2 and the dated AFP2026-08-07 archive,
registered as an Isabelle component. \code{REPRODUCING.md} gives the download
URL, SHA-256 checksum and installation commands. From the repository root the normal
validation command is:
\begin{lstlisting}[language={}]
/home/dm656/Isabelle2025-2/bin/isabelle build -D .
\end{lstlisting}
Use the corresponding launcher if Isabelle is installed elsewhere. This checks
the twelve daily proof sessions. To generate \code{output/document.pdf}, request
the separate native publication session explicitly:
\begin{lstlisting}[language={}]
/home/dm656/Isabelle2025-2/bin/isabelle build -d . -D Publication
\end{lstlisting}
The daily build retains the Core heap; the optional publication session replays
the required sources from an earlier parent so that all document theories can
be selected natively. Execution is a separate session and is not included in
the proof PDF. The
authored technical report is another separate TeX artifact: successfully
typesetting this report is not evidence that an Isabelle proof was checked.

After an accepted build, extract the current statement and trust manifest
without initiating another build:
\begin{lstlisting}[language={}]
/home/dm656/Isabelle2025-2/bin/isabelle export -n -d . \
  -O /tmp/stark-publication-exports \
  -x '*Publication_Manifest:publication/*' Stark
\end{lstlisting}
Choose an empty output directory for a new check. The \code{-n} flag prevents
the export command from starting a build. The current checkout does not contain
the older project-specific publication/source-audit scripts; commands and
successful logs from archived revisions must not be represented as current
checking machinery.

The manifest exports thirteen actual typed statements, with their sort
hypotheses and explicit premises. It checks the expected exported premise
counts, absence of hidden hypotheses and residual flex--flex constraints, and
absence of proof oracles and skipped proofs for those statements. A generic
locale predicate is itself an exported premise: it retains the inherited
protocol and verifier obligations. The concrete square interpretation
discharges these conditions rather than silently deleting them.

The imported-axiom inventory includes the HOL basis and definitional, type and
class declarations. It describes the imported environment, not a minimal or
transitive per-theorem axiom-use certificate. An empty oracle list does not make
HOL axiom-free or establish fidelity to a deployed implementation.

The normal execution-session build checks native correspondence theories, SML
generation and code-dependency checks, and the in-theory GF(5) runtime checks.
The standalone prime192 driver and fixtures are separate validation material;
their presence as \code{external_file} declarations does not execute the full
prime192 runtime suite. Run \code{tools/run-stark-execution.sh} for both examples,
or select one with \code{--example gf5} or \code{--example prime192}. This requires
Bash and Python~3; the runner uses Isabelle's bundled Poly/ML and the launcher
on \code{PATH}, or the path supplied by \code{ISABELLE}. The release-preparation
checks and their external dependency-cache reuse are recorded in
\code{VALIDATION.md}; they are distinct from independent clean-machine
reproduction. Neither runtime testing nor deterministic SHA-256
execution instantiates random-oracle security with a concrete hash function.
The companion's cache-erasure theorem retains its cache-consistency premise.

To audit the current claims, read the following interfaces in order:
\begin{enumerate}[leftmargin=*]
  \item The native interface and protocol definitions: confirm state, failure,
    transcript absorption, typed hash inputs and modulo sampling.
  \item Adversary syntax and native execution: confirm finite private support,
    the all-continuation query restriction and repeated-call counting.
  \item Private normalization and the compiled experiment: check the exact
    observation correspondence and residual-transcript qualification.
  \item Actual event coverage and assembly: require the supported accepting-run
    cover and every component probability proof.
  \item First-root and optimized soundness: retain parameter guards, fallbacks,
    locale premises and the separate query-accounting definitions.
  \item Concrete field/workload interpretation, numerical certificate, global
    cutoff comparison and honest absorbing-RO execution: check the endpoint
    premises and distinguish published parameters from the smaller runtime.
\end{enumerate}

Current sources and fresh build/export results determine current claims.
Historical appendices document archived proofs and diagnostics, not additional
build roots. A successful local build is not independent clean-machine
reproduction, SML compiler verification or an audit of every historical theory.

\section{Conclusion}

The current development connects fixed-workload semantics, honest correctness
and query-bounded cryptographic soundness to the specified STARK verifier.
Algebra/FRI mathematics, generic Merkle collision reduction and native
execution/replay/accounting interfaces support the concrete FS compilation
and composition proof. Its final event bound is supplied by actual coverage
and component proofs, rather than assumed as an abstract ledger obligation.

For the certified square profile, correct endpoints have actual absorbing-RO
honest acceptance one; incorrect endpoints have acceptance at most $2^{-137}$
under the original uniform syntactic cap $2^{20}$. The first-root route and
its automatic cutoff interface preserve the parameter fallbacks. Separate
proofs establish global comparison within that error family and the selected
pair $(10581,10581)$. Exact modulo-fiber accounting retains the sampler's bias.

These guarantees retain their locale assumptions, adversary restrictions and
classical field-valued random-oracle scope. The deterministic companion proves
a different, fixed-hash execution correspondence and uses smaller runtime
examples. No bit-hash hardness, compiler-correctness, zero-knowledge,
knowledge-extraction or QROM theorem follows. The historical appendix preserves
earlier approaches and unresolved diagnostic questions without presenting them
as current build dependencies or additional certified security profiles.

\section*{Acknowledgements}
\addcontentsline{toc}{section}{Acknowledgements}

Parts of this research were carried out using OpenAI's
\textbf{ChatGPT for Academic Researchers} programme. The formalization work
uses the Isabelle PIDE MCP server~\cite{pide_mcp_code}.

Parts of the soundness proof, in particular the MCA approach, were inspired
by \href{https://github.com/zksecurity/simple-rbr-fri}{zkSecurity's
\code{simple-rbr-fri} project}.

\appendix

\section{Main Isabelle Statements}
\label{app:statements}

\subsection{Current thirteen-statement manifest}

The current publication interface is
\theorysource{Soundness/Publication_Manifest.thy}. It checks and exports these
thirteen existing statements with their types, sorts and explicit premises.
The counts below are exported premise counts. For generic
\code{soundness.*} facts, the locale predicate contributes one premise in
addition to theorem-local conditions; it includes inherited \code{verifier}
and \code{stark} assumptions. The concrete square statements discharge those
locale obligations.

\begingroup
\scriptsize
\begin{longtable}{
  >{\raggedright\arraybackslash}p{0.28\linewidth}
  >{\raggedright\arraybackslash}p{0.50\linewidth}
  >{\raggedright\arraybackslash}p{0.10\linewidth}}
\toprule
Claim & Exported theorem & Premises \\
\midrule
\endfirsthead
\toprule
Claim & Exported theorem & Premises \\
\midrule
\endhead
Square workload semantics & \code{square_192_valid_trace_iff} & 0\\
Honest RO acceptance & \code{square_ro_honest_experiment_acceptance_one} & 1\\
Shared final verifier & \code{soundness.staged_fs_verifier_factorization} & 1\\
Fixed-private-strategy acceptance transport & \code{soundness.fs_compile_replay_acceptance} & 3\\
Private-randomness mixture transport & \code{soundness.fs_compiled_mixture_acceptance} & 2\\
Sufficient replay allowance & \code{soundness.fs_replay_total_allowance} & 1\\
Generic first-root soundness & \code{soundness.fs_first_root_soundness} & 3\\
Exact first-root/prefix ledger comparison & \code{square_fs_first_root_ledger_balance} & 0\\
Certified numerical target & \code{square_fs_137_prefix_target} & 0\\
Concrete false-endpoint bound & \code{square_192_fs_137_incorrect_endpoint} & 2\\
Generic optimized soundness & \code{soundness.fs_optimized_soundness} & 3\\
Global cutoff comparison & \code{soundness.fs_optimized_parameter_error_le} & 1\\
Optimized concrete false-endpoint bound & \code{square_192_fs_optimized_incorrect_endpoint} & 2\\
\bottomrule
\end{longtable}
\endgroup

The fixed-private-strategy acceptance theorem has locale, fixed-strategy and
query-bound premises. The private mixture removes the fixed-strategy premise
by the proved normalization and probability correspondence. First-root and
optimized soundness have locale, query-bound and false-statement premises.
The cutoff comparison retains the locale predicate but introduces no attacker
premise. The two concrete false-endpoint theorems have just incorrect endpoint
and query restriction. The ledger identity and numerical target are arithmetic
facts, not equalities for actual acceptance probability.

For each listed theorem the manifest rejects hidden hypotheses, residual
flex-flex constraints, proof oracles and skipped proofs, and checks the expected
premise count. It also exports the imported theory axiom inventory. That inventory
includes foundational and definitional/type/class declarations: it is not a
minimal per-theorem axiom-use certificate, and empty oracle lists do not make
HOL axiom-free. These checks are evidence about the named formal statements;
typesetting the report is not itself proof validation.

\subsection{Generic completeness and the canonical FS bound}

In locale \code{verification}, the retained generic completeness result is:

\begin{lstlisting}
lemma completeness:
  assumes "honest_trace_valid"
  shows "wp_event exec (\<lambda>a. Option.is_none a) init_state \<le> (0::real)"
\end{lstlisting}

The premise \code{honest_trace_valid} supplies the honest trace's algebraic
constraint validity; the locale retains degree, domain and encoding obligations.
This is the pure-concat combined prover/verifier execution \code{exec}. It is
separate from honest acceptance in the absorbing-RO experiment below.

In locale \code{soundness}, the uniform first-root theorem is:

\begin{lstlisting}
theorem fs_first_root_soundness:
  assumes bound: "fs_query_bound Q P" and false_statement: "\<not>exists_valid_trace"
  shows "fs_acceptance_probability P \<le> fs_first_root_parameter_error N rT rC Q"
\end{lstlisting}

The canonical optimized theorem instantiates it at the existing selected cutoffs:

\begin{lstlisting}
theorem fs_optimized_soundness:
  assumes bound: "fs_query_bound Q P" and false_statement: "\<not>exists_valid_trace"
  shows "fs_acceptance_probability P\<le>fs_optimized_parameter_error N Q"
\end{lstlisting}

The separate optimality proof establishes the global comparison in the same locale:

\begin{lstlisting}
lemma fs_optimized_parameter_error_le:
  "fs_optimized_parameter_error N Q \<le> fs_first_root_parameter_error N rT rC Q"
\end{lstlisting}

Here \code{N} is the analysis parameter of the existing error family. The
comparison quantifies over all natural cutoff pairs; its proof includes finite
clipping and the complete minimum/fallback structure. Neither the optimized
soundness theorem nor this comparison changes the adversary class, verifier
or mathematical premises.

\subsection{Concrete false-endpoint and honest-acceptance statements}

For arbitrary fixed \code{a,z :: field_192}, the ordinary concrete endpoint is:

\begin{lstlisting}
corollary square_192_fs_137_incorrect_endpoint:
 assumes incorrect: "z\<noteq>a^(2^1023)" and bound: "fs_query_bound (2^20) P"
 shows "(soundness.fs_acceptance_probability
   field_generator_192 field_generator_192 64 1024 2 stark_mod_ring_decode
   field_cardinality_192 (square_workload_spec 1024 a z) 640
   (square_workload_spec2 1024 a z) P)\<le>(1/2^137::real)"
\end{lstlisting}

The optimized concrete interface gives the same certified bound and premises:

\begin{lstlisting}
corollary square_192_fs_optimized_incorrect_endpoint:
  assumes incorrect: "z\<noteq>a^(2^1023)" and bound: "fs_query_bound (2^20) P"
  shows "(soundness.fs_acceptance_probability
    field_generator_192 field_generator_192 64 1024 2 stark_mod_ring_decode
    field_cardinality_192 (square_workload_spec 1024 a z) 640
    (square_workload_spec2 1024 a z) P)\<le>(1/2^137::real)"
\end{lstlisting}

The field cardinality, generator, trace length 1024, scale 64, two shifted
powers and 640 repetitions are fixed by the concrete interpretation. The
scalar is already \code{real}; no custom probability conversion is required.
The syntactic cap counts repeated queries and bounds every continuation. Its
native GPV interaction-bound implication is proved, not assumed to be an
equivalence of attacker classes. The sufficient replay allocation is $664Q$ in
this profile, distinct from the original cap and from exact execution costs.

The matching honest theorem is:

\begin{lstlisting}
lemma square_ro_honest_experiment_acceptance_one:
  assumes endpoint: "z=a^(2^1023)"
  shows "wp_event
    (s.ro_absorb_checked_staged_security_experiment (square_ro_honest_callbacks a z))
    (\<lambda>out. \<not>Option.is_none out) s.adversary_initial_state=1"
\end{lstlisting}

As in its source, \code{s} is the instantiated square-workload soundness locale.
The correct endpoint is the sole premise. The theorem concerns the actual final
absorbing verifier from its original adversary initial state and explicitly
observes non-failure. It adds no claim that these honest callbacks satisfy the
malicious producer's syntactic query restriction or an all-state staged allowance.

\section{Selected Proof Patterns}

This appendix gives a more detailed view of the recurring proof patterns used
throughout the development.  The aim is to make the Isabelle source easier to
navigate without reproducing long proof scripts.

\subsection{Weakest-precondition normalization}

The WP calculus is used as a normalization tool.  A goal about the probability
of a verifier event is unfolded until the structure of the program is visible.
For straight-line code, bind rules expose the intermediate result and state.
For assertions, the goal splits into a deterministic side condition and the
continuation.  For list traversals, the proof proceeds by induction over the
list or by specialized no-failure rules for map and fold combinators.

In the completeness proof, this pattern turns the final probabilistic
statement into a collection of replay and algebra obligations.  In the
soundness proof, it is used less aggressively at the endpoint: many low-level
WP facts are proved in smaller theories and then composed.  This prevents the
final public theorem from depending on large unfolded verifier states.

\subsection{Support-to-probability reasoning}

Many zero-probability arguments have the same shape.  First prove that no
outcome in the support of a distribution satisfies the bad event.  Then use
the native probability semantics to conclude that the event has probability
zero.  This is particularly useful for honest completeness, where the honest
execution may sample random challenges but no sampled value can cause rejection
once all deterministic verifier checks are known to succeed.

For soundness, the same support reasoning appears inside conditional branches.
Some events are proved impossible under clean side conditions.  The remaining
branches are charged to collision or target-hit bounds.  Separating impossible
branches from unlikely branches keeps the final bound more meaningful than a
single coarse upper bound.

\subsection{Event inclusion and union bounds}

The soundness proof repeatedly establishes inclusions of the form
\[
  E \subseteq E_1 \cup \cdots \cup E_n.
\]
In WP form, this yields
\[
  \Pr[E] \le \Pr[E_1] + \cdots + \Pr[E_n].
\]
The Isabelle development phrases these facts as lemmas about
\code{wp_event}.  This allows the proof to combine deterministic reductions
and probabilistic bounds without changing the underlying experiment.

The decomposition is intentionally fine-grained.  Hash collisions, Merkle
inconsistencies, composition randomization failures, query misses, and FRI
failures are separate events because they are bounded by different arguments.
The final theorem is therefore a sum of semantically meaningful terms rather
than an opaque bound.

\subsection{Prefix-fixed target accounting}

Fiat-Shamir soundness requires attention to timing.  A target set for a
challenge may depend on what the adversary has already committed to, but it
must not depend on the challenge value being sampled.  The development models
this by using prefix states.  A deterministic extraction lemma identifies the
target set after a transcript prefix, and a dynamic target-budget lemma bounds
the probability that the subsequent random-oracle output lands in that set.

This pattern is used for composition alpha challenges and for query-index
events.  It is also the reason broad existential events are kept away from the
final public route.  If an event existentially chooses witnesses after the
challenge is known, it is usually too coarse for direct target accounting.

\subsection{Partial-opening reasoning}

Merkle proofs in this development are evidence-local.  An authenticated
opening says that a concrete value at a concrete index checks against a
concrete root under the current hash map.  A partial table candidate is a
function or table compatible with the sampled authenticated openings.  The
proof never assumes that sampled openings determine unopened entries.

This is the key to the publication-facing proof architecture.  Completeness
can use created-tree correctness because the prover is honest and the full
tables are available in the construction.  Soundness cannot use that route for
malicious transcripts.  It must instead show that any inconsistency in the
opened evidence implies a collision, and otherwise reason only about the
partial candidates actually supported by the verifier's checks.

\subsection{FRI evidence, products, and fibers}

The FRI proof layers extract challenge lists, query lists, layer openings, and
final checks from accepted verifier executions.  Trace and composition FRI are
separated because they use different low-degree predicates, but the extracted
evidence has the same shape.  Actual-query lemmas then count repeated choices
by exact products.  Raw-list preimage lemmas connect those products to the
concrete sampler, while staged prequery and relation-fiber lemmas account for
adaptive oracle history.

The endpoint composes these facts into closed parameter expressions.  The
current coefficients may still be conservative, but tightening them should
preserve the products and fibers rather than replacing them with trivial
probability bounds.  The verifier-local evidence layer and public theorem shape
can remain unchanged.

\section{Notation and Terminology}

The report uses the following informal terms consistently with the Isabelle
development.

\begin{description}[leftmargin=3.2cm,style=nextline]
  \item[Trace table]
  The committed evaluations associated with the computation trace.  In the
  prover locale these are generated from the honest trace polynomial; in the
  soundness proof they are represented by verifier-observed openings and
  candidate predicates.

  \item[Composition table]
  The committed evaluations of the randomized composition polynomial.  The
  verifier checks consistency between trace openings, composition openings, and
  the alpha challenges.

  \item[Partial opening]
  A sampled authenticated opening checked against a Merkle root.  Partial
  openings constrain only the sampled indices and do not determine complete
  tables.

  \item[Candidate]
  A table or polynomial object compatible with the available partial evidence.
  Candidate predicates are used to state algebraic disagreement without
  assuming full reconstruction from Merkle proofs.

  \item[Bad event]
  An event that explains why a false statement could be accepted.  Soundness
  bounds the probability of the union of these events.

  \item[Prefix target]
  A target set determined by a transcript prefix and oracle state before the
  next Fiat-Shamir output is sampled.  Prefix targets are used for valid
  adaptive random-oracle accounting.

  \item[Transcript absorption]
  A domain-separated random-oracle query at
  \code{TranscriptAbsorb (PState s) x} that updates the Fiat-Shamir transcript
  state and is included in the staged hash-query budget.

  \item[Earlier staged parameter error]
  The adversary-independent closed right-hand side of
  \code{ro_absorb_stark_soundness_conditioned_explicit}, selected by the
  zero/nonzero \code{ceil_log clength} branch.
\end{description}

\section{Current Source Anchor Index}
\label{app:source-index}
This index collects the current source anchors from the main narrative.
The historical appendix uses its own pinned source revision.
\formalanchors{
\anchorrow{Protocol definitions}{Core/Stark.thy}{stark; prover; verifier}
\anchorrow{Generic completeness}{Completeness/Completeness_Verifier.thy}{verification.completeness}
\anchorrow{Generic optimized soundness}{Soundness/FS_Optimized_Soundness.thy}{soundness.fs_optimized_soundness}
\anchorrow{Concrete 137-bit endpoint}{Soundness/FS_Square_192_137.thy}{square_192_fs_137_incorrect_endpoint}
\anchorrow{Publication statement checks}{Soundness/Publication_Manifest.thy}{thirteen exported statements}
\anchorrow{Exact multiplicative order}{Core/Stark.thy}{exact_order}
\anchorrow{Trace-domain generator}{Core/Stark.thy}{g}
\anchorrow{Evaluation-domain generator}{Core/Stark.thy}{h}
\anchorrow{Shifted evaluation domain}{Core/Stark.thy}{eval_domain}
\anchorrow{Trace powers}{Core/Stark.thy}{trace_powers_of}
\anchorrow{Native state and probability interfaces}{Probability/Native_State_Interface.thy}{state_monad; wp; wp_event; failure_prob; finite_program}
\anchorrow{Private sampling and uniform syntactic cap}{Soundness/FS_Program.thy}{fs_program; fs_query_bound}
\anchorrow{Native oracle interpretation}{Soundness/FS_Native_Execution.thy}{fs_gpv; fs_run; fs_gpv_interaction_bound}
\anchorrow{State-monad assertion}{Probability/Native_State_Interface.thy}{assert}
\anchorrow{Event WP}{Probability/Native_State_Interface.thy}{wp_event}
\anchorrow{Failure probability}{Probability/Native_State_Interface.thy}{failure_prob}
\anchorrow{WP rules}{Probability/Native_State_Interface.thy}{wp_return; wp_bind}
\anchorrow{Transcript send}{Core/Channel_Core.thy}{send}
\anchorrow{Transcript read}{Core/Channel_Core.thy}{read}
\anchorrow{Domain-separated absorption}{Core/Channel_Core.thy}{TranscriptAbsorb}
\anchorrow{Staged message absorption}{Soundness/Staged_Security_Experiment_Core.thy}{ro_record_staged_message}
\anchorrow{Absorbing verifier}{Soundness/Staged_Security_Experiment_RO_Checked.thy}{ro_verify_monad}
\anchorrow{Transcript query budget}{Soundness/Staged_Security_Experiment_RO_Transcript_Budgets.thy}{ro_checked_staged_transcript_hash_query_budget_for}
\anchorrow{Merkle creation}{Core/Merkle_Tree.thy}{create}
\anchorrow{Protocol locale}{Core/Stark.thy}{stark}
\anchorrow{Prover locale}{Core/Stark.thy}{prover}
\anchorrow{Verifier locale}{Core/Stark.thy}{verifier}
\anchorrow{Prover program}{Core/Stark.thy}{prover_monad}
\anchorrow{Verifier program}{Core/Stark.thy}{verify_monad}
\anchorrow{Core protocol locale}{Core/Stark.thy}{stark}
\anchorrow{Maximum declared degree}{Core/Stark.thy}{maxDegree}
\anchorrow{Composition polynomial}{Core/Stark.thy}{cp}
\anchorrow{Query raw-preimage envelope}{Soundness/Soundness_Core_Base.thy}{query_raw_preimage_card_envelope}
\anchorrow{Exact modulo preimage arithmetic}{Mathematics/FRI/Modulo_Arithmetic.thy}{card_modulo_residue_preimage; card_modulo_set_preimage}
\anchorrow{Raw/natural sampler correspondence}{Soundness/Soundness_Core_Events.thy}{card_query_index_raw_preimage_eq_nat_preimage}
\anchorrow{Honest trace algebra}{Completeness/Completeness_Core.thy}{honest_trace_algebra}
\anchorrow{Honest trace validity}{Completeness/Completeness_Core.thy}{honest_trace_valid}
\anchorrow{Verification locale}{Completeness/Completeness_Core.thy}{verification}
\anchorrow{WP completeness reduction}{Completeness/Completeness_Verifier.thy}{completeness_wp_reduction}
\anchorrow{Completeness theorem}{Completeness/Completeness_Verifier.thy}{completeness}
\anchorrow{Producer syntax and restriction}{Soundness/FS_Program.thy}{fs_program; fs_query_bound}
\anchorrow{Native GPV execution and replay}{Soundness/FS_Native_Execution.thy}{fs_native_gpv_replay; fs_gpv_interaction_bound}
\anchorrow{Original attack and acceptance}{Soundness/FS_Adversary_Model.thy}{fs_security_experiment; fs_acceptance_probability}
\anchorrow{Staged transcript absorption}{Soundness/Staged_Security_Experiment_Core.thy}{ro_record_staged_message}
\anchorrow{Actual absorbing verifier}{Soundness/Staged_Security_Experiment_RO_Checked.thy}{ro_verify_monad}
\anchorrow{Uniform first-root endpoint}{Soundness/FS_First_Root_Soundness.thy}{fs_first_root_soundness}
\anchorrow{Independent folding and agreement}{Mathematics/FRI/Multiround_Fold_Soundness.thy}{multiround folding contracts}
\anchorrow{Certified-path collision extraction}{Evidence/Lookup_Path_Collision.thy}{lookup_path_collision_complete}
\anchorrow{Oracle-execution Merkle reduction}{CryptHOL/Merkle_Collision_Reduction.thy}{merkle_attack_collision_reduction; merkle_reduction_queries}
\anchorrow{Native adversary execution}{Soundness/FS_Native_Execution.thy}{fs_run; fs_gpv_interaction_bound}
\anchorrow{Private strategy normalization}{Soundness/FS_Private_Randomness_Normalization.thy}{fs_normalize_security; fs_normalize_fixed; fs_normalize_bound}
\anchorrow{Compiled experiment and correspondence}{Soundness/FS_Adaptive_Staged_Embedding.thy}{fs_compile_replay; fs_compiled_mixture_acceptance}
\anchorrow{Replay allowances}{Soundness/FS_Replay_Allowance.thy}{fs_replay_budgets; fs_compile_replay_controlled}
\anchorrow{Direct experiment cost}{Soundness/FS_Probability_Accounting.thy}{fs_direct_hash_budget}
\anchorrow{Actual event coverage and ledger}{Soundness/Stark_Soundness_Composition.thy}{stark_acceptance_components; stark_component_bound}
\anchorrow{Concrete bound assembly}{Soundness/FS_Soundness_Assembly.thy}{fs_composed_security_bound; fs_compiled_witnessed_bound_suffices}
\anchorrow{Optimized final bound}{Soundness/FS_Optimized_Soundness.thy}{fs_optimized_soundness}
\anchorrow{Trace low-degree predicate}{Soundness/Soundness_Core_Base.thy}{trace_table_low_degree}
\anchorrow{Composition low-degree predicate}{Soundness/Soundness_Core_Base.thy}{composition_table_low_degree}
\anchorrow{Decoded all-query agreement event}{Soundness/Soundness_FRI_Robust_Security_Classification.thy}{ro_absorb_checked_staged_security_robust_decoded_all_queries_consistent}
\anchorrow{Authenticated partial-table evidence}{Soundness/Soundness_Partial_Merkle.thy}{authenticated_opening_in; partial_authenticated_table}
\anchorrow{Compatibility with observed values}{Soundness/Soundness_Partial_Merkle.thy}{table_agrees_with_authenticated_openings}
\anchorrow{Inconsistent openings imply collision}{Soundness/Soundness_Partial_Merkle.thy}{inconsistent_authenticated_openings_imp_hash_collision}
\anchorrow{Authenticated FRI query evidence}{Soundness/Soundness_FRI_RO_Actual_Query_Accepted_Evidence.thy}{ro_recorded_query_fri_accepted_evidence}
\anchorrow{Exact modulo query fibers}{Mathematics/FRI/Modulo_Arithmetic.thy}{card_modulo_residue_preimage; card_modulo_set_preimage}
\anchorrow{Concrete raw-preimage envelope}{Soundness/Soundness_Query_Index_Modulo_Bounds.thy}{card_query_index_raw_preimage_le_query_envelope}
\anchorrow{Query-list probability envelope}{Soundness/Soundness_FRI_Query_List_Exact_Product.thy}{query_index_raw_list_preimage_probability_bound}
\anchorrow{Query relation fibers}{Soundness/Soundness_FRI_Query_List_Prequery.thy}{query_index_raw_list_relation_fiber_bound}
\anchorrow{Hash collision budget}{Soundness/Soundness_Oracle_Budgets.thy}{hash_collision_budget_value}
\anchorrow{Hash target budget}{Soundness/Soundness_Oracle_Target.thy}{hash_target_budget_value}
\anchorrow{Authenticated partial-table evidence}{Soundness/Soundness_Partial_Merkle.thy}{partial_authenticated_table; table_agrees_with_authenticated_openings}
\anchorrow{Authenticated accepted evidence}{Soundness/Soundness_FRI_RO_Actual_Query_Accepted_Evidence.thy}{ro_recorded_query_fri_accepted_evidence}
\anchorrow{Recorded fold-chain evidence}{Soundness/Soundness_FRI_RO_Actual_Query_Fold_Evidence.thy}{ro_fri_fold_recorded_chain_evidence}
\anchorrow{Supported authenticated query evidence}{Soundness/Soundness_FRI_RO_Actual_Query_Accepted_Evidence.thy}{ro_verifier_query_round_program_recorded_fri_accepted_evidence}
\anchorrow{Explicit security classifier}{Soundness/Soundness_FRI_Conditioned_Explicit_Security_Classification.thy}{ro_absorb_checked_staged_security_clean_conditioned_explicit_classification}
\anchorrow{Verifier specification obligations}{Examples/Square_Sequence_Workload.thy}{square_workload_mod_ring_verifier}
\anchorrow{Actual valid-trace implication}{Soundness/Soundness_Square_Sequence_Workload.thy}{square_workload_valid_trace_endpoint}
\anchorrow{Reference metadata}{Soundness/Soundness_Square_Sequence_Workload.thy}{square_workload_reference_metadata}
\anchorrow{Shared-domain true and false cases}{Soundness/Soundness_Square_Sequence_Workload.thy}{square_workload_prime_field_nonvacuity}
\anchorrow{Exact primality certificate}{Examples/Prime_192_Certificate.thy}{certificate_prime_192_is_prime}
\anchorrow{Concrete field cardinality}{Examples/Prime_Field_192.thy}{CARD_field_192}
\anchorrow{Fixed proper evaluation domain}{Examples/Prime_Field_192.thy}{field_192_domain}
\anchorrow{Concrete soundness-locale obligations}{Soundness/Soundness_Square_Sequence_192.thy}{square_192_soundness}
\anchorrow{Actual valid-trace and nonvacuity results}{Soundness/Soundness_Square_Sequence_192.thy}{square_192_semantics}
\anchorrow{Witness-preserving replay}{Soundness/FS_Prefix_Witness_Replay.thy}{fs_prefix_security_witness_elimination; fs_prefix_head_budget; fs_prefix_tail_budget}
\anchorrow{Composition-prefix and query-phase charges}{Soundness/FS_Prefix_Target_Accounting.thy}{fs_compiled_composition_prefix_bound; fs_compiled_query_phase_bound}
\anchorrow{Direct producer/verifier allowance}{Soundness/FS_Probability_Accounting.thy}{fs_direct_hash_budget; fs_direct_range_target_collision_budget}
\anchorrow{Total staged replay allowance}{Soundness/Publication_Reduction_Facts.thy}{fs_replay_total_allowance}
\anchorrow{Actual first-root target bounds}{Soundness/FS_First_Root_Accounting.thy}{fs_first_root_builder_bound; fs_first_root_security_bound}
\anchorrow{Nested-alpha target bound}{Soundness/FS_First_Root_Alpha.thy}{fs_compiled_first_root_alpha_security}
\anchorrow{Complete native FS theorem}{Soundness/FS_First_Root_Soundness.thy}{fs_compiled_first_root_nonempty_bound; fs_first_root_soundness}
\anchorrow{Eligible full-sum comparison}{Soundness/FS_Cutoff_Optimality.thy}{fs_first_root_nonempty_error_le_prefix}
\anchorrow{Exact source-connected ledger and acceptance bound}{Soundness/FS_Square_First_Root_Ledger.thy}{square_fs_first_root_error_ledger; square_fs_first_root_soundness}
\anchorrow{Native ledger comparison}{Soundness/FS_Square_First_Root_Comparison.thy}{square_fs_first_root_ledger_balance; square_fs_first_root_ledger_le_prefix}
\anchorrow{Retained small power certificates}{Soundness/FS_Square_135_Arithmetic.thy}{fs_135_block; fs_135_blocks; fs_135_power}
\anchorrow{Closed rational certificate}{Soundness/FS_Square_137_Arithmetic.thy}{fs_137_total}
\anchorrow{Exact specialization and numerical target}{Soundness/FS_Square_192_137.thy}{fs_137_ledger; square_fs_137_prefix_target}
\anchorrow{Concrete false-endpoint theorem}{Soundness/FS_Square_192_137.thy}{square_192_fs_137_incorrect_endpoint}
\anchorrow{Optimized definitions and canonical soundness theorem}{Soundness/FS_Optimized_Soundness.thy}{fs_optimal_cutoffs; fs_optimized_parameter_error; fs_optimized_soundness}
\anchorrow{Clipping the complete error}{Soundness/FS_Cutoff_Optimization.thy}{fs_first_root_parameter_error_clip; fs_first_root_parameter_error_fallback}
\anchorrow{Global comparison and lexicographic rule}{Soundness/FS_Cutoff_Optimality.thy}{fs_optimized_parameter_error_le; fs_optimal_cutoffs_lex_least}
\anchorrow{Concrete complete-error optimum}{Soundness/FS_Square_Optimized.thy}{fs_square_optimal_cutoffs; fs_square_optimized_ledger}
\anchorrow{Certified code shortcuts and endpoint}{Soundness/FS_Square_Optimized.thy}{square_192_optimized_cutoffs_code; square_192_optimized_error_code; square_192_fs_optimized_incorrect_endpoint}
\anchorrow{Recorded hash-only replay}{Soundness/RO_Honest_Shadow_Replay.thy}{shadow_absorb_known_wp; replayable_ro_bind}
\anchorrow{Commitment replay}{Soundness/RO_Honest_Query_Reconstruction.thy}{shadow_commitment_prefix_replay}
\anchorrow{Actual honest composition}{Soundness/Square_Sequence_192_RO_Composition.thy}{square_ro_composition_chain_authenticated}
\anchorrow{Actual opening witnesses}{Soundness/Square_Sequence_192_RO_Openings.thy}{square_ro_query_opening_authenticated_trees}
\anchorrow{Created-path authentication}{Soundness/RO_Honest_Verifier_Openings.thy}{protocol_created_path_exact_wp}
\anchorrow{Indexed FRI replay}{Soundness/RO_Honest_Verifier_FRI_Chains.thy}{mfold_indexed_exact_wp}
\anchorrow{Concrete polynomial fold}{Soundness/Square_Sequence_192_RO_FRI_Algebra.thy}{square_ro_fold_expression_next_value}
\anchorrow{Whole query replay}{Soundness/Square_Sequence_192_RO_Query_Verifier.thy}{square_ro_actual_query_round_wp}
\anchorrow{Header replay}{Soundness/Square_Sequence_192_RO_Verifier_Header.thy}{square_ro_actual_header_wp}
\anchorrow{All queries and actual acceptance}{Soundness/Square_Sequence_192_RO_Completeness.thy}{square_ro_all_queries_replay_wp; square_ro_honest_experiment_acceptance_one}
\anchorrow{Pure instruction interpreter}{Execution/Stark_Execution_Program.thy}{execution_program; run_det}
\anchorrow{Native probabilistic interpretation}{Execution/Stark_Execution_Semantics.thy}{run_prob; run_prob_bind}
\anchorrow{Cache and fixed-hash correspondence}{Execution/Stark_Execution_Cache.thy}{run_fixed_pointmass; fixed_hash_erases; inconsistent_cache_counterexample}
\anchorrow{Prover interpretation}{Execution/Stark_Execution_Prover.thy}{run_ro_prover_program}
\anchorrow{Verifier interpretation}{Execution/Stark_Execution_Verifier.thy}{run_ro_verifier_program}
\anchorrow{Explicit large-field generator}{Execution/Stark_Execution_Generator_192.thy}{execution_generator_192_order}
\anchorrow{Large-field executable instance}{Execution/Stark_Execution_Square_192.thy}{prove_square_192; verify_square_192}
}

\section{Historical proof routes and bounded investigations}
\label{app:history}

This appendix preserves the mathematics and research history of earlier proof
routes. The current publication claims and their source anchors are in
Appendices~\ref{app:statements} and~\ref{app:source-index}.

The retained account distinguishes archived formal arguments, external
arithmetic diagnostics, and unresolved construction or certificate obligations.
It does not claim a fresh rebuild of every archived proof. In particular,
algebraic or relational feasibility is not automatically a supported accepting
execution, and a numerical diagnostic is not an Isabelle-certified security
profile. Older typed listings use their historical probability API; they are
not signatures of the current native PMF/SPMF interface. Words such as
``new'', ``current'', ``public'' and ``retained'' inside these historical
accounts refer to the stage of the research being described.

\subsection{Historical contribution index}

\noindent\textbf{Historical status.} This index records earlier contributions and investigations, including archived implementation and proof interfaces. It is not the list of results exported by the current manifest.

The following index also preserves earlier formal contributions and proof routes;
their bounds and restrictions should not be confused with the current endpoint.

\begin{itemize}[leftmargin=*]
  \item A reusable probabilistic state monad and weakest-precondition calculus
    for finite probabilistic programs with failure.
  \item An executable Isabelle model of a STARK prover and verifier, including
    transcript operations, Merkle commitments, random-oracle calls, query
    checks, and FRI folding.
  \item A zero-failure honest-completeness theorem showing that an honest trace
    satisfying the protocol's algebraic validity predicate is accepted with
    probability one.
  \item A domain-separated random-oracle experiment in which transcript
    absorption is an accounted oracle operation on \code{TranscriptAbsorb}
    inputs.
  \item A staged soundness theorem bounding the acceptance probability of an
    adversarially supplied transcript by a closed arithmetic expression with
    adaptive prefix-conditioned FRI challenge, residual-query, and
    composition-padding contributions.
  \item A symbolic arithmetic-feasibility corollary showing acceptance
    probability below one from eleven component allocations of at most
    \(1/12\), together with primitive field-size and sampler-envelope
    sufficient conditions for the new sampled terms.
  \item Exact quotient/remainder accounting for the existing modulo query
    sampler, including exact raw fibers, query-list products, and a closed
    maximum-fiber envelope without a divisibility assumption.
  \item A positive-query robust refinement for the nonempty FRI branch.  It
    replaces complete-list enumeration by a decode-or-reject classification
    and transition-specific products with exponent equal to the number of
    verifier rounds minus the unchanged staged attacker query budget.
  \item A generic exact-rate list-capacity interface that retains the actual
    Reed--Solomon layer length, padded degree, reconstruction radius, strict
    Johnson positivity condition, and natural-number list-size quotient.
  \item A ceiling-balanced radius schedule with zero terminal radius, followed
    by an authenticated selected-boundary theorem that replaces the residual
    union over all possible folding layers by one canonical pre-query layer.
  \item An arbitrary-schedule audit that proves a matching non-strict optimum
    for the scale-64 diagnostic under the same exact recurrence and cap-two
    interface, followed by an authenticated global-chain audit that retains the
    intersection of every layer agreement preimage.  A bounded realization
    audit uses strict bad-challenge membership to rule out the exact 64,296
    envelope and certify a 64,294 ceiling.  A subsequent algebraic construction
    attains that tighter ceiling with the actual fold equations, canonical
    decoders, all layer transitions, and the constant terminal table.  The
    authenticated-execution support audit rules out the literal all-zero
    challenge presentation but proves that only the second challenge must be
    zero, leaving the same 64,294 residual.  A bounded semantic-coupling audit
    then proves that a trace-side use fixes both composition-check trace inputs
    on 64,166 residual indices, while a composition-side use fixes the
    composition equation on all 64,294.  A companion-chain construction then
    supplies a different global composition family with full fold agreement and
    distinct nonzero challenges.  A prefix-causal refinement commits its initial
    table before the first challenge, consumes that challenge online, and retains
    the residual on a support slice fixing the next six challenges.  The
    execution-feasibility audit then proves that the still-separate explicit
    trace family, which requires a zero second challenge, necessarily triggers
    the already charged initial-state target event.  This rules out that witness
    on the clean residual branch without proving a smaller universal envelope;
    a subsequent nonzero prefix-causal trace companion retains the exact first
    agreement set and 64,294 raw residual indices while reaching a fresh
    nonzero constant terminal layer.  A bounded joint header-challenge audit
    then follows the exact domain-separated root-before-challenge keys.  It
    proves that a clean finite assignment for both ten-fold chains exists
    exactly under the degree and field-capacity conditions
    $512\leq\mathit{maxDegree}$ and
    $21+|\mathit{spec}|\leq|\mathbb F|$, and that the same joint values retain
    both residual witnesses.  The follow-up Merkle/root audit proves supported
    construction and canonical openings for the concatenated companion table
    family, then isolates the exact deduplicated joint hash-key capacity
    condition.  The exact capacity follow-up decomposes this footprint into
    $20+|\mathit{spec}|$ header inputs, the union of committed leaf values, and
    the union of internal ordered-root pairs.  It improves the safe structural
    envelope from $523{,}776+|\mathit{spec}|$ to
    $456{,}457+|\mathit{spec}|$, but proves no corresponding lower bound.
    Existing public hypotheses therefore neither discharge that safe gate nor
    prove that the exact footprint is too large, so the unconditional capacity
    question remains underdetermined.  A new conditional chronological
    construction closes the positive relational direction under the explicit
    local gate $456{,}457+|\mathit{spec}|\leq|\mathbb F|$: it constructs the
    trace Merkle/challenge chain, then alpha lookups, then the composition
    Merkle/challenge chain in root-before-challenge order in one collision-free
    shared map.  This realizes the existing relational target without adding
    the gate to the public soundness premises.  A bounded follow-up proves full
    support for the default random-oracle output, prescribed support for fresh
    hashes and deterministic replay, and a converse from relationally created
    Merkle trees to supported executions.  An audit program built from the
    existing create, send, and counted challenge operations then realizes the
    chronological header fragment with positive support and a collision-free
    actual state below the relational target map.  A further bounded audit
    embeds those prescribed tables in the actual staged malicious-adversary
    prefix and proves the complete zero-query header has positive support with
    the exact verifier-visible transcript, state, and challenge counters.  It
    does not construct query openings or acceptance.  Budget admissibility is
    also separate: all fixed-message adversary stages cost zero, while the
    Merkle-root stages still need controlled-program budget proofs that are
    not supplied by the distinct-key capacity theorem.
  \item A strict selected-residual endpoint that carries this invariant through
    the unchanged actual-query experiment, together with formal inequalities
    showing its residual and final tightened errors cannot exceed the earlier
    selected-boundary errors.
  \item A comparable-minimum endpoint and generic $2^{-\lambda}$
    security-target theorem, together with a kernel-checked audit of the
    modulo-preimage envelope.  No concrete field or parameter profile is
    hidden in these generic results.
  \item A small executable square-sequence example over GF(5), together with a
    separate concrete soundness-locale witness proving that false statements
    and a genuinely non-singleton, biased query sampler coexist.
\end{itemize}

\subsection{Positive-query robust, exact-rate and selected-boundary investigations}

\noindent\textbf{Historical status.} The following sequence records the earlier robust and selected-boundary routes and their bounded realization audits. Proven local or algebraic constructions, necessary conditions, untrusted diagnostics and unresolved execution/certificate questions remain distinguished in the text. These routes are outside the current retained publication interface.

The earlier all-round endpoint serves as the baseline. For the genuine
nonempty FRI branch, the development follows the actual checked query-head
program and counts future \code{QueryIndexChallenge} keys already present in
the live random-oracle map.  A staged attacker with total query budget $B$ can
account for at most $B$ such keys.  Every adaptive product therefore has
exponent

\[
  \mathit{rounds}-\mathit{staged\_attacker\_query\_budget}(\mathit{budgets}),
\]

using the existing truncated natural subtraction.  No opening-stage budget is
set to zero, and the adversary, transcript experiment, and staged budget
semantics are unchanged.

An earlier positive-query refinement removed the artificial enumeration of
all complete residual lists by covering the actual verifier-generated family
with one rectangle per active transition.  The robust development strengthens
the deterministic FRI input to those products.  Canonical nearest-code
decoders are applied to the committed trace and composition layers.  The
multiround theorem proves a decode-or-reject alternative: the first table is
within the certified decoding radius, an online challenge lies in a
prefix-dependent bad family, or a first far-to-close transition has an exact
successor-layer agreement set bounded away from the complete domain by
\code{fri_linear_margin}$(K,i)$.  Merkle authentication, transcript-prefix
dependence, and fresh-update drift are then transported through the unchanged
security experiment.

The exact-rate interface does not replace this argument by a rate slogan.  For
each active layer, \code{fri_linear_exact_list_cap_at} records the actual
domain length $n$, folded padded degree $d_i$, and reconstruction radius $t_i$.
It requires

\[
  n d_i < (n-t_i)^2
\]

and retains the exact natural-number list envelope

\[
  \frac{n(n-d_i)}{(n-t_i)^2-n d_i}
\]

with Isabelle's natural division.  The quantified predicate
\code{fri_linear_exact_list_cap} requires this envelope at every active layer.
A cap of two yields the robust challenge and multiround classifications used
by the endpoint.  The separate factor-two repair also proves that
\(8\,\operatorname{Suc}(m)\le 2^K\) is sufficient for the older quarter-radius/cap-two
specialization; the exact-rate endpoint itself uses the stronger per-layer
predicate.

The robust nonempty error has twelve displayed components: collision, decoded
semantic rectangle, first-root prefix target, trace/composition prefix target,
FRI-builder target, initial target, separate trace and composition challenge
terms, separate trace and composition residual products, query-phase target,
and the robust alpha-pivot term.  The residual products retain the exact
quotient/remainder modulo envelope and have the form

\[
  \sum_i
  \left(
    \frac{\mathit{query\_raw\_preimage\_card\_envelope}(E_i)}
         {|\mathbb F|}
  \right)^{\mathit{rounds}-B}.
\]

The exact-rate endpoint above still uses a fixed linear margin schedule and
sums over every possible active folding layer.  The later balanced refinement
removes both proof artifacts without changing the verifier or experiment.
For a free denominator $C$, it assigns layer $i$ the ceiling of the successor
domain length divided by $C$, sets the terminal radius to zero, and defines
earlier radii by the exact backwards folding recurrence.  Its
\code{fri\_balanced\_exact\_list\_cap} predicate retains the same Johnson and
natural-division conditions at every layer.

For an authenticated accepted chain, the roots, challenges, final value, and
recorded builder state fix the set of far-to-close boundaries before any query
location is sampled.  If that set is nonempty, its least element selects a
single residual layer.  Thus each trace or composition residual becomes

\[
  \left(
    \frac{\mathit{query\_raw\_preimage\_card\_envelope}
      (\max_i E_i)}{|\mathbb F|}
  \right)^{\mathit{rounds}-B},
\]

rather than a sum of such powers over all layers.  Trace and composition remain
separate summands: the current verifier equations do not prove that their two
bad-index events overlap or contain one another.

The earlier exact-rate safe endpoint is a comparable minimum of the preceding
complete-list tightened error and the linear-schedule robust error:

\begin{lstlisting}
lemma ro_absorb_stark_soundness_robust_exact_rate_tightened_nonempty:
  assumes false_statement: "\<not> exists_valid_trace"
    and wf: "staged_budget_wellformed budgets"
    and controlled: "staged_adversary_controlled budgets A"
    and nonempty: "0 < ceil_log clength"
    and eval_power: "clength * scale = 2 ^ N"
    and trace_exponent_fit: "ceil_log clength + K \<le> N"
    and composition_exponent_fit:
      "ceil_log (Suc maxDegree) + K \<le> N"
    and K_pos: "0 < K"
    and trace_exact:
      "fri_linear_exact_list_cap
        (clength - 1) (ceil_log clength) K 2"
    and composition_exact:
      "\<And>d. d \<le> maxDegree \<Longrightarrow>
        fri_linear_exact_list_cap d (ceil_log (Suc d)) K 2"
  shows
    "ro_absorb_checked_staged_adversary_acceptance_probability A
      \<le>
      ro_absorb_checked_staged_robust_exact_rate_tightened_nonempty_parameter_error
        K budgets"
\end{lstlisting}

The balanced selected endpoint takes a further minimum with the
selected-residual error.  Isabelle proves both the same unchanged acceptance
experiment bound and that this error is no greater than the prior balanced
tightened endpoint. The strict refinement described below supplies the final
endpoint of this selected-residual staged route.

The exact-rate, balanced-list, and exponent conditions are explicit arithmetic/profile
discharge data.  They do not change the verifier, sampler, protocol, transcript
experiment, staged budget semantics, adversary, or the three existing public
soundness premises.  The target
\code{soundness_security_target}$(\mathit{security\_bits})$ is $2^{-\lambda}$
in Isabelle's nonnegative probability type.  A separate
composition theorem derives acceptance at most that target from an arithmetic
proof about the exact comparable error.  No concrete field, budget, security
level, or parameter profile is certified by these generic theorems.

The modulo tightness audit remains applicable: the set \code{{..<k}} attains
\code{modulo_preimage_card_envelope}$(N,q,k)$ whenever $k\le q$, and every
uniform preimage bound that knows only $\operatorname{card}(B)=k$ must be at least that
large.  The robust proof improves the verifier-derived bad-set cardinality; it
does not pretend that the exact modulo envelope itself can be uniformly
reduced.

The later schedule audit works below the public endpoint, without changing its
experiment.  It exposes arbitrary layer margins, good radii, and radii already
accepted by the generic decode-or-reject theorem.  For the scale-64 diagnostic
degree it proves that the non-strict cardinality interface cannot make every
layer envelope smaller than 64,296 of 65,472 raw query indices.  A matching
positive-margin, zero-terminal-radius arithmetic schedule is checked.  The
balanced envelope is 64,322, so the remaining schedule artifact is small.
These numerals certify an optimization decision for that proof interface; they
are not accepted-execution lower bounds, a concrete security profile, or an
inherent lower bound for FRI.

The authenticated-chain audit proves that an accepted recorded chain places
every sampled query in the intersection of all layer-specific agreement
preimages.  The realization audit then retains strict non-membership in the
good-challenge set: at the scale-64 first boundary this forces one additional
successor-domain disagreement and lowers the exact modulo ceiling to 64,294.
Therefore the 64,296 conceptual configuration cannot satisfy the actual
residual premises.  The follow-up algebraic realization constructs explicit
tables whose canonical decoding gives the required near/far distances, whose
actual binary folds leave exactly 32,147 first-successor agreements and full
agreement at every later layer, and whose terminal table is constant zero.  The
exact modulo sampler then leaves exactly 64,294 raw query indices, so the strict
envelope is attained by the algebraic fold chain.

The literal construction used ten zero challenges.  The authenticated-execution
support audit proves that this exact presentation is incompatible with a
collision-free oracle map: challenge calls at different counters use distinct
oracle inputs and therefore cannot all return the same output.  The same audit
also proves that the residual construction does not need that repetition.
Because the first layer has identical halves and the layers after the second
fold are zero, the first and all later challenges are unrestricted; only the
second challenge must be zero.  Every ten-entry list satisfying that one
condition retains the exact 64,294 residual.  Collision freedom alone therefore
does not produce a smaller cardinality bound.

This does not construct an accepted execution.  The bounded
trace/composition semantic-coupling audit follows the executable verifier one
step further.  The trace commitment root and first trace-FRI root are separate,
so their whole tables need not coincide.  At an accepted index their base values
must coincide, however, and the composition check also opens the trace value at
$q+64$.  Of the 64,294 explicit residual indices, exactly 64,166 have both $q$
and $q+64$ inside the residual.  If the family is used on the trace side, those
indices therefore satisfy the composition equation with both trace inputs fixed
by the explicit first table.  Only 128 boundary indices retain freedom in the
shifted opening.  If the family is used on the composition side, its first-table
entry equals the evaluation of one global trace table and one alpha vector on
all 64,294 indices.

The companion-chain solvability theorem now constructs a different global
composition-table family for the explicit global trace table and an arbitrary
alpha vector.  Its first table satisfies the verifier's composition equation on
all 64,294 residual indices, every canonical FRI transition has full agreement,
and the terminal tail is zero.  The required ten composition challenges exist
and may be chosen pairwise distinct and nonzero: at most 4,238 values are
forbidden in a field of at least 65,537.  This settles the bounded semantic
companion-chain gate without specializing the locale-parametric constraint
evaluator.

The prefix-causal follow-up now applies the executable FRI timing discipline:
layer $k$ is committed before challenge $k$ is returned, so it may depend only
on the preceding $k$ challenges.  The original backwards recipe can expose a
future-challenge dependency at layer six.  The refined construction instead
commits a fixed initial composition table, consumes the first challenge online,
and fixes only challenges one through six for the remaining steering steps.
Provided the online first challenge avoids 621 first-layer domain values,
challenges seven through nine are unrestricted and the same 64,294 residual
indices satisfy the trace/composition equation and every conceptual FRI
transition.  This describes a nonempty semantic support slice; it is not a
probability bound for that slice.

The authenticated-execution feasibility audit traces the still-separate
explicit trace-FRI family into the checked staged experiment.  Its 64,294
saturation proof requires the second trace challenge to be zero.  Since the
initial transcript state is zero and the initial oracle map is empty, every
recorded trace-challenge lookup with that output is a fresh hit on the
initial-state target already charged by the soundness theorem.  Thus this
particular trace witness cannot inhabit the no-initial-target residual branch.
This conclusion is stronger than the earlier repeated-output collision
observation and does not depend on collision freedom.

The result is nevertheless witness-specific and does not reduce the universal
residual envelope.  The nonzero prefix-causal trace-companion construction now
supplies the previously missing conceptual alternative.  It keeps the same
initial trace table and exact first agreement set, uses a pairwise-distinct
nonzero ten-challenge instance, reaches a fresh nonzero constant terminal
layer, and retains exactly 64,294 raw residual indices.  Its support slice fixes
challenges one through six; challenge zero and challenges seven through nine
remain algebraically free.  Thus a zero trace challenge is not a universal
algebraic or prefix-causal requirement.

The bounded joint header-challenge audit now combines the two companion
families at the next layer.  It instantiates the verifier's exact
domain-separated trace-FRI, alpha, and composition-FRI challenge inputs in
root-before-challenge order.  With ten roots supplied for each chain, it proves
that a collision-free finite assignment with pairwise-distinct nonzero outputs
exists if and only if
$512\leq\mathit{maxDegree}$ and
$21+|\mathit{spec}|\leq|\mathbb F|$.  The degree condition is precisely what
permits an encoded degree claim selecting ten composition folds.  The field
condition counts the initial zero target, twenty FRI outputs, and every alpha
output.  A strengthened endpoint uses the same jointly selected vectors and
map while retaining the trace companion's exact 64,294 residual, first
agreement set, and nonzero constant terminal layer, as well as the composition
companion's residual support.

The follow-up Merkle/root audit closes the local part of this gap.  The actual
Merkle monad has a supported outcome for the concatenated twenty committed
companion tables, and canonical paths authenticate every created power-of-two
table.  It also defines the exact replay-aware union of distinct leaf,
internal-node, and chronological header inputs.  Any collision-free shared
target must fit that union injectively into the field.  Conditional on such a
target, all committed conceptual layers are exactly the prescribed companion
tables, the composition support survives, and the trace residual remains
exactly 64,294.

The exact shared-map capacity audit sharpens this gate.  Any target must fit
exactly
\[
  20+|\mathit{spec}|+
  |K_{\mathrm{leaf}}|+|K_{\mathrm{node}}|
  \leq |\mathbb F|,
\]
where $K_{\mathrm{leaf}}$ is the union of distinct values across all twenty
committed tables and $K_{\mathrm{node}}$ is the union of distinct ordered
child-root pairs.  A position-wise count gives
$523{,}776+|\mathit{spec}|$.  Using that the fixed current trace table has at
most three values improves the safe envelope to
$456{,}457+|\mathit{spec}|$.  At the minimum proved field size $65{,}537$,
the two envelope gaps are respectively $458{,}239$ and $390{,}920$.

These gaps are not lower bounds on an execution: the exact formula preserves
cross-table leaf replay and internal-root-pair replay.  The existing
$21+|\mathit{spec}|\leq|\mathbb F|$ hypothesis counts only the fixed header
outputs.  Thus the unconditional question remains \emph{underdetermined} under
the public assumptions: they prove neither the sufficient structural gate nor
a lower bound forcing failure.

The conditional chronological shared-map construction closes the positive
relational direction under the explicit local assumption
\[
  456{,}457+|\mathit{spec}|\leq|\mathbb F|.
\]
Its generic layer maintains a collision-free finite map while reserving all
fixed challenge outputs and preserving replay for existing inputs.  Starting
from the empty map, the concrete construction creates each trace companion
Merkle tree and only then installs that root's exact domain-separated trace-FRI
challenge lookup.  It next installs the alpha challenge lookups, then creates
the composition companion trees and their composition-FRI challenge lookups in
the same root-before-challenge order.  The endpoint
\code{ex_chronological_joint_companion_realization} therefore produces both
tree families and one final shared map; its projection satisfies the already
defined \code{joint_companion_merkle_root_target}.  Consequently the
composition support and exact 64,294 trace residual survive at the relational
authenticated-table layer.

This gate is a local premise of the diagnostic construction theorem, not a
locale assumption or public soundness premise, and the current public
hypotheses do not prove it.

The probabilistic-support follow-up proves the converse at the next bounded
layer.  First, \code{hash_default_full_support} shows that every field element
belongs to the support of the existing uniform default hash distribution.
Theorems \code{hash_prescribed_extension_support} and
\code{create_prescribed_extension_support} lift this to prescribed fresh hash
outputs, deterministic replay, and relationally created Merkle trees.  They
produce an actual executable state between the starting map and the prescribed
target map; equality with the target is neither needed nor claimed.

The audit program \code{supported_header_chain} composes the existing
\code{create}, \code{send}, and counted trace-FRI, alpha, and composition-FRI
receive operations in the chronological order used by the relational
construction.  From an empty live oracle map, the endpoint
\code{ex_chronological_joint_companion_realization_support} returns the
prescribed roots and challenge values with support membership, keeps the actual
final state collision-free and below the relational target map, and retains the
existing \code{joint_companion_merkle_root_target}.  Support membership gives
positive mass in the finite distribution, but the theorem derives no useful
closed-form probability or lower bound.

This is not an honest-prover execution theorem.  The audit program is supplied
the conceptual companion tables, whereas the honest prover's FRI commitment
routine constructs successive tables through \code{next_fri_layer}.

The staged-header follow-up instead asks the verifier-compatibility question on
the actual malicious-adversary path.  The fixed adversary
\code{companion_header_adversary} creates each prescribed trace or composition
Merkle table before the verifier records its root and derives the corresponding
counted challenge.  The endpoint
\code{ex_chronological_joint_companion_staged_header_support} executes
\code{checked_staged_query_challenge_prefix_program} with zero query chunks and
proves support for the exact staged proof data.  It also proves the forward
transcript equation, all three header challenge-counter equations, collision
freedom of the actual oracle state, and preservation of the existing
\code{joint_companion_merkle_root_target}.  Thus the conceptual companion is
structurally realizable through the complete verifier-consumed staged header.

The initial trace root and the first trace-FRI root are distinct messages in
this execution.  No header assertion equates them.  Their tables may be coupled
later by equality of authenticated values at sampled query locations, so the
zero-query theorem does not discharge that opening-level consistency gate.

Staged budget admissibility is likewise distinct from structural support.
The rule \code{companion_header_adversary_controlledI} shows that the initial
root, terminal values, degree claim, and empty query stages are return programs
and hence controlled at zero attacker-oracle cost.  It reduces the remaining
obligation exactly to controlled-program proofs for the trace and composition
Merkle-root stages.  The earlier capacity bound counts distinct inputs in a
compatible collision-free target map; it does not count sequential hash calls.
The ten committed table lengths sum to $130{,}944$ entries per family.  The
executable binary Merkle recursion makes $2n-1$ hash calls on a nonempty
$n$-entry table, so a source-level count gives $261{,}878$ calls per family and
$523{,}756$ across the twenty root stages.  Replay-based deduplication can
reduce the final map footprint but does not reduce this sequential call count.
The retained lemma \code{controlled_ro_create} in
\theorysource{Soundness/RO_Honest_Replay_Primitives.thy} now proves
\code{controlled_ro_program (2 * length xs - 1) (protocol_merkle.create xs)}.
It supplies the generic Merkle-control bridge that was absent at this
historical checkpoint. The capacity result still does not establish admission
under arbitrary staged budgets: the sequential call allowances and the
remaining companion obligations must also be discharged.
Query openings, full budget admissibility, verifier acceptance, positive
acceptance probability, and a false-statement witness remain outside the
bounded result.  No concrete cryptographic profile or general protocol flaw is
inferred.

The fixed-root/lazy-authentication audit separates this eager witness cost from
what the staged verifier interface actually requires.  The adversary
\code{fixed_root_lazy_adversary} simply returns every announced root and
opening chunk.  It makes no oracle calls, and
\code{fixed_root_lazy_adversary_controlled} proves admissibility for
arbitrary configured attacker budgets.  Root provenance is not checked at the
root-return stage; the verifier enforces the root only later by executing the
Merkle authentication recursion on a supplied opening.

\begin{sloppypar}
The theorem
{\small\code{chronological_joint_companion_fixed_root_header_lazy_trace_opening_support}}
combines the complete fixed-root zero-query header with one actual canonical
authentication path for a committed trace-FRI layer, while keeping both
outcomes inside the same prescribed oracle-map extension.
\end{sloppypar}
For the scale-64
domain the path has sixteen siblings.  Recording its leaf and path costs 17
transcript hashes; verifier replay costs a further 17 transcript hashes and 17
Merkle hashes, giving a source-level count of 51 calls for those operations.
The fixed header eliminates the eager witness's 523,756 attacker calls.
These figures cover different operations: header framework calls, query
challenges, and other openings are excluded from the 51-call count.  The
support theorem itself uses the non-absorbing staged prefix and raw Merkle
checker, under the existing local realization/capacity conditions; it does not
yet compose the absorbing operations.

This proves that the eager creation cost is witness-artifactual, not that a
complete accepted query is cheap or realizable.  The initial trace-root
openings, all other query openings, the absorbing random-oracle query
transcript, simultaneous consistency, collision-free capacity for every later
key, verifier acceptance, false-statement relevance, and useful probability
remain outside the theorem.  No soundness parameter is changed.

The subsequent absorbing-transcript audit isolates a necessary capacity
condition from the executable alpha phase.  The theorem
\code{ro_alpha_program_collision_free_capacity} proves that a supported,
collision-free execution of $n$ alpha steps requires $2n\leq |\mathbb{F}|$.
The increasing counters distinguish the alpha keys; collision freedom then
makes the coefficients distinct.  Their absorption keys are distinct in their
message component and are disjoint from the alpha keys.  Thus the result
counts distinct inputs, not merely sequential calls.  It also holds in any
later collision-free map extension.

The old local gate $456457+\lvert\mathsf{spec}\rvert\leq|\mathbb{F}|$ does
not imply this alpha-only condition by arithmetic alone: for $m>2\cdot456457$,
the choice $n=m-456457$ satisfies the old inequality but violates $2n\leq m$.
This arithmetic counterfamily is not a concrete finite-field/protocol
interpretation or a false-statement witness.  Nor does it rule out the
companion in an adequately sized field.

Absorption changes the state component of later challenge keys.  Keeping the
old keys while assigning the same prescribed outputs to different new keys
would violate collision freedom.  The audit proves that the Merkle-only
restriction of a relational target preserves all created-tree witnesses and
collision freedom, while freeing its non-Merkle outputs.  This is a proof-side
reconstruction option, not deletion of entries from an executed oracle map.
The audit's conditional support rules alone did not construct a common
absorbing target.  The subsequent fresh-map construction now does so for a
bounded execution fragment.  Its endpoint
\code{ex_scale64_absorbing_companion_first_path} starts from the standard empty
adversary state, constructs both companion Merkle families, supports the actual
absorbing staged header and one exactly sampled, whole shape-valid query chunk,
and supports verifier-header replay followed by the first authentication under
the initial trace root.  The unchanged sampler is \code{index (to_nat raw)}.
The root is chosen as the root of the first trace table, which is an allowed
adversary message, not a new protocol equation.  The chosen raw field value
zero represents one sampled query; it is not a zero-query profile.

The explicitly sufficient local construction gate, under
$\mathsf{clength}\cdot\mathsf{scale}=2^{16}$, $\mathsf{scale}=64$, and
$512\leq\mathsf{maxDegree}$, is
\[
456982 + 2\lvert\mathsf{spec}\rvert + 17\mathsf{powers}
  \leq |\mathbb{F}|.
\]
The sum counts at most 456,437 entries for both Merkle families,
$20+\lvert\mathsf{spec}\rvert$ reserved header challenge outputs, one
reserved raw query output, $24+\lvert\mathsf{spec}\rvert$ absorbed header
fields, and $500+17\mathsf{powers}$ absorbed query-chunk fields.  Each ten-layer
FRI family contributes 250 query fields.  These are sufficient construction
charges, not a lower bound on verifier work or a claimed optimal capacity.
Already known absorption keys replay consistently; fresh keys extend the
target while preserving unused output reservations and counter freshness.

The target map containing all Merkle witnesses is a ghost map.  The actual
staged map is only a submap and is the map transferred into verifier replay;
authentication may reveal missing entries by supported lazy oracle extension.
The replay theorem constructs an outcome rather than assuming one.  Exact
decomposition lemmas identify the header and first initial-root operation of
the existing verifier.  When the configured repetition count is positive,
these are its first header/query operations.  The fixed return-stage adversary
satisfies arbitrary staged attacker budgets and imposes no zero opening-stage
budget requirement.

The result stops after that first path: the remaining openings in the recorded
chunk satisfy only its shape constraints and may reject.  It neither constructs
a full accepted query nor extends support to all 64,294 residual indices.
False-statement relevance, residual realizability, and useful probability mass
remain open.  In particular, existential support on a prescribed oracle branch
is not a quantitative lower bound on acceptance.  The local gate is not a new
public soundness premise; the verifier, transcript experiment, sampler, model,
and budget semantics are unchanged.

\paragraph{Residual-branch relevance of the fixed-return witness.}
The bounded relevance audit does not invalidate the preceding support theorem.
It instead identifies why that witness cannot establish the residual case
after the other charged events have been excluded.  These event categories
need not be disjoint.

From the standard empty map and initial transcript state zero, the
initial-target predicate holds exactly when zero occurs among the actual
oracle outputs.  The recorded raw-zero query therefore already meets this
charge.  Independently, the actual first-root prefix of
\code{fixed_root_lazy_adversary} contains only two root-message absorptions,
before the first FRI challenge.  Such hashes cannot install Merkle-leaf or
Merkle-node entries.  Any later successful initial-root authentication must
supply a previously absent Merkle input whose output is the announced root.
For a collision-free prefix, that root belongs to the existing first-root
prefix target set.  The audit proves the resulting target hit for both an
actual individual opening and a successful actual verifier query round.
These are necessary-condition implications; no accepting execution is constructed.

The sampler does not force the zero-output obstruction.  For the scale-64
domain and $\mathsf{powers}=2$, the existing encoding range supplies a nonzero
raw field value whose exact modulo index is zero or one, within the
companion's first agreement set.  No equation between field zero and its
natural-number encoding is assumed.  A conditional support theorem uses
reservations $\{0,\mathsf{raw}\}\to\{0\}$ through the actual sampled query and
whole-chunk absorption.  Given the existing reserved-target invariant and
$\operatorname{card}(\operatorname{dom}M)+2+
\operatorname{length}(\mathsf{chunk})\leq|\mathbb F|$, it constructs a supported
outcome avoiding the initial target.  Zero-valued messages are allowed.
This is not an integrated alternative companion header, and the earlier
numerical capacity gate is not advertised as certifying one.

The actual prefix-fixed conceptual tables are constant lists of the arbitrary
Hilbert-choice value $\mathsf{THE}\ v.\ \mathsf{False}$ when no values are
authenticated; that default is not assumed to be zero.  They cannot simply
be identified with the companion tables in the larger ghost map.  Nor does
the algebraic 64,294 agreement intersection alone imply membership in the
experiment's strict selected residual.  The latter has an online bad-challenge
exclusion, which distinct nonzero steering challenges do not establish.
Composition-prefix, builder-target, decoded-semantic, and false-statement
obligations remain separate.

For a collision-free prefix extended by the later state, Merkle pullback proves
that excluding the first-root prefix-target event requires the entire later
authenticated path to be present already in the saved prefix map.  This is not
a lower bound on distinct attacker hashes:
sharing and replay require separate accounting.  This motivates the bounded
early-authentication investigation below, rather than merely filling the
same placeholder query.
The witness-specific obstruction does not rule out near-worst-case behavior
for other adversaries, tighten the universal residual bound, or establish
a flaw in STARK/FRI.

\paragraph{Early-prefix authentication feasibility and cost.}
The follow-up constructs a supported outcome of the actual absorbing first-root
prefix from the original empty oracle state. A fixed list of ordinary attacker hash queries
populates a clean Merkle map before either root message is absorbed.  Both
messages announce the same root of the fixed scale-64 initial companion
table.  The theorem \code{ex_scale64_early_prefix_authenticated} identifies
both actual prefix conceptual tables with that table and excludes the
initial-target event.  These are actual authenticated-map equalities, not
identifications with an unqueried ghost map.

The fixed initial table has identical 32,768-element halves and at most
three distinct leaf values.  Building one half and sharing it yields a
sufficient seed envelope of 32,771 distinct inputs.  The existing field-size
facts leave space for reserved zero and both absorption calls; no additional
capacity premise is needed for this prefix.  Querying a list without repeated
keys gives a certificate equal to its length.  The witness wrapper charges
that certificate to the existing initial root-stage budget and retains all
later supplied attacker stages.  No zero opening-stage budget is imposed.
The number is an upper bound for this construction, not an optimal cost,
an amplification parameter, or a concrete security estimate.  For the
converse accounting direction, collision freedom injects an actual key
footprint into the output range bounded by the existing budget calculus.

Canonical paths for this whole table have supported replay with the hash map
unchanged, hence no fresh-output target hit from that raw checker.  Its hash
invocations remain unchanged.  This does not cover intervening transcript
absorption or an entire verifier query.  Hard-coded internal-node inputs may
contain guessed hashes; support requires the oracle to return those guesses,
and no useful probability is proved.  The result settles only the
early-authentication feasibility gate.  Later companion layers, complete
query acceptance, all other charged-event exclusions, and false-statement
relevance remain separate.  In particular, the conceptual 64,294 agreement
count is not promoted to an authenticated strict-residual realization.

The subsequent bounded transition audit checks challenge relevance before
constructing more of that authenticated execution.  The actual continuation
receives challenge $b_0$ after the two root absorptions, then obtains and
absorbs the layer-one root before receiving $b_1$.  The equal initial halves
make the first folded word independent of $b_0$, with exact degree-511 distance
6,773.  Thus the fold-good set is empty for any good radius below 6,773,
including the balanced first radius 6,719 and optimized first radius 6,772.
The theorem \code{fri_first_root_balanced_online_bad_empty} also establishes
the empty first online target under identification of the actual conceptual
table.  It does not claim that an entire continuation avoids every bad event.

Let $\Delta_i$ be the exact canonical code distance of companion layer $i$
at the degree-1023 diagnostic.  The fixed entries, including arbitrary values
in the remaining suffix, force
\[
  6152 \leq \Delta_1 \leq 6773,
  \qquad
  b_1\neq 0 \Longrightarrow 3076 \leq \Delta_2 \leq 3697.
\]
The code degrees are respectively 511 and 255 and the domain sizes 32,768
and 16,384.  The proof uses the existing unique-zero sparse decoder theorem,
not a Hamming-weight estimate assumed to equal code distance.  The public
balanced schedule has $r_1=6112$, so layer one is far.  The optimized schedule
has $r_2=2766$, so layer two is far when $b_1\neq0$, as in the previously
constructed nonzero companion.  No challenge is newly restricted in the model.

Exact agreement at every later fold and a constant terminal layer force a
far-to-close boundary.  At such a boundary the folded word is the close
successor, so $r_{i+1}\leq t_i$ forces $b_i$ into the existing fold-good set.
The balanced schedule identifies this with its existing online bad-challenge
target.  In
\code{fri_nonzero_companion_builder_online_bad}, identification of the actual
builder's conceptual layers with the whole companion implies the actual
balanced online event.  Consequently
\code{fri_nonzero_companion_builder_strict_residual_empty} proves that both
existing strict-selected residual sets are empty.  The optimized result
\code{fri_nonzero_companion_optimal_bad_boundary} is a canonical far-to-close
challenge obstruction for the existing arbitrary schedule; it is not a new
optimized online endpoint.

This is a conditional exclusion of this fully identified companion from a
clean residual branch, not a universal residual improvement or verifier
rejection theorem.  The builder-table equality is an explicit diagnostic
hypothesis, not a locale or public soundness premise, and no execution
satisfying it is constructed here.  Partial authentication is not silently
identified with full tables.  The original 64,294 agreement count is
unchanged, but it is not evidence of clean residual membership.  A different
family, a tail with controlled disagreements, or a different partial-table
identification requires a further bounded investigation.  No useful
acceptance probability, security profile, or limitation of STARK/FRI generally
follows.  All verifier, transcript, sampler, adversary, and budget semantics
remain unchanged.

The controlled-tail follow-up keeps the scale-64, degree-1023 diagnostic
($m=10$, $\mathit{powers}=2$), the fixed initial table, and the exact first
agreement set, but no longer fixes the later tables to the companion recipe.
Write
\[
 I=[0,32147)\ \cup\ \bigl(32768+[0,32147)\bigr).
\]
These are natural query indices in the existing sample space, not raw field
outputs of the random oracle. For a later successor size $n\mid32768$ and a
disagreement set $D\subseteq[0,n)$, Isabelle proves the exact retained-lift count
\[
 \left|\{q\in I:q\bmod n\in D\}\right|
 =2\left(\left\lfloor\frac{32147}{n}\right\rfloor |D|
       +|D\cap[0,32147\bmod n)|\right).
\]
Thus a later exclusion cannot simply be hidden among the original rejected
indices. This formula accounts for overlap with the first rejection set;
it makes no independence assumption and does not sum exclusions across
different layers.

For the balanced schedule, the locked layer-one entries still force distance
at least 6,152, exceeding radius 6,112, regardless of the remaining suffix.
A constant terminal layer and avoidance of the existing online bad-challenge
event force a later far-to-close transition with at least
$1+\mathit{margin}_i$ disagreements. For layers 1 through 9, the retained-lift
lower bounds are respectively
\[
 (610,918,1078,1170,1240,1364,1500,2008,3012).
\]
Consequently \code{fri_scale64_builder_balanced_global_card} proves a global
agreement cardinality of at most $64294-610=63684$ for the actual builder
definitions, conditional on the explicit initial-table and first-agreement
identification. It does not prescribe a replacement tail. The existing
authenticated-chain-to-global-agreement theorem supplies the necessary
query intersection; this is not a converse execution construction.

The optimized diagnostic requires its own canonical challenge exclusions,
not an identification with the public balanced event. With a nonzero second
challenge, the fold of any successor retaining the locked prefix has
degree-255 distance at least 3,076. If layer two is within the optimized
radius 2,766, at least 310 disagreeing residues exclude at least 620 retained
queries. If layer two is far, a later strict boundary excludes at least 936.
Thus \code{fri_scale64_optimal_nonzero_tail_global_card} gives the conditional
ceiling $64294-620=63674$. A recorded zero challenge already hits the existing
initial-target charge; the execution-level nonzero corollary uses the actual
challenge outcome and hash-map extension, not a new adversary restriction.
No optimized public online endpoint is introduced.

The lift formula is exact, but these ceilings discard its nonnegative
remainder-intersection term and use only one necessary transition.
They do not prove that actual FRI tables can place all disagreements in the
least-populated fibers, that either ceiling is attainable or optimal, or that
other first-agreement layouts have the same overlap behavior. The results
exclude preservation of the old 64,294 set under the stated fixed-family
clean conditions, not acceptance on already charged branches or every
near-envelope witness. They are neither a universal soundness improvement
nor an acceptance-probability estimate.

\paragraph{Alternative first-agreement layouts.}
The next bounded audit keeps the initial table and the scale-64, degree-1023,
ten-fold diagnostic, but changes the placement of the first agreement set.
For any $A\subseteq[0,32768)$, the exact sampler-index balance is
\[
 \left|\{q<65472:q\bmod32768\in A\}\right|
 +|A\cap[32704,32768)|=2|A|.
\]
Thus retaining 64,294 queries from 32,147 agreement positions must account
for the shortened final block; equal agreement cardinalities alone are not
interchangeable.

The paired layout $A=[0,16073)\cup[16384,32458)$ retains 64,294 queries
and completely hides the 310 later residues $[16074,16384)$. The theorem
\code{complete_pair_hiding_capacity} proves that 310 is the largest number
of complete pairs that can fit in a 621-position complement. Successor
tables using the actual folding functions realize these first two local
agreement tests. Nevertheless, their fixed second-layer prefix forces
degree-255 distance at least 3,076 when the second challenge is nonzero.
The theorem \code{paired_builder_balanced_global_card} derives a cap of
63,682 under the explicit first-agreement and layer-two prefix identification,
with arbitrary later tables and the existing balanced online event avoided.
It is a necessary-condition bound for that family, not an attainment result.

The periodic nested layout demonstrates a more substantial counting
alternative. Define
\[
 \begin{gathered}
 D_n=\{a<n:a\bmod4096\in[4019,4096)\},\\
 R=D_{32768}\cup[32686,32691),\qquad A=[0,32768)\setminus R.
 \end{gathered}
\]
Here $|R|=616+5=621$, $|A|=32147$, and the query lift set again has
64,294 indices. Its fully hidden later sets compare as follows:
\begin{center}
\begin{tabular}{rrrr}
\toprule
Successor size & Hidden residues & Balanced strict threshold & Optimized strict threshold\\
\midrule
16384 & 308 & 305 & 311\\
8192 & 154 & 153 & 156\\
4096 & 77 & 77 & 79\\
\bottomrule
\end{tabular}
\end{center}
Consequently the balanced boundary cardinalities alone do not force another
retained-query exclusion at these three positions. The optimized thresholds
are different: the displayed hidden sets are insufficient for them. This
comparison does not identify the two canonical challenge families.

The theorem \code{nested_four_fold_query_set} goes beyond arbitrary sets:
it constructs four consecutive local folding transitions preserving exactly
those 64,294 query indices, with the correct canonical table lengths.
The first successor is challenge-independent; the next three use respectively
$b_1$, $(b_1,b_2)$, and $(b_1,b_2,b_3)$. This matches root-before-current-challenge
data dependence locally, but is not a shared Merkle/random-oracle transcript
construction or a proof that the challenges avoid charged events. In particular,
it is not a complete protocol repetition or a new amplification scheme.

The same nested query set covers every residue in every remaining successor
domain, whose sizes are at most 2,048. If the whole set survives globally,
all those later folds must agree everywhere. The existing far-to-close
argument then yields the actual-builder necessary condition
\code{nested_builder_retention_requires_degree63_close}: full retention and
avoidance of the existing balanced online event force the 4,096-position layer
to have distance at most 536 from degree-63 code. The constant terminal layer
comes from the existing builder definition, not a new public premise.

That audit isolated, but did not discharge, the algebraic completion bridge.
Its classification was local counting and folding compatibility, not a
realizable near-worst-case execution. The following audit resolves clean
full retention negatively for this layout. Merkle authentication, shared
hash-map/transcript consistency, and staged execution support remain separate
requirements; no public error or protocol/model premise changed.

\paragraph{Nested-layout clean-completion obstruction.}
Retaining the entire nested query set forces every first-layer ancestor
$i+4096k$, for $i<3076$ and $k<8$, to agree with the unchanged initial fold.
All these positions avoid both the periodic holes and the five extra
rejected positions. The next three actual folding equations consequently
force the fourth-layer value
\[
 T_4[i]=\frac{b_1(z_i^2+b_2)(z_i^4+b_3)}{4z_i^7},
 \qquad z_i=\mathit{fri\_canonical\_domain\_at}(1)[i],\quad i<3076.
\]
The corresponding fourth-layer domain point is $z_i^8$. This argument
allows arbitrary hidden-entry values and arbitrary later tables; it does
not assume that the displayed add-one recipe is the only construction, or
assume prefix causality to obtain the obstruction.

For any polynomial $p$ of degree at most 63, equality with $p(z_i^8)$
implies that $z_i$ is a root of
\[
 R(X)=4X^7p(X^8)-b_1(X^2+b_2)(X^4+b_3).
\]
This polynomial has degree at most $7+8\cdot63=511$. If $b_1\ne0$, its
degree-six numerator cannot equal the first term: that term is either zero
or has degree at least seven. Thus $R$ is nonzero, independently of $b_2$
and $b_3$. Distinctness of the canonical domain bounds agreement on the
3,076 forced coordinates by 511 and yields
\[
 \operatorname{dist}(T_4,\mathrm{RS}_{\le63})\ge3076-511=2565>536.
\]
This is distance from every allowed codeword, not Hamming weight relative
to zero. The theorem \code{nested_displayed_layer4_distance} also establishes
this lower bound directly for the displayed fourth table, without assuming
full retention, a terminal layer, or clean completion.

The existing-builder theorem \code{nested_builder_retention_forces_charge}
therefore derives, from fixed-initial-table identification and full retention,
either $b_1=0$ or the existing balanced online bad-challenge event. Its
constant terminal layer is still supplied by the unchanged builder. In
\code{nested_builder_clean_retention_impossible}, an actual supported
challenge outcome, extension to the final hash map, and avoidance of the
existing initial-target charge exclude $b_1=0$. No adversary restriction,
locale assumption, or public soundness premise is added. These are explicit
conditional diagnostic bridges, not a construction of an authenticated run.

The classification is \emph{not realizable as clean full retention for the
identified nested layout and fixed initial table}. It does not rule out
verifier acceptance on already charged branches, other layouts, or every
near-envelope set. In particular, the difference $2565-536=2029$ is a
code-distance gap, not a proved number of additional rejected queries.
Losing some queries can break the forced ancestor cones; the next audit
quantifies that partial-retention tradeoff. No universal
residual cap, useful acceptance-probability estimate, security profile, or
general STARK/FRI limitation is claimed.

\paragraph{Nested partial-retention tradeoff.}
Let $I$ be the identified 64,294-element nested query set and let $S$ be
any subset of the actual global agreement intersection. Define
$m=|I\setminus S|$ and let $d$ be the fourth-layer distance from
degree-at-most-63 code. For each $i<3076$, a cone remains forced whenever,
for each $k<8$, at least one of $i+4096k$ and $i+4096(k+8)$ belongs to $S$.
All sixteen indices belong to $I$. Both lifts enforce the same relevant
fold equations under the unchanged modulo projections. Thus breaking a
cone requires losing both lifts of an ancestor; different cones have
disjoint ancestors. If $g$ cones remain forced, Isabelle proves
\[
 2(3076-g)\le m,\qquad g\le511+d,
 \qquad\text{hence}\qquad m+2d\ge5130\quad(b_1\ne0).
\]
The second inequality applies the same nonzero comparison polynomial to
an arbitrary subset of still-forced coordinates, not to a full-prefix
assumption. A close fourth layer has $d\le536$, so this branch requires
$m\ge4058$.

If layer four is far, the unchanged builder's constant terminal layer and
the complement of the balanced online event force a strict boundary at
some $i\in\{4,\ldots,9\}$. For successor length $n$, every missing
successor index excludes all its modulo lifts in $I$. The proof explicitly
counts the subgrid $j+4096k$, $j<2048$, $k<16$, which lies inside $I$ and
has exactly $16(2048/n)$ points per residue when $n$ divides 2048. The
following table makes this lower-count relaxation visible:
\begin{center}
\begin{tabular}{rrrrr}
\toprule
$n$ & Strict holes & Proved subgrid fiber & Audited full minimum & Proved exclusions\\
\midrule
2048 & 39 & 16 & 16 & 624\\
1024 & 20 & 32 & 48 & 640\\
512 & 11 & 64 & 112 & 704\\
256 & 6 & 128 & 240 & 768\\
128 & 4 & 256 & 496 & 1024\\
64 & 3 & 512 & 992 & 1536\\
\bottomrule
\end{tabular}
\end{center}
The full minimum column is a direct finite enumeration of the unchanged
nested set, not an additional Isabelle theorem. Its stronger layerwise
products would be $624,960,1232,1440,1984,2976$; both lists have minimum
624. Omitting the other nested queries therefore does not change this
minimum, although neither optimality of the resulting bound nor its
attainment is proved. No uniform-modulo sampling premise is used.

The two branches give the existing-builder result
\[
 |I\setminus S|\ge624,\qquad
 |I\cap\mathit{GlobalAgreement}|\le64294-624=63670.
\]
The hypotheses fix the 65,536-point evaluation domain, identify the fixed
initial table and ten roots/challenges,
require $b_1\ne0$, and exclude the existing balanced online event.
\code{nested_builder_executed_clean_overlap_bound} derives the nonzero
condition from an actual supported challenge outcome, hash-map extension,
and the existing initial-target complement. These are diagnostic conditions,
not new locale or public soundness premises, and they do not assume an
accepting execution into a validity predicate.

Crucially, this is primarily an \emph{intersection bound}. The separate
\code{nested_builder_clean_first_agreement_bound} bounds the entire global
agreement set by 63,670 only when its actual first agreement is contained
in the identified nested agreement set, with scale 64 and powers 2 fixing
the unchanged query space to 65,472. Queries outside $I$ cannot otherwise
be silently discarded. The result permits arbitrary hidden entries and
later tables, but does not exclude other layouts, other initial tables,
or already charged accepting branches. It proves neither realizability of
the remaining cap nor a new public probability/security estimate.

\paragraph{Layout-independent fixed-initial-table exclusions.}
The next bounded audit removes the first-agreement placement and cardinality
hypotheses altogether, keeping the fixed initial table, scale 64, powers 2,
65,536 evaluation points, ten roots/challenges, and the existing clean balanced
builder branch. Let $Q=[0,65472)$ and let $S$ be any subset of the actual
global agreement intersection. The forced-cone theorem already quantifies
over arbitrary $S$: its nested set is a counted subgrid, not an assumption
that the first agreement has that layout. Since $I\subseteq Q$, the same
distance argument gives
\[
  |Q\setminus S|+2d\ge5130\quad(b_1\ne0).
\]
Thus a close fourth layer still costs at least 4,058 whole-space queries.
If it is far, clean constant completion forces a strict later boundary.
For its successor length $n$ and disagreement set $D\subseteq[0,n)$,
the exact count is
\[
 |\{q\in Q:q\bmod n\in D\}|=
 \left\lfloor\frac{65472}{n}\right\rfloor|D|
       +|D\cap[0,65472\bmod n)|.
\]
Keeping the quotient term gives the formally checked exclusion vector
\[
 (1209,1260,1397,1530,2044,3069)
\]
for $n=(2048,1024,512,256,128,64)$ respectively. No independence, uniform modulo
distribution, or sum of overlapping boundary exclusions is assumed.
Both branches therefore imply
\[
 |Q\setminus S|\ge1209,\qquad
 |\mathit{GlobalAgreement}|\le65472-1209=64263.
\]
The theorem \code{layout_builder_clean_global_bound} uses the existing
builder's conceptual tables and constant terminal. Its execution-facing
corollary \code{layout_builder_executed_clean_global_bound} obtains
$b_1\ne0$ from a supported actual challenge, hash-map extension, and the
existing initial-target complement. It adds no public or locale premise
and does not assume an accepting execution.

The broader ceiling is only 31 below the 64,294 diagnostic benchmark and
is weaker than the layout-specific nested ceiling 63,670. In particular,
the nested 624 additional losses have not been generalized to all layouts.
The balanced clean-event guard is not identified with the optimized
schedule's guard, and the public parameter error is unchanged.

The remaining limitation is explicit. At the counting level,
$D=[2009,2048)$ has exactly 39 residues and exactly 1,209 lifts in $Q$;
\code{layout_late_fiber_counting_sharp} checks this in Isabelle. Thus the
discarded remainder alone cannot raise that local minimum without further
restrictions on $D$. This abstract set is not a construction of compatible
FRI tables, an authenticated execution, or clean challenges. Simultaneous
folding, code-distance, commitment, and transcript constraints could still
force a larger loss. Neither attainment nor optimality of 64,263 is proved.
Near-ceiling clean executions, other initial tables, and already charged
accepting branches remain unresolved. This is a fixed-family necessary
condition, not a new probability/security profile or a general STARK/FRI
limitation.

\paragraph{Bounded near-saturation feasibility at layer four to five.}
The next audit tests simultaneous folding compatibility rather than optimizing
the schedule again. It keeps the fixed initial table and existing clean
balanced branch. Write $m=|Q\setminus S|$, let $D\subseteq[0,2048)$ be the
fourth-to-fifth transition disagreements, and let $d_5$ be the fifth-layer
distance from the degree-31 code. Every global survivor avoids every lift of
$D$. Define the retained paired cones by
\[
 G(S)=\{i<1028:\ \forall k<16,\quad
     i+2048k\in S\ \lor\ i+2048(k+16)\in S\}.
\]
Each such cone retains both fourth-layer ancestors $i$ and $i+2048$ from
the already proved 3,076-position rational patch. A lost whole residue class
costs at least 31 global queries but can destroy at most one of these cones.
Breaking a further cone outside those classes requires a missing pair of
queries. Counting these disjoint losses proves, for \emph{every} eligible
placement of $D$,
\[
 2056+29|D|\le m+2|G(S)|.
\]
This does not assume an interval-shaped agreement or disagreement set.

The earlier actual fold equations give more than a cardinality envelope.
Put $z=\mathit{Domain}_1[i]$, $w=h^{4096}$, and
\[
 P(X)=b_1(X^2+b_2)(X^4+b_3).
\]
On a retained paired cone the fifth-layer value is
\[
 \frac{N(z)}{8w^7z^{15}},\qquad
 N(X)=w^7(X^8+b_4)P(X)+(X^8-b_4)P(wX).
\]
The existing domain facts give $w^8=-1$, hence $w\ne0,-1$; the existing
initial-target complement supplies $b_1\ne0$. Consequently $N$ has degree
14 for every $b_4$. For any degree-31 codeword polynomial $p$, agreement
implies that $z$ is a root of
\[
 8w^7X^{15}p(X^{16})-N(X).
\]
This polynomial is nonzero and has degree at most 511. Distinct canonical
points therefore give $|G(S)|\le511+d_5$. Combining the two statements gives
the joint feasibility invariant
\[
 \boxed{m+2d_5\ge1034+29|D|.}
\]
This is a necessary consequence of the actual global folding equations, not
an arbitrary conceptual-table realization assumption.

If layer four is far and layer five is close, the unchanged clean challenge
condition forces $|D|\ge39$, while closeness means $d_5\le230$. Thus
\[
 m\ge1705,\qquad |S|\le63767.
\]
If layer four is already close, the earlier 4,058-exclusion theorem is
stronger. Hence \code{near_builder_close5_global_bound} covers every
clean completion with a close fifth layer, not only a hypothesized first
boundary. Its execution-facing corollary derives the nonzero challenge from
the existing executed-challenge support and initial-target complement.
At the old $m=1209$ counting minimum the new invariant would require
$d_5\ge478$. Even $|S|>64000$, or $m\le1471$, requires $d_5\ge347$ when
$|D|\ge39$. Both contradict $d_5\le230$. Therefore the proposed near-64k
fourth-to-fifth geometry is \emph{not realizable on this clean fixed-table
branch}, before any additional authentication or transcript restrictions.

\paragraph{What this does not settle.}
The number 63,767 is \emph{not} a bound for all clean completions. Above
64,000 survivors, layer five must remain far. Clean constant completion then
forces a later strict boundary, and the exact whole-space fiber theorem
reduces the necessary alternatives to
\[
\begin{array}{c|c|l}
 \text{transition} & \text{disagreements} & \text{necessary exclusion inequality}\\
 5\to6 & 20\le|D|\le23 & 63|D|+|D\cap[0,960)|\le m\\
 6\to7 & |D|=11 & 1397+|D\cap[0,448)|\le m.
\end{array}
\]
Strict boundaries at seven or later already cost at least 1,530 exclusions.
The theorem \code{near_builder_high_survival_obligations} establishes
these necessary alternatives; it does not prove them jointly feasible. In
the second case there are at most 74 missing queries outside the minimal
1,397 disagreement-fiber losses (fewer after the remainder term). In the
first case the corresponding allowance is at most 211 when there are 20
disagreements, and shrinks with additional disagreements.

Simply repeating the same uniform-patch argument does not settle those
cases: requiring all four fourth-layer ancestors $i+1024k$, $k<4$, to lie
in the 3,076-position patch leaves only $i<4$, as formally checked by
\code{near_next_uniform_patch}. A root bound of order 511 on this
four-position patch is ineffective. The precise remaining bridge is joint
compatibility of the \emph{piecewise} values forced by the fixed initial
table with the shared low-degree decoder, later disagreements, and clean
challenges. It calls for further proof work, not a new axiom or model change.

The bounded audit therefore establishes proof slack at the formerly weakest
transition, but neither a substantially smaller whole-family maximum nor a
near-64k accepting execution. No positive acceptance probability, attractive
security profile, universal initial-table statement, or general STARK/FRI
limitation follows. The subsequent joint investigation addresses the two
remaining transitions.

\paragraph{Joint later-transition feasibility: the whole-word gate.}
The follow-up retains the same fixed initial table, ten challenges, balanced
clean guard, nonzero second challenge, and exact query space. It does not
allow the rational pieces to choose independent decoders. Define $U_j$ as
the unmodified recursive fold of the fixed initial table under the same
challenge list. The auxiliary definition \code{later_full_fold} uses exactly
\code{fri_folded_received} and the existing canonical domains; it changes no
execution experiment. Write $D_j$ for the canonical domain and $V_j$ for
the actual conditioned table.

For a successor length $n$ dividing $32768$, let $G_n(S)$ contain those
$i<n$ for which every ancestor $a<32768$ with $a\bmod n=i$ retains at least
one of $a$ and $a+32768$ in $S$. Induction through the actual global query
agreement equations proves
\[
 i\in G_{|D_j|}(S)\quad\Longrightarrow\quad V_j(i)=U_j(i).
\]
This uses all earlier folds, not an arbitrary conceptual-table realization.
The initial case is the stipulated fixed table; each successor uses both
parent cones and the existing folded-received equation.

Let $B\subseteq[0,n)$ be the actual late disagreement set, $h=|B|$,
$K=65536/n$, and $m=|[0,65472)\setminus S|$. Counting entire disagreement
fibers and broken ancestor pairs outside them yields
\[
 2n+Kh\le m+64+2|G_n(S)|+2h.
\]
The 64 term explicitly allows the positions outside the actual query space.
The proof does not charge every absent ancestor pair twice within the
truncated sampler. It applies to all eligible placements of $B$, not just
an interval example. Comparison with a closest codeword for $V_j$ gives
\[
 \operatorname{dist}_d(U_j)+|G_n(S)|
 \le \operatorname{dist}_d(V_j)+n,
\]
and hence the joint necessary inequality
\[
 2\operatorname{dist}_d(U_j)+(K-2)h
 \le 2\operatorname{dist}_d(V_j)+m+64.
\]

The strict-boundary argument now retains the successor's closeness witness.
Above 64,000 survivors, $m\le1471$, layer five is far, and boundaries seven
or later are already excluded. The two remaining cases therefore require
\[
\begin{array}{c|c|c|c}
 \text{boundary}&(n,K,h\text{ minimum})&\operatorname{dist}_d(V_j)
   &\text{necessary full-word distance}\\
 5\to6&(1024,64,20)&\operatorname{dist}_{15}(V_6)\le96
   &\operatorname{dist}_{15}(U_6)\le243\\
 6\to7&(512,128,11)&\operatorname{dist}_{7}(V_7)\le38
   &\operatorname{dist}_{7}(U_7)\le112.
\end{array}
\]
For example the respective doubled upper bounds are
$192+1471+64-62\cdot20=487$ and
$76+1471+64-126\cdot11=225$. Thus one codeword must agree with at least
781 of all 1024 sixth-word positions, or at least 400 of all 512 seventh-word
positions. The theorem \code{later_builder_high_survival_full_fold_obstruction}
proves this disjunction. Its executed-challenge corollary derives nonzero
from the same existing initial-target complement, rather than assuming a
new public soundness premise.

A further formal reduction, \code{later_distance_error_locator}, constructs
the product of linear factors at the error positions. Proximity at radius
$e$ to a degree-$d$ codeword therefore implies polynomials $E\ne0$ and $Q$
such that
\[
 \deg E\le e,\qquad\deg Q\le d+e,\qquad
 E(x_i)U_j(i)=Q(x_i)\quad\text{at every domain position }i.
\]
The specialized gate has degree limits $(243,258)$ at layer six or
$(112,119)$ at layer seven. It is necessary only: no converse decoding or
execution-completeness claim is made. For fixed field and challenges these
are homogeneous linear systems in the coefficients of one shared pair
$(E,Q)$; challenge-uniform exclusion of nonzero solutions remains unproved.

\paragraph{Bounded diagnostics and unresolved exceptions.}
The untrusted diagnostic checked exact modular folds for 48 challenge tuples
over primes $998244353$ and $2013265921$, using respective generators and shifts
$3$ and $31$. It checked root orders and the shift's exclusion from the evaluation
subgroup. In each field, eight structured tuples included sparse and zero later
challenges, and sixteen additional tuples used a fixed pseudorandom seed. All
had nonzero $b_1$.
These arithmetic choices are not asserted to instantiate every assumption
of the soundness locale or satisfy the clean guard.

The script computes power-sum syndromes
$s_k=\sum_i U_j(i)x_i^k$, $1\le k<n-d$. A degree-$d$ word contributes zero
to these sums on the multiplicative coset. A word with at most $e$ errors
therefore gives a sum of at most $e$ geometric sequences. This explains
the diagnostic recurrence test and the independent Hankel-minor checks;
the algorithm and numerical certificates are not kernel-verified here.
Self-tests compare the transform with direct summation and recover known
injected error counts; selected array-fold values are also checked against
a separate recursive even/odd calculation. Every tested tuple has syndrome linear complexity
504 at layer six and 252 at layer seven, exceeding the required 243 and
112. Selected 244- and 113-dimensional leading Hankel minors for the
only-$b_1$ and all-ones tuples are independently nonzero in both fields.
These are bounded exact-arithmetic diagnostics, not random-oracle success
probabilities, clean executions, concrete profiles, or a uniform theorem.

\emph{Classification remains unresolved.} The formal development now
identifies a joint algebraic gate over every piece of the actual forced
word. The bounded tests support searching for an obstruction, but do not
exclude exceptional challenges or establish a substantially smaller true
maximum. No near-64k execution is supported by a construction. The strongest
proved whole-family fixed-table ceiling remains 64,263, with the earlier
63,767 cap when layer five is close. The next bounded question is whether
the whole-word gate can hold under the existing clean guard, uniformly in
the model's field and challenge parameters, or whether an exceptional
solution exists. Only a positive algebraic candidate would justify testing
the remaining clean-chain, authentication, and transcript bridges. No new
assumption, protocol change, or public security interpretation is introduced.

\paragraph{Shared-polynomial representation and decoder-degree gate.}
The next bounded development identifies the actual full folds with shared
recursive polynomial numerators. For $j\in\{6,7\}$, put
$M=2^{j-1}$, $n=32768/M$, and $z_i=(h^i\mathit{shift})^2$.
The target domain point is $z_i^M$. A recursive numerator $N_a$ of degree
at most $M-1$ represents the full-word value as $N_a(z_i)/z_i^{M-1}$.
Its anchor is $a=0,4,403$ according as $i$ lies in $[0,4)$, $[4,403)$,
or $[403,n)$. These intervals describe fixed orbit patterns, not a
restriction on disagreement placements. Every numerator is generated by
the same folding challenges, and the representation is proved from
\code{later\_full\_fold}, including the actual sibling and domain indices.

For any one decoder polynomial $p$, agreement is exactly the root equation
\[
 z_i^{M-1}p(z_i^M)-N_a(z_i)=0.
\]
When $d=\deg p>0$, the comparison polynomial is nonzero and has degree
at most $M-1+Md$. The resulting whole-word agreement bounds are
\[
\begin{aligned}
 A_6(p)&\le4+\min(399,31+32d)+\min(621,31+32d),\\
 A_7(p)&\le4+\min(399,63+64d)+\min(109,63+64d).
\end{aligned}
\]
Thus positive degrees at most 11 give at most 770 sixth-word agreements;
positive degrees at most 3 give at most 368 seventh-word agreements.
Constants are treated separately. The coefficient at degree 29 of the
sixth-word anchor-403 numerator is $-b_2/(32(h^{2048})^{14})$; the
coefficient at degree 61 of the seventh-word anchor-4 numerator is
$-b_2/(64(h^{1024})^{30})$. They are nonzero if $b_2\ne0$ and cannot be
cancelled by a constant decoder's degree-$M-1$ term. Root counting on
these pieces alone bounds total constant-decoder agreement by 434 and
176 respectively. These auxiliary bounds suffice for exclusion and are
not asserted optimal.

Using attainment of the existing code-distance minimum gives the formal
necessary disjunction
\[
 \bigl(\exists p:\ 12\le\deg p\le15,\ A_6(p)\ge781\bigr)
 \quad\lor\quad
 \bigl(\exists p:\ 4\le\deg p\le7,\ A_7(p)\ge400\bigr).
\]
Here $A_j(p)$ counts agreements with the actual full word and domain.
The auxiliary execution theorem combines this with the fixed-table
high-survival gate, using executed outcomes for $b_1,b_2$ extending into
the same final hash state. The existing no-initial-target branch supplies
their nonzeroness; no new public premise or adversary restriction is added.
The theorem is not a construction of a complete chronological execution.

A separate untrusted exact-arithmetic diagnostic fixes
$(b_1,\ldots,b_5)=(2,3,5,7,11)$ and reconstructs two degree-113 Hankel-minor
polynomials in $b_6$. Their monic gcd is 1 in each test field of prime order
998244353 and 2013265921, using generator/shift 3 and 31 respectively.
Exact interpolation checks and a polynomial Bezout identity therefore
exclude every $b_6$ for those particular field/prefix diagnostics.
This is not an Isabelle Hankel-rank theorem, a full model instance, or an
exclusion for all earlier prefixes. The sixth-layer alternative remains.
At the largest remaining decoder degrees, the comparison degree reaches
511; root counting alone cannot decide the gates. A uniform result must
exploit the jointly constrained coefficients and common decoder, rather
than treating the pieces independently. Joint realizability is still open;
the fixed-table ceilings 64,263 and conditional 63,767 are unchanged.

The bounded earlier-prefix follow-up adds a formal coordinate-affinity lemma
for the same recursive numerator: with all other challenges fixed,
$N(b)=N(0)+b(N(1)-N(0))$. Prefix independence of the children and linearity
of subsequent folds justify this identity. It is not joint affinity in all
challenges; the shared products remain essential.

Exact but untrusted diagnostics then exclude all parameter values in 22
single-coordinate fibers through $(2,3,5,7,11,13)$, counting both layers and
both test fields. They also test simultaneous curves $b_r=c_ra$ for
$c=(1,1,1,1,1,1)$ and $c=(2,3,5,7,11,13)$. At layer six the two minor
polynomials have degree 1,220 and gcd 1; at layer seven they have degree 678
and gcd $a^4$, in each field and on each curve. Thus all nonzero parameters
on these eight curves are excluded. At $a=0$, the seventh word really has
403 agreements with the constant decoder $1/64$ and 109 disagreements,
but those zero challenges violate the existing no-initial-target branch.
This control is not an authenticated execution or a survivor construction.
The certificates use exact interpolation and checked Bezout identities,
with independent determinant checks. They remain diagnostic, not Isabelle
syndrome/rank theorems. Arbitrary jointly exceptional prefixes and other
allowed fields remain unresolved; no stronger survivor ceiling follows.

\paragraph{Exact joint-syndrome bridge and projection audit.}
For the actual canonical domain $x_i$, define
$S_k=\sum_{i<n}U(i)x_i^k$. Exact multiplicative order and the existing coset
formula prove the positive power-sum identities; moment zero is not assumed
to vanish. From a nonzero locator $E$ of degree at most $r$ and
$E(x_i)U(i)=Q(x_i)$ with $\deg Q\le d+r$, Isabelle proves
\[
 \sum_{t=0}^{r}\operatorname{coeff}(E,t)S_{k+t}=0,
 \qquad 1\le k\le n-d-r-1.
\]
The resulting nonzero coefficient vector lies in the kernel of the full
$765\times244$ system for $(n,d,r)=(1024,15,243)$ or the $392\times113$
system for $(512,7,112)$. Their entries use moments $1,\ldots,1008$ and
$1,\ldots,504$, respectively. The actual full words and syndromes inherit
separate coordinate affinity from the shared numerator. This does not make
the joint challenge equations linear. The auxiliary executed high-survival
gate transports to this necessary kernel disjunction with unchanged premises.

The proof also retains a stronger locator form $Q=E p$, $\deg p\le d$.
On a distinct domain it is equivalent to at most $r$ disagreements with one
polynomial decoder. Forgetting this factorization is a genuine projection:
the control word $U(i)=1/x_i$ has locator $E=X,Q=1$ and passes the two
respective kernel conditions, but agreement with $p$ is bounded by the
at most $d+1$ roots of the nonzero polynomial $Xp-1$. Thus it has at most 16 or 8
agreements, far below the required 781 or 400. This control is not proved
to arise from the fixed initial FRI word or an authenticated execution.
Separately, the syndrome control $(1,0,0,1,0)$ has two vanishing adjacent
$2\times2$ minors but no nonzero kernel in its full $4\times2$ system.

These controls prevent reversing necessary conditions; they do not refute
the sufficient route of proving uniform full column rank for the actual
word family. No uniform joint certificate for that full family or compatible
factorized decoder is yet obtained. Both layer alternatives, the existing field family, and the actual
clean/executed challenge conditions still require coverage. The fixed-table
64,263 and conditional 63,767 ceilings remain unchanged. No verifier, protocol,
transcript, budget, adversary, or public soundness premise is changed.

\paragraph{Bounded joint-certificate feasibility.}
Write the actual separately affine syndrome matrix as
$H(b)=\sum_a w_a(b)H_a$, retaining the squarefree monomial weights.
A fixed row combination $L$ annihilating all $H_c$ for $c\ne a$
gives $w_a(b)LH_a v=0$ for every kernel vector $v$. When $w_a(b)\ne0$,
this yields a necessary locator constraint. The formal argument proves this rule,
rejection of the positive-moment kernel from explicit isolation and left-inverse
identities, and an explicit free-coordinate reconstruction identity for
soundly restricting the remaining locator space before another pass.
These are certificate implications, not new rank assumptions in a public
soundness theorem or assertions that the full family has such certificates.

In each of the same two diagnostic fields, let $b_1,b_2$ range independently
over all nonzero values, fixing $(b_3,b_4,b_5)=(5,7,11)$ and also $b_6=13$
for the seventh word. Exact, untrusted coefficient isolation rejects both
kernel alternatives on these entire two-dimensional families. For the sixth
word the first pass gives 132 independent constraints on 244 coordinates;
the second gives rank 112 on the remaining 112-dimensional space. For the
seventh word one pass gives rank 113. Nullspace products and independent
square-minor condensation checks support these arithmetic certificates.
They have not been reconstructed as numerical Isabelle theorems and do not
cover arbitrary other challenges, roots, cosets, fields, or clean executions.

For the full five- or six-variable family in either diagnostic field, two
disjoint groups of four coefficient matrices each have columns spanning
the entire row-coordinate space, of dimension 765 or 392. Every target
monomial is absent from a spanning group. Consequently a constant row
combination annihilating all other coefficients must be zero; Isabelle
proves the generic explicit-certificate version of this obstruction.
This prevents this particular single-coefficient method from starting on
the full family. It does not imply that their weighted sum has a nonzero
kernel, that a decoder exists, or that a near-64k execution is realizable.
Challenge-dependent combinations or grouped-coefficient certificates
remain open routes. The 64,263 and conditional 63,767 ceilings are unchanged;
the diagnostic evidence is insufficient to decide the true extremal scale.
The existing model, clean guards, factorized-decoder requirement, and public
soundness premises are unchanged.

\paragraph{Challenge-dependent polynomial-certificate pilot.}
The bounded follow-up keeps $b_1,b_2$ free and takes $x=b_3-5$ as one
additional symbolic coordinate, fixing $(b_4,b_5,b_6)=(7,11,13)$ (the last
coordinate is irrelevant at layer six). Each of the four coefficient
matrices is now affine in $x$. For a target coefficient, stack the other
three transposes to obtain a pencil $A+xB$ with dimensions $732\times765$
or $339\times392$. A constant change of coordinates, computed at $x=0$,
reduces its polynomial kernel equations to
\[
 q+x(Dq+Ey)=0,\qquad
 q_0=0,\quad q_{j+1}=-Dq_j-Ey_j,\quad Dq_d+Ey_d=0.
\]
Thus a degree-$d$ solution is governed by
$[(-D)^dE,\ldots,E](y_0,\ldots,y_d)^T=0$.
This avoids a large unstructured polynomial system. Only constant matrices
are inverted when constructing the isolators; their identities are checked coefficient by
coefficient for all $x$, without assuming the pencil is nonsingular there.

In both diagnostic fields, the first nonzero polynomial isolators appear
at degree 22 for layer six and degree 6 for layer seven. Degree 23 and
degree 7 respectively suffice to span each target's reference nullspace,
of dimension 33 or 53. At layer seven the polynomial constraint stack has
full reference rank 113. Two square minors, reconstructed under the
algebraic degree envelope $113\cdot8=904$, have degrees 833 and 904 and
gcd 1 in each field. A checked Bezout identity, together with independent
off-grid determinant checks, therefore removes every $b_3$ exception for
this diagnostic family. It excludes the seventh-layer kernel for all
nonzero $b_1,b_2$ and arbitrary $b_3$ at the displayed fixed remaining
coordinates, field, root, and coset. It is not a universal-prefix theorem.

At the end of that pilot, layer six remained essential: its degree-23
isolators yielded a degree-at-most-24
constraint stack with reference rank 132 on 244 columns. The remaining
112 locator coordinates required a second parameter-dependent reduction.
The earlier second pass at fixed $b_3=5$ cannot be applied uniformly without
proving coverage of its varying parametrization and handling exceptional
values. That pilot did not construct the bridge, and the degree cost is
a cost of this certificate method, not a demonstrated protocol or
adaptive-adversary limitation.

The formal argument proves the exact affine-row coefficient equations, transfer
of scaled polynomial certificates to the condition $\Delta(x)=0$, a finite-root
bound when $\Delta$ is known nonzero, and elimination of simultaneous roots
by an explicit Bezout identity. A scalar Hankel control shows why a nonzero
reference value alone does not remove exceptions. These are generic
certificate implications; the numerical matrices, minors, and field/model
instances are not reconstructed in Isabelle. The finite-root bound is not
a new public probability estimate. Both fixed-table ceilings and all
protocol/model/public-premise boundaries remain unchanged.

\paragraph{Division-free U6 lifted reduction.}
The next bounded audit resolves this restricted diagnostic branch without
constructing a challenge-dependent inverse. Write the actual sixth-word
syndrome matrix as
$H=H_0(x)+b_1H_1(x)+b_2H_2(x)+b_1b_2H_3(x)$, with each $H_i$ affine
in $x=b_3-5$ and $(b_4,b_5)=(7,11)$ fixed. If $Hv=0$, the six blocks
\[
 w=(v,\ b_2v,\ b_1v,\ b_1b_2v,\ b_1^2v,\ b_1^2b_2v)
\]
satisfy the original equation and its $b_1$ multiple. These form an affine
pencil $(A+xB)w=0$ with 1,530 rows and 1,464 columns. The original $v$ is
one block, so no nonzero locator is lost, including at zero challenges.
Allowing arbitrary lifted vectors only enlarges this necessary test; a
lifted solution would not by itself construct a decoder or execution.

In each diagnostic field, a constant 1,464-row selection $I$ makes $B_I$
invertible. For the remaining 66 rows $J$, define
$T=B_I^{-1}A_I$ and $C=A_J-B_JT$. A kernel of the pencil then satisfies
$(T+x\mathrm{Id})w=0$ and $Cw=0$. Consequently $CT^kw=0$ for every $k$.
The stack $C,CT,\ldots,CT^{23}$ has dimensions $1584\times1464$ and full
column rank in both fields. Independent square-minor checks give residues
453771633 and 1451302670 respectively. This excludes every $x$, without
dividing by a challenge-dependent pivot or leaving an exceptional set.
Thus the U6 diagnostic excludes all $b_1,b_2,b_3$, including zeros, at these
fixed remaining coordinates, roots, cosets, and fields. Combined with the
prior U7 result, both alternatives are excluded for nonzero $b_1,b_2$ and
arbitrary $b_3$ with $(b_4,b_5,b_6)=(7,11,13)$ fixed.

The formal argument proves scaled reconstruction and transport without assuming
the scale nonzero, monomial lifting, constant pencil normalization, and the explicit finite
observation-certificate implication. The numerical matrices, inverse and
rank certificates, and field/model instances are not reconstructed in
Isabelle. These are therefore still untrusted numerical instances of a
proved algebraic method, not a universal high-survival exclusion. The
64,263 and conditional 63,767 ceilings, all clean-execution guards, and the
three public soundness premises are unchanged.

The subsequent formal argument proves a denominator-cleared
observation-certificate implication for polynomial pencils. It uses scaled
identities throughout, without assuming a
challenge-dependent denominator nonzero. A nonzero kernel therefore forces
a root of the certificate polynomial, including at zero or singular
parameter values; a nonzero reference evaluation bounds the number of
exceptional values by its degree. A row-weighted determinant degree lemma
supports later instantiation. These are generic certificate implications,
not new public soundness premises. No concrete U7/U6 rank witness,
exceptional-set count, or improved survivor ceiling has thereby been
certified.

The subsequent modular cyclic-certificate layer proves that bounded integer
checks of $ZM-MZ=FQ^{\mathsf T}$, $MU=F$, and $Mc=e_0$ imply a trivial
kernel over a finite commutative ring with the matching characteristic.
The separate control theory reconstructs three tiny positive certificates
and six rejection controls through the inference kernel, without evaluation
oracles, and derives concrete residue-ring kernel corollaries. No primality
or protocol-field instance is asserted by those controls. Identification of
the actual U7/U6 coefficient and observation arrays, their large certificate
checks, and the field/domain bridges remain open. These generic checking
results change no survivor ceiling or public soundness premise.

\subsection{The strict selected-residual endpoint}

\noindent\textbf{Historical status.} This is the conclusion of the archived selected-boundary route. It is separate from the current MCA, weighted, first-root and optimized-cutoff composition.

The pre-existing strict selected-residual endpoint is unchanged by these audits.
It charges the already existing online
bad-challenge family as one branch and, on the complementary branch, replaces
the selected successor agreement cap at margin $s$ by the strict cap at
$\operatorname{Suc}(s)$.  The exact modulo sampler then maps this tighter
cardinality through the same quotient/remainder envelope and adaptive product.
For the ceiling-balanced scale-64 diagnostic at $C=54$, this changes the public
endpoint's residual envelope from 64,322 to 64,320; the separate optimized-
schedule audit is the one whose ceiling changes from 64,296 to 64,294.
The trace and composition transports, acceptance classification, staged
budgets, and three public soundness premises are unchanged.  Isabelle proves
the strict residual, combined residual, and final tightened parameter errors
are no larger than the corresponding earlier selected-boundary errors.  The
public endpoints are
\code{ro\_absorb\_stark\_soundness\_balanced\_strict\_selected\_tightened\_nonempty}
and
\code{ro\_absorb\_stark\_soundness\_balanced\_strict\_selected\_tightened\_security\_target\_nonempty}.

\subsection{Symbolic nontriviality and arithmetic feasibility}

\noindent\textbf{Historical status.} The archived feasibility corollaries show symbolic conditions sufficient for an error below one; they do not instantiate a cryptographic profile for that route.

The branch predicates
\code{ro_absorb_checked_staged_conditioned_explicit_*_parameter_nontrivial}
allocate at most $1/12$ to each of the eleven expanded components.  Their sum
is at most $11/12<1$.  Isabelle first proves the corresponding explicit
parameter error below one and then the following public corollary:

\begin{lstlisting}
corollary ro_absorb_stark_soundness_conditioned_explicit_nontrivial:
  assumes false_statement: "\<not> exists_valid_trace"
    and wf: "staged_budget_wellformed budgets"
    and controlled: "staged_adversary_controlled budgets A"
    and parameter_conditions:
      "ro_absorb_checked_staged_conditioned_explicit_parameter_nontrivial
        budgets"
  shows
    "ro_absorb_checked_staged_adversary_acceptance_probability A < 1"
\end{lstlisting}

The stronger predicate
\code{ro_absorb_checked_staged_conditioned_explicit_arithmetic_feasible}
replaces the new sampled-component probability clauses by primitive natural
inequalities.  If (q) is the staged transcript hash-query budget, it requires
the field size to dominate the initial target, (q(7q+3)) challenge terms,
the remainder-aware residual raw-fiber bound plus $q^2+7q+2$, and the
branch-specific padding fiber plus $5q+2$, all with the $1/12$ allocation.
It also requires strict challenge-round, residual-envelope, and padding-index
inequalities.  Isabelle proves these relations imply the component predicate
and proves the adaptive challenge, residual-query, and padding sets are not the
full spaces under the same conditions.

This is symbolic attainability, not a cryptographic-scale instantiation: no
concrete large-field parameter set is claimed to satisfy the arithmetic
predicate.

\subsection{Small-field nonvacuity and modulo-bias witness}

\noindent\textbf{Historical status.} The GF(5) witness and its full soundness-locale interpretation are archived-only. Current general modulo-fiber bounds do not depend on restoring this auxiliary example.

The auxiliary example served as a consistency check for the soundness
assumptions. It uses GF(5), trace length two, scaling factor one, one trace
power, and one query round, and installs a global interpretation of the full \code{soundness}
locale.  Its single constraint is the constant-one polynomial at root zero,
with an aligned list constraint, so Isabelle proves
\code{modulo_bias_false_statement}: no valid trace exists.

This instance also exposes the modulo sampler's bias.  The query space is
exactly \(\{0,1\}\), while the five raw field encodings split into fibers of
cardinalities three and two.  The repaired singleton envelope is therefore
three, and both exact fibers are proved below it.  This is a deliberately small
logical and combinatorial witness, not a secure parameter choice.  In
particular, it does not establish the explicit arithmetic-feasibility predicate
needed by the active nontrivial soundness corollary.

\subsection{Earlier staged ledger, allocation maximization and 128-bit certificate}

\noindent\textbf{Historical status.} This is the older staged-budget interpretation and its own allocation-uniform certificate. Its allowance cap is a total declared staged allowance, not the original FS producer cap; it is not a separately supported current security profile.

The reference square workload now has an exact bridge from the existing staged
budget definitions to the complete weighted alternative. Let $R$ be the
positive repetition count, $F$ the field cardinality, and partition the
attacker's declared staged budget as $Q=s+j+o$. Here $s$ contains the trace
root and first trace-FRI root stages, $j$ contains all remaining pre-query
attacker stages, and $o$ contains all query-opening budgets. No allocation is
set to zero. The proved budget envelopes are
\[
 q=Q+50+535R,\qquad v=50+1429R,\qquad n=q+v=Q+100+1964R.
\]
They bound builder, verifier-replay and complete-experiment hash work,
respectively; they are not exact execution call counts or distinct-key counts.
The per-query checked opening transcript bound is 534, with 250 entries for
each of the two ten-layer FRI chains. The verifier retains its original
maximum-path-depth envelope, giving 1429 hash calls per query in its budget.

For the prefix charges, put
\[
 h=s+j+50,\quad u=j+o+48+535R,\quad t=o+535R,\quad b=2s+6.
\]
The source identities prove $s+2\leq q$ and $u=q-s-2$, so natural subtraction
cannot silently truncate a charge. With these abbreviations the complete
common numerator is
\begin{align*}
 C={}&n^2+(u+v)b+(t+v)(1+2h)+v(20+2q)\\
     &+2q(16384+7q+2)+t(20+2h)\\
     &+q^2+q+ub+q(6q+3).
\end{align*}
The final line retains all nested alpha charges; no apparently repeated
collision or target charge is canceled.

For the reference radii $r_T=r_C=10581$, define the exact modulo envelope
\[
 p(B)=\frac{\lfloor F/65472\rfloor B+
                 \min(B,F\bmod65472)}{F}.
\]
The real value of the source weighted alternative is proved equal to
\[
 E_{\mathrm{weighted}}=
 \frac C F+\frac n F+
 n\bigl(p(54953)^R+2p(54954)^R\bigr)+\frac{5n(n-1)}{2F}.
\]
The final summand is the existing weighted connection charge. The modulo
quotient and remainder have not been replaced by ideal uniform index sampling.
The workload satisfies the existing weighted-regime gate, so the unchanged
public minimum is at most this expression; it need not be equal to it.
The acceptance-probability theorem still uses precisely its original false
statement, budget wellformedness and adversary-control premises.

These identities are proved for every wellformed staged allocation in the
existing soundness locale with the reference workload metadata. The previously
certified 192-bit field/workload supplies that metadata. This is an accounting
bridge, not a proof of the diagnostic script. The following layer independently
proves allocation maximization and at-most-budget control. No named bit-security
level or attainability of the residual envelopes follows from these results.

\subsection{Allocation-uniform, at-most-budget soundness}

Now let $Q$ be an upper limit on the total declared staged allowance, with
$s+j+o\leq Q$. Write $E(F,R,s,j,o)$ for the complete weighted ledger above and
$C(R,s,j,o)$ for its common numerator. The exact identity
\[
 C(R,s+1,j,o)+C(R,s,j,o+1)
   =2C(R,s,j+1,o)+8j+401
\]
shows that one of the two one-unit transfers out of a positive middle
coordinate increases the numerator. Induction eliminates that coordinate
for maximization without forbidding any adversary allocation. At fixed total
$Q$, the remaining identity is
\[
 C(R,s,0,Q-s)+8s^2=C(R,0,0,Q)+(8Q+159+9996R)s.
\]
The exact integer maximum is obtained at
\[
 k=\min\left(Q,\left\lfloor\frac{8Q+167+9996R}{16}\right\rfloor\right),
 \qquad U(F,R,Q)=E(F,R,k,0,Q-k).
\]
The clipped-quadratic proof handles zero budget, clipping and ties. All other
ledger terms depend only on the total allocation. Monotonicity in the opening
coordinate then permits padding unused allowance; for the connection charge,
$n\geq1$ follows from the constant offset 100 before multiplying the
nonnegative factors $n$ and $n-1$. Thus Isabelle proves
\[
 s+j+o\leq Q\quad\Longrightarrow\quad E(F,R,s,j,o)\leq U(F,R,Q),
\]
and monotonicity of $U$ in $Q$. No modulo-preimage approximation or cancellation
of exceptional charges is introduced.

\code{square_soundness_at_most_budget} composes the arithmetic result with
the unchanged public soundness theorem. Its displayed schema and geometry
specialize existing parameters; the cap bounds the total declared staged
allowance, not an exact execution call count. The original false-statement,
wellformedness and adversary-control requirements are retained. Padding is an
arithmetic comparison, not a modification of a budget record or execution.
The public minimum is bounded by $U$, not identified with it. This is an exact
maximum of the weighted ledger over aggregate allocations, not a proof that
an adversary attains the bound, a further numerical tightening, or a selected
bit-security profile. The following separate layer supplies the approved
numerical target.

\subsection{Certified staged 128-bit acceptance target}

The approved profile fixes the certified field cardinality $F$ above, trace
length $1024$, scale $64$, and exactly $R=640$ query repetitions. The total
declared staged adversary allowance is at most $Q=2^{20}$. Isabelle proves
the closed, source-connected inequality
\[
 U(F,640,2^{20})\leq 2^{-128}.
\]
No field/domain, budget-allocation maximum, or numerical-target premise is
left in this arithmetic certificate.

The exact maximizing aggregate allocation is $(924138,0,124438)$, with
$n=2305636$ and common numerator $C=55372749872256$. This is an arithmetic
witness, not a restriction on an actual adversary's middle or opening budgets.
The full original ledger is proved equal to
\[
 \frac{68662639825042}{F}
 +2305636\bigl(p(54953)^{640}+2p(54954)^{640}\bigr),
\]
where $p$ retains its exact quotient/remainder definition. Thus every common,
initial-target and connection charge is accounted for before simplification.

Standard HOL arithmetic proves both modulo bases are at most $27/32$ and
\[
 (27/32)^{128}\leq 2^{-31},\qquad
 (27/32)^{640}\leq 2^{-155}.
\]
The sampling contribution is at most $2^{-132}$, while all non-sampling terms
together are at most $2^{-144}$. Their sum is at most $2^{-128}$. The dyadic
estimate sacrifices numerical margin to make certification small; it neither
replaces the original sampler nor changes the general public error expression.
No external numerical artifact or arithmetic oracle is trusted.

The theorem \code{square_192_128_soundness} combines the closed inequality
with the proved field and workload interpretation and the original
acceptance theorem. Its
four displayed requirements are false statement, staged-budget wellformedness,
adversary control, and the approved total allowance cap. The original public
theorem's three premises are unchanged, and the cap only specifies this
profile's scope. Every permitted allocation with total at most $2^{20}$ is
covered. The companion \code{square_192_128_incorrect_endpoint} uses the proved
valid-trace implication to give the same guarantee for $z\ne a^{2^{1023}}$.
The previously proved same-domain true and false instances remain available.

This is a scoped false-statement acceptance upper bound in the existing
adaptive random-oracle experiment, not an unrestricted 128-bit security
work-factor claim. The cap concerns declared staged allowances, not only calls
observed on one path. It is not automatically a theorem for larger budgets or
all $R\geq640$, since non-sampling charges also grow with $R$. No optimality,
near-$64000$ realizability, quantum-oracle security, zero knowledge, knowledge
extraction or deployment-ready implementation follows. The generator remains
chosen from proved existence. The earlier diagnostic value near $2^{-138.96}$
is not advertised as a certified target.

\subsection{Earlier conventional-query ledger with the replay allocation}

\noindent\textbf{Historical status.} The generic replay allocation also appears in the current proof. The named replay-soundness carrier, concrete ledger and diagnostic query thresholds described here belong to the older archived accounting route, not the current first-root numerical endpoint.

The allowance layer allocates $Q$ to every declared callback position of the
actual compiler. For $T=\lceil\log_2(\mathit{clength})\rceil$,
$C=\lceil\log_2(\mathit{maxDegree}+1)\rceil$ and $R$ query repetitions, its
wellformed total is exactly
\[
 f(Q)=(T+C+R+4)Q.
\]
Every supported fixed program in the oracle-independent private mixture is
controlled by this allocation. Composing with the unchanged public theorem
gives \code{fs_replay_soundness}: the only attacker requirements are its ordinary
query cap and the original false-statement condition, within the existing
soundness locale. No validity, freshness, successful-run or completed-map
hypothesis remains. The field-valued lazy oracle and exact modulo sampler are
unchanged. Every lookup, including a cached repeat, counts. This covers the
existing terminating finite-support classical syntax, not quantum queries,
countably supported unbounded private sampling or a bit-hash conversion theorem.

The public statement and workload are fixed before the experiment; the attacker
outputs a proof transcript, not a statement selected from oracle answers.
Uniformity over fixed endpoints does not itself cover adaptively chosen
statements. That would require a separate execution/selection bridge.

This is a sufficient allocation, not its proved minimum or an exact full-driver
call count. Shorter composition depth, early failure and programs using fewer
than $Q$ queries may cost less. The compiler really reruns the same fixed
program; the factor is not a theorem that every possible embedding requires
replay. Acceptance preservation is through a mixture of entire staged records,
not equivalence of the attacker classes at equal budgets.

For the square geometry, $T=C=10$ and the actual early/middle/opening allocation
is $(s,j,o)=(2Q,22Q,RQ)$. Early includes both the trace root and the first trace
FRI root. The builder and whole-experiment hash envelopes respectively become
$(R+24)Q+50+535R$ and $(R+24)Q+100+1964R$; their protocol/verifier overhead is
not adversarial FS queries. Retaining this allocation gives a no-larger bound
than maximizing the existing ledger over all allocations of $(R+24)Q$.
The existing allocation maximum already respected its joint total constraint;
no independent-per-stage maximization error is being repaired.

The concrete interpretation is discharged in
\code{square_192_fs_replay_incorrect_endpoint}. For $z\ne a^{2^{1023}}$ and
\code{fs_query_bound Q P}, its FS acceptance probability is at most
\[
 \mathit{square\_weighted\_ledger}(F,640,2Q,22Q,640Q),
 \qquad F=\mathit{field\_cardinality\_192}.
\]
The source-connected HOL polynomial identity rewrites this ledger as
\[
 \frac{21737376Q^2+33425487832Q+17262492169912}{2F}
 +(664Q+1257060)(p_{54953}^{640}+2p_{54954}^{640}),
\]
where $p_B=((F\mathbin{\mathrm{div}}65472)B+\min(B,F\bmod65472))/F$.
All existing common, initial-target and connection charges remain; sampling
retains exponent $640$, not $640-Q$.

The following evaluations are external exact-rational diagnostics with rounded
logarithmic displays, not Isabelle-certified numerical instances:
\begin{center}
\begin{tabular}{rr}
\toprule
Conventional query cap & $-\log_2(\text{sufficient ledger})$ \\
\midrule
$2^{20}$ & $127.618626$ \\
$2^{32}$ & $103.796643$ \\
$2^{40}$ & $87.796689$ \\
$2^{64}$ & $39.796689$ \\
\bottomrule
\end{tabular}
\end{center}
The largest integer satisfying this ledger's $2^{-128}$ inequality is $910767$;
the adjacent value $910768$ fails. The coarse compiled-uniform threshold is
$880537=\lfloor584676569/664\rfloor$, so retaining the allocation adds $30230$
queries, or about $3.433\%$. The prior staged-only threshold $584676569$ is
unchanged, not conventional $Q$. These thresholds concern sufficient bounds,
not actual acceptance or the public minimum over all routes. Failure of the
inequality is not an attack. This older ledger does not certify conventional
$Q=2^{20}$ at 128 bits; the refined certificate below does.

The remaining quadratic dependence includes generic collision envelopes,
nested union bounds and terminal-map drift, all inflated by this replay
construction. Generic ROM collision probabilities do not prove that every
charged event or coefficient is unavoidable for false-statement acceptance.
A bounded lower-cost embedding investigation is a separate decision; free
cached reads or oracle-dependent postselection of a staged record are not valid
shortcuts. The refined conventional-query certificate below is a separate
result; the honest-producer allowance obstruction remains unchanged.

\subsection{Historical conventional-FS 128-bit target}

\noindent\textbf{Historical status.} The closed 128-bit certificate and named endpoint in this subsection are archived. Their diagnostics are not substituted for the current 137-bit result.

The new source equality \code{square_fs_accounted_error_ledger} identifies
the weighted alternative with
\[
 \begin{aligned}
 E(Q)={}&\frac{18651111Q^2+21752664601Q+12056674456762}{2F}\\
 &+(Q+914610)\bigl(p(54953)^{640}+2p(54954)^{640}\bigr),
 \end{aligned}
\]
where $F=\mathit{field\_cardinality\_192}$ and the unchanged exact modulo base is
\[
 p(B)=\frac{\lfloor F/65472\rfloor B+\min(B,F\bmod65472)}{F}.
\]
The theorem \code{square_fs_accounted_soundness} connects conventional-FS
acceptance to this ledger. The equality is not for the public minimum or
actual acceptance, and no classification branch has been discarded.

The closed numerical certificate proves both bases lie between zero and
$215/256$. The small block inequality $(215/256)^4\leq 1/2$ then gives
$p(B)^{640}\leq 2^{-160}$. At $Q=2^{20}$ the remaining exact inequality is
\[
 \frac{20529934794151974074}{2F}
 +1963186\cdot 3\cdot 2^{-160}\ \leq\ 2^{-128}.
\]
All these steps are proved in HOL; external arithmetic output is not trusted.
The final theorem \code{square_192_fs_128_incorrect_endpoint} states
\[
 z\ne a^{2^{1023}}\ \wedge\ \mathit{fs\_query\_bound}(2^{20},P)
 \quad\Longrightarrow\quad
 \Pr[\mathit{FSVerify}(P)\text{ accepts}]\leq 2^{-128}.
\]
The concrete interpretation uses the certified field/domain, trace length
1024, scale 64, two shifted powers and 640 query repetitions. These are exactly
two displayed premises; all locale, control and staged-budget obligations
are discharged. The conventional cap counts original adversarial oracle calls,
including cached repetitions. Verifier calls are accounted separately. The
producer may adapt to earlier answers and use private randomness, within the
existing terminating finite-support classical field-valued ROM semantics.

For comparison, the preserved external exact-rational diagnostics give:
\begin{center}
\begin{tabular}{lr}
\toprule
Conventional query cap & Approximate $-\log_2 E(Q)$ \\
\midrule
$2^{20}$ & $128.015387$ \\
$2^{32}$ & $104.017605$ \\
$2^{40}$ & $88.017605$ \\
$2^{64}$ & $40.017605$ \\
\bottomrule
\end{tabular}
\end{center}
The diagnostic threshold is $1054187$, with the next integer failing,
versus $910767$ for the older conventional ledger. These decimal evaluations
and threshold claims are not Isabelle-certified here. The earlier staged-only
threshold $584676569$ does not have the conventional-query interpretation.
Only the displayed $Q=2^{20}$ target is certified: no larger-query 128-bit
guarantee, optimality or attack claim follows. The public statement is fixed
before the experiment; this is not a QROM, adaptively chosen-statement,
bit-hash implementation, zero-knowledge or knowledge-extraction theorem.
The operational allocation still totals $664Q$. MCA and nested-alpha
costs dominate this historical diagnostic ledger. Honest-producer resource
accounting is a separate issue.

\subsection{Historical conventional-FS 135-bit target}

\noindent\textbf{Historical status.} The 135-bit endpoint and refined ledger here are archived. A shared 135-labelled arithmetic prerequisite remains in the current 137-bit proof; that does not make this older endpoint a current supported publication result.

The same field, square workload and 640 repetitions now admit a certified
false-acceptance bound $2^{-135}$ at the same conventional cap $Q=2^{20}$.
The theorem \code{square_192_fs_135_incorrect_endpoint} retains exactly the
incorrect-endpoint premise $z\ne a^{2^{1023}}$ and \code{fs_query_bound (2^20) P}.
The fixed-statement, classical field-valued ROM and adaptive/private attacker
model are unchanged. This is not an unrestricted 135-bit work-factor claim.

\code{square_fs_refined_error_ledger} proves equality between the complete
refined weighted alternative and
\[
\begin{aligned}
 E_{\mathrm{ref}}(Q)
  &=\frac{133493Q^2+2618301025Q+8773058386762}{2F}\\
  &\quad +(Q+914610)\bigl(p(54953)^{640}+2p(54954)^{640}\bigr),
\end{aligned}
\]
where $p$ retains the original modulo quotient and remainder. This is neither
an equality for actual acceptance nor for the complete public minimum.
Relative to the preceding accounted ledger, the removed numerator over $F$ is
$21s^2+32776s-7b^2-32776b$, with $s=664Q+342450$ and $b=Q+342450$.
All saved-prefix exceptions remain; the operational compiler allowance is 664Q.

At $Q=2^{20}$ the complete numerator over $2F$ is 149531366400678730.
Both exact modulo bases are at most $68/81$, and the kernel-checked blocks
\[
 (68/81)^{16}\le125/2048,\qquad
 (125/2048)^{40}\le\frac{2}{5\cdot2^{160}}
\]
give the closed total inequality
\[
 \frac{149531366400678730}{2F}
 +1963186\cdot3\cdot\frac{2}{5\cdot2^{160}}\le2^{-135}.
\]
No external arithmetic oracle or new public assumption is used.

Separate exact-rational diagnostics give approximate negative binary logarithms
135.035203, 111.143941, 95.143959 and 47.143959 at $Q=2^{20},2^{32},2^{40},2^{64}$.
The sufficient ledger's largest diagnostic 128-bit query cap is 12,442,508;
the next integer fails. These decimals and the adjacent-query threshold are
not additional HOL certificates, attack thresholds, or optimization over all
public alternatives. At $Q=2^{20}$ the retained composition-prefix and query-phase
target terms account for about 85.47\% of the error. Their necessity remains
unproved. The older certificates and tables are preserved, and no QROM,
adaptive-statement, bit-hash implementation, zero-knowledge or extraction claim
is added.

\subsection{Earlier honest-producer allowance obstruction}

\noindent\textbf{Historical status.} The following semantic obstruction concerns the specific archived honest callbacks and all-state staged-control interface. Its source owners are not retained in the current normal build; the argument is not presented as a newly proved current resource lower bound.

For the unchanged \code{square_ro_honest_callbacks}, the valid workload
$a=-1$, $z=1$ rules out uniform fit within the profile's declared allowance:
\[
 Q_{\mathrm{declared}}
 \;\geq\;\sum_i q_{\mathrm{opening},i}
 \;\geq\;640\cdot2048
 \;=\;1{,}310{,}720
 \;>\;2^{20}=1{,}048{,}576.
\]
This is a semantic lower bound, not an inference from the conservative
$21{,}364{,}931$-call sufficient envelope.

Polynomial uniqueness identifies the actual trace interpolant of
$-1,1,1,\ldots$ as
\[
 f(X)=1-\frac{2}{1024}\sum_{i<1024}X^i.
\]
For the existing full-order generator $\gamma$ and trace generator $g$,
at $x_i=\gamma g^i$, $i<1024$, the geometric identity gives
\[
 (x_i-1)(f(x_i)-1)=-\frac{2}{1024}(\gamma^{1024}-1)\ne0.
\]
The nonvanishing and exact-order conditions follow from the existing
certified field and domain. Equal values of $f$ force equal $x_i$, hence
equal indices. Domain alignment $g=h^{64}$ therefore gives 1024 distinct
actual table values at indices $64i$. These are tree-construction indices,
not additional allowed verifier query locations.

Choose consistent answers $F(\mathrm{MerkleLeaf}(x))=x$ and zero on every
other tagged input. The shared fresh/known support lemmas and a hash-only
induction supply a supported outcome of the \emph{whole} unchanged opening
callback, from any empty oracle map at raw argument zero, consistent with
$F$. This is a trajectory of the existing oracle, not oracle programming or
a new assumption. Repeated keys retain their answers. Existing callback
outcome facts supply an authenticated initial tree in the final map.
Balanced-tree induction extracts 1024 distinct selected leaf keys and 1024
distinct bottom-parent keys. The latter are distinct by their first
coordinates, and the two constructors give disjoint key sets. No
injectivity of internal outputs or sibling values is required.

The original \code{controlled_ro_program_new_key_count_bound} bounds new
keys by every semantic control allowance. Each opening slot must control
all states and raw arguments, and the callback ignores the slot index.
The same empty-map witness therefore forces each separate slot allowance
to be at least 2048. This does not assert that 640 empty-map openings occur
in one honest execution. Known-prefix replay or a denotationally equal
rewrite cannot remove the existing all-state obligation.

The witness allows zero challenges and collisions. It is not a clean
near-64k execution, a typical runtime estimate, or an attack on soundness.
The result rules out uniform fit for this producer across valid inputs,
not fit for every input or every possible producer. It is not an optimum,
and a 25\% allowance increase is not proved sufficient. Endpoint-only honest
acceptance and the earlier staged false-statement $2^{-128}$ theorem remain valid
as stated. No callback, verifier, exact sampler, transcript, budget,
adversary, protocol/model or public premise changes. A different producer
or revised allowance/profile remains a separate decision.

\subsection{Historical typed core and staged interfaces}

\noindent\textbf{Historical status.} These verbatim earlier interfaces retain their old probability-scalar notation and additional selected-route premises. They are historical statements, not the current native theorem signatures.

The generic core completeness statement is:

\begin{lstlisting}
lemma completeness:
  assumes "honest_trace_valid"
  shows "wp_event exec (\<lambda>a. Option.is_none a) init_state \<le> (0::prob)"
\end{lstlisting}

The baseline all-round public soundness statement is:

\begin{lstlisting}
theorem ro_absorb_stark_soundness_conditioned_explicit:
  assumes false_statement: "\<not> exists_valid_trace"
    and wf: "staged_budget_wellformed budgets"
    and controlled: "staged_adversary_controlled budgets A"
  shows
    "ro_absorb_checked_staged_adversary_acceptance_probability A
      \<le>
        ro_absorb_checked_staged_conditioned_explicit_parameter_error budgets"
\end{lstlisting}

The parameter error and its branch condition are:

\begin{lstlisting}
definition ro_absorb_checked_staged_conditioned_explicit_parameter_error
  :: "staged_budgets \<Rightarrow> prob"
where
  "ro_absorb_checked_staged_conditioned_explicit_parameter_error budgets =
    (if ceil_log clength = 0
     then
       ro_absorb_checked_staged_conditioned_explicit_zero_round_parameter_error
         budgets
     else
       ro_absorb_checked_staged_conditioned_explicit_nonempty_parameter_error
         budgets)"

definition ro_absorb_checked_staged_conditioned_explicit_arithmetic_feasible
where
  "ro_absorb_checked_staged_conditioned_explicit_arithmetic_feasible
      budgets \<longleftrightarrow>
    (if ceil_log clength = 0
     then
       ro_absorb_checked_staged_conditioned_explicit_zero_round_arithmetic_feasible
         budgets
     else
       ro_absorb_checked_staged_conditioned_explicit_nonempty_arithmetic_feasible
         budgets)"
\end{lstlisting}

The resulting strict soundness statement is:

\begin{lstlisting}
corollary ro_absorb_stark_soundness_conditioned_explicit_arithmetic_nontrivial:
  assumes false_statement: "\<not> exists_valid_trace"
    and wf: "staged_budget_wellformed budgets"
    and controlled: "staged_adversary_controlled budgets A"
    and feasible:
      "ro_absorb_checked_staged_conditioned_explicit_arithmetic_feasible
        budgets"
  shows
    "ro_absorb_checked_staged_adversary_acceptance_probability A < 1"
\end{lstlisting}

For genuine FRI rounds, the strict selected staged route supplies this
positive-query endpoint under its additional profile premises:

\begin{lstlisting}
lemma ro_absorb_stark_soundness_balanced_strict_selected_tightened_nonempty:
  assumes false_statement: "\<not> exists_valid_trace"
    and wf: "staged_budget_wellformed budgets"
    and controlled: "staged_adversary_controlled budgets A"
    and nonempty: "0 < ceil_log clength"
    and eval_power: "clength * scale = 2 ^ N"
    and trace_rounds_fit: "Suc (ceil_log clength) \<le> N"
    and composition_rounds_fit:
      "Suc (ceil_log (Suc maxDegree)) \<le> N"
    and trace_exact:
      "fri_balanced_exact_list_cap
        (clength - 1) (ceil_log clength) C 2"
    and composition_exact:
      "\<And>d. d \<le> maxDegree \<Longrightarrow>
        fri_balanced_exact_list_cap d (ceil_log (Suc d)) C 2"
  shows
    "ro_absorb_checked_staged_adversary_acceptance_probability A
      \<le>
      ro_absorb_checked_staged_balanced_strict_selected_tightened_nonempty_parameter_error
        C budgets"
\end{lstlisting}

\begin{sloppypar}
The corresponding generic target composition has the same model and profile
facts and adds the arithmetic target. The theorem is
\code{ro\_absorb\_stark\_soundness\_balanced\_strict\_selected\_tightened\_security\_target\_nonempty}.
\end{sloppypar}

\subsection{Reading the earlier staged and selected-residual bounds}

\noindent\textbf{Historical status.} This reference guide concerns the earlier staged families. The main report and current theorem appendix give the current acceptance guarantees and native interfaces.

This appendix explains the earlier staged and selected-residual routes. The
current FS interfaces are listed in Appendix~\ref{app:statements}; their
conventional producer cap is related to staged control by the replay bridge,
not by equating the two allowances.

The baseline endpoint chooses between explicit zero- and nonzero-round definitions
according to whether \code{ceil_log clength = 0}.  This split is logically
necessary because the zero-round experiment has no trace-FRI challenge chain,
although composition FRI may still have positive round count.  After the joint
sampled term is expanded, each branch uses an eleven-component allocation.
Every component is a function of protocol parameters and staged budgets; none
depends on the adversary \code{A} or retains an event probability or semantic
bad-set cardinality.  The robust endpoint is deliberately nonempty-round and
has twelve components because it displays trace and composition residual
products separately.  Its comparable error uses a minimum with the preceding
balanced error, while a separate theorem proves the strict selected error no
larger than the earlier selected-boundary error.  Thus the strict route cannot
weaken any preceding endpoint.

\subsubsection{Budget parameters}

The staged theorem is parameterized by a budget record.  The budget record
does not describe a particular implementation of a hash function.  It records
semantic upper bounds for the adversary and verifier random-oracle interaction,
including every \code{TranscriptAbsorb} query generated while constructing and
checking the transcript.  Collision terms count the complete combined query
budget, while target and relation terms bound fresh outputs landing in selected
bad fibers.  The endpoint instantiates these through the random-oracle and
staged-experiment budget layers.

The advantage of this representation is modularity.  A cryptographic
instantiation can later prove that a concrete hash model satisfies the budget
record.  The STARK proof itself does not need to be rewritten when the hash
model is refined, provided the same budget interface is preserved.

\subsubsection{Collision terms}

Hash collisions enter the proof in three main places.  First, an adversary may
cause a collision while constructing the transcript.  Second, verifier-side
Merkle checking may expose an inconsistency that is charged to a collision in
the shared hash map.  Third, some staged reductions classify unclean prefixes
by showing that the prefix already contains an output collision.  The final
bound includes these terms through the hash-collision budget values and the
staged bounds derived from them.

The formalization treats collision events as ordinary events in the
probabilistic monad.  This is useful because it allows collision reasoning to
compose with WP rules and union bounds.  It also prevents hidden assumptions:
when a Merkle inconsistency is charged to a collision, the corresponding lemma
states that implication explicitly.

\subsubsection{Target-hit terms}

Target-hit terms account for Fiat-Shamir challenges, Merkle prefixes, and
relation fibers.  A target is fixed from the relevant transcript/oracle prefix
before the fresh output is sampled.  For composition, alpha values may hide
constraint failure; for query sampling, indices may miss a disagreement; for
FRI, product and fiber bounds account for choices across the rounds.

The subtle point is that target sets may be state-dependent.  The adversary
chooses a transcript prefix, and that prefix may determine the set of bad
values.  The proof therefore establishes target bounds at the prefix point:
after the prefix has been fixed, but before the relevant random-oracle output
is sampled.  This is the reason for the dynamic target-budget layer in the
soundness development.

\subsubsection{Query terms}

The query term has the familiar low-degree-testing shape.  If two candidate
tables correspond to different low-degree polynomials, then they can agree on
only a bounded number of evaluation points.  A random query catches the
disagreement except when it lands in this agreement set.  The formal bound
therefore depends on a query agreement bound and the size of the query sample
space.

The current sampler is implemented through field sampling followed by a
projection modulo the query-range size.  If \(N\) is the field-encoding size,
\(q\) the query-space size, and \(B\) a bad set, the proof counts its raw
preimage exactly as
\(\lfloor N/q\rfloor |B|+|B\cap[0,N\bmod q)|\).  The closed query error uses
the envelope \(\lfloor N/q\rfloor |B|+\min(|B|,N\bmod q)\), divided by \(N\).
Thus the bound is valid for the modeled sampler without a divisibility premise.

\subsubsection{Actual-query products and relation fibers}

The FRI/query terms combine three kinds of proved quantities.  First, repeated
independent query-index choices contribute an exact raw-list preimage count
divided by \(|\mathbb{F}|^{\mathit{rounds}}\), or a closed product of
maximum-fiber envelopes with the same denominator.  This specializes to
\((1/|Q|)^{\mathit{rounds}}\) only in the locally proved uniform divisible
case.  Other field challenges contribute their ordinary inverse-field-size
powers.  Second, the positive-query refinement counts future challenge keys
already in the live oracle state. For the experiment's initially empty oracle
map, the live pre-query count is at most the declared pre-query allowance.
The declared pre-query allowance and opening-stage allowances sum to
\code{staged_attacker_query_budget budgets}. Substituting the count inequality
therefore gives the coarser public exponent
\code{rounds - staged_attacker_query_budget budgets}; actual counts need not
exhaust their allowances. Third,
fresh-query targets are bounded by the cardinality of a relation fiber and the
corresponding hash-relation budget.  This structure is preserved all the way to
the endpoint instead of being collapsed to an upper bound of one.

Trace and composition contributions remain distinct because their candidate
predicates and degree bounds differ.  Their evidence layer is parallel:
accepted executions supply only authenticated openings, challenge lists, local
fold equations, and final checks.  The closed actual-query and fiber estimates
bound precisely that verifier-observed evidence, without reconstructing
unopened tables.

\bibliographystyle{alpha}
\bibliography{references}
\end{document}